\documentclass[extra, mreferee]{gji}

\usepackage{epsfig}
\usepackage{amssymb, amsmath}
\usepackage{color}
\usepackage{ulem}
\def\rev#1{\textcolor{black}{#1}}
\def\revv#1{\textcolor{black}{#1}}

\title[Unified matrix-vector wave equation]
{Unified matrix-vector wave equation, reciprocity and representations}

\author[Wapenaar]
{\small Kees Wapenaar\\
  Department of Geoscience and Engineering, Delft University of Technology,\\ P.O. Box 5048, 2600 GA Delft, The Netherlands}
\pagerange{001--025}
\volume{999}
\pubyear{2017}

\begin{document}
\label{firstpage}

\def\bfc{{\bf c}}
\def\bfC{{\bf C}}
\def\bfu{{\bf u}}
\def\bfU{{\bf U}}
\def\R{{{\mbox{\boldmath $\rho$}}}}
\def\bfmu{\mbox{\boldmath $\mu$}}
\def\bfpartial{\nabla}
\def\bfpartiala{\bfpartial_{\!1}}
\def\bfpartialb{\bfpartial_{\!2}}
\def\bfk{{\bf k}}
\def\m{m}
\def\sigmaa{p}

\def\bq{{\bf q}}
\def\bd{{\bf d}}

\def\bqa{{\bf q}_1}
\def\bqb{{\bf q}_2}

\def\bx{{\bf x}}
\def\bxh{{\bf x}_{\rm H}}
\def\bfS{{\bf S}}
\def\bfT{{\bf T}}
\def\bfV{{\bf V}}

\def\bN{{\bf N}}
\def\bK{{\bf K}}
\def\bG{{\bf G}}
\def\bJ{{\bf J}}
\def\bO{{\bf O}}
\def\bI{{\bf I}}
\def\bA{{{\mbox{\boldmath ${\cal A}$}}}}
\def\bAb{{\,\,\bar{\mbox{\boldmath $\!\!{\cal A}$}}}}
\def\bU{{{\mbox{\boldmath ${\cal U}$}}}}
\def\bV{{{\mbox{\boldmath ${\cal V}$}}}}
\def\bW{{{\mbox{\boldmath ${\cal W}$}}}}
\def\bfO{\bO}
\def\bfAhat{\bA}
\def\bfahat{\bA}
\def\bbA{{\,\,\,\bar{\!\!\!\bA}}}
\def\tbA{{\,\,\,\tilde{\!\!\!\bA}}}

\def\u{u}
\def\k{k^{-1}}
\def\c{c}
\def\s{\vartheta}
\def\H{X}
\def\q{Y}
\def\e{c'}
\def\a{a}
\def\b{b}
\def\J{{j}}
\def\h{h^b}
\def\hh{{\bf h}^b}
\def\qf{q^f}
\def\chia{{\chi}}
\def\bftau{{{\mbox{\boldmath $\tau$}}}}

\def\i{i}
\def\bfdelta{{\bf i}}
\def\bfgamma{{\bf j}}

\def\setA{\mathbb{A}}
\def\setD{\mathbb{D}}
\def\setdD{{\partial\setD}}
\def\setddD{{\setdD_0\cup\setdD_1}}
\def\half{\begin{matrix}\frac{1}{2}\end{matrix}}
\def\quarter{\begin{matrix}\frac{1}{4}\end{matrix}}
\def\bfcalE{\mbox{\boldmath ${\cal E}$}}

\def\eep{{\cal E}_{33}''}
\def\eepp{{\cal E}_{33}'}

\def\beep{\bfcalE_3''}
\def\beepp{\bfcalE_3'}

\def\beepo{\bfcalE_1''}

\maketitle

\begin{summary}
{\small The matrix-vector wave equation is a compact first-order differential equation. It was originally used for the analysis of elastodynamic plane 
waves in laterally invariant media. It has been extended by various authors for laterally varying media.
Other authors derived a similar formalism for other wave phenomena. 
This paper starts with a unified formulation of the matrix-vector wave equation for 3D inhomogeneous, dissipative media. 
The wave vector, source vector and operator matrix are specified in the appendices for 
acoustic, quantum mechanical, electromagnetic, elastodynamic, poroelastodynamic, piezoelectric and seismoelectric waves. 
It is shown that the operator matrix obeys unified symmetry relations for all these wave phenomena. 
Next, unified matrix-vector reciprocity theorems of the convolution and correlation type are derived, utilising the symmetry properties of the operator matrix. 
These theorems formulate mathematical relations between two wave states in the same spatial domain.
A unified wave field representation is obtained by replacing one of the states in the convolution-type reciprocity theorem by a Green's state.  
By replacing both states in the correlation-type reciprocity theorem by Green's states, a unified representation of the homogeneous Green's matrix is obtained.
Applications of the unified reciprocity theorems and representations for forward and inverse wave problems are briefly indicated.
}
\end{summary}


\section{Introduction}
\revv{The basic equations for wave propagation in an inhomogeneous medium can be organised in a compact matrix-vector wave equation.
This equation expresses the vertical derivative of a wave vector in terms of an operator matrix acting on this wave vector.
This specific form of the wave equation is useful, for example, to evaluate wave problems in media of which the medium parameters vary more 
rapidly in the vertical direction than in the lateral directions. 
It is also particularly useful for situations in which the vertical direction is the preferred direction of wave propagation.
However, the theoretical treatment of the matrix-vector wave equation in this paper is not limited to these special situations.}
%

The  matrix-vector wave equation finds its roots in early work on the analysis of plane waves in laterally invariant media.
\cite{Thomson50JAP} introduced a matrix formalism for the analysis of elastodynamic plane waves propagating through a stratified solid medium.
\cite{Haskell53BSSA} used the same formalism to analyse the dispersion of surface waves in layered media.
\cite{Backus62JGR} used similar concepts to derive long-wave effective anisotropic parameters for stratified media. This approach has become known as Backus averaging \citep{Mavko2009Book}.
\cite{Gilbert66GEO} used the matrix-vector wave equation to derive so-called propagator matrices for elastodynamic wave problems in stratified media.
\cite{Woodhouse74GJR} extended the formalism for arbitrary anisotropic inhomogeneous media and used it for the study of surface waves in  laterally varying layered media.
\cite{Frasier70GEO}, \cite{Kennett78GJRAS}, \cite{Frazer89GJI} and \cite{Chapman94GJI} used the matrix-vector wave equation to derive symmetry properties 
of reflection and transmission responses of  laterally invariant media.
\cite{Haines88GJI}, \cite{Kennett90GJI}, \cite{Koketsu91GJI} and \cite{Takenaka93WM} exploited the symmetry properties of the matrix-vector wave equation to 
derive so-called propagation invariants for laterally varying layered media and used this for modelling of reflection and transmission responses of such media.
Using the same symmetry properties, \cite{Haines96JMP} and \cite{Wapenaar96JASA} derived  reciprocity theorems and representations for the acoustic wave vector. 

The matrix-vector wave equation has been used by many authors as the starting point for decomposition into 
coupled wave equations for downgoing and upgoing waves, 
for example for modelling in horizontally layered media  \citep{Kennett79GJRAS, Kennett81GJRAS}, 
wide-angle propagation in laterally variant media \citep{Fishman84JMP, Weston89JMP, Fishman92JMP}, 
reciprocity theorems for coupled down- and upward propagating waves \citep{Wapenaar96GJI1, Thomson2015GJI1, Thomson2015GJI2}, 
generalised Bremmer series representations for reflection data \citep{Corones75JMAA, Haines96JMP, Wapenaar96GJI2, Hoop96JMP}
and seismic interferometry \citep{Wapenaar2003GEO}.

This paper discusses the matrix-vector wave equation and its symmetry properties for a range of wave phenomena in a unified way (section \ref{secR}
and Appendices \ref{secAC} to \ref{secSE}).
The treatment builds on earlier systematic treatments of different wave phenomena by \cite{Auld73Book}, \cite{Ursin83GEO}, \cite{Kennett83Book}, \cite{Muller85JG},
\cite{Wapenaar89Book}, \cite{Hoop95Book, Hoop96GJI}, \cite{Gangi2000GJI}, \cite{Carcione2007Book} and \cite{Mittet2015GEO}.
The matrix-vector wave equation forms the basis for the derivation of unified
matrix-vector reciprocity theorems (section \ref{secRR})  and representations (section \ref{secRRR}), 
analogous to those for the acoustic wave vector \citep{Haines96JMP, Wapenaar96JASA}.

\section{The unified matrix-vector wave equation and its symmetry properties}\label{secR}

\subsection{The matrix-vector wave equation}

\begin{table}
\caption{Wave field vectors $\bqa$ and $\bqb$ for the different wave phenomena considered in this paper.
Labels \ref{secAC} to \ref{secSE} refer to the appendices in which these wave phenomena are discussed in more detail.}\label{table1}
\begin{center}
\begin{tabular}{lcc}
\hline\hline
& $\bqa$ & $\bqb$  \\
\hline\hline
\ref{secAC}: Acoustic & $p$ & $v_3$ \\
\hline
\ref{secQM}: Quantum mechanic & $\psi$ & $\frac{2\hbar}{m\i}\partial_3\psi$ \\
\hline
\ref{secEM}: Electromagnetic &${\bf E}_0=\begin{pmatrix}  E_1 \\  E_2\end{pmatrix}$& ${\bf H}_0=\begin{pmatrix}  H_2 \\ - H_1\end{pmatrix}$ \\
\hline
\ref{secED}: Elastodynamic & $-\bftau_3=-\begin{pmatrix}\tau_{13}\\ \tau_{23}\\ \tau_{33}\end{pmatrix}$ & ${\bf v}=\begin{pmatrix}v_1\\v_2\\v_3\end{pmatrix}$ \\
\hline
\ref{secPOE}: Poroelastodynamic & $ \begin{pmatrix}- \bftau_3^b \\ p^f  \end{pmatrix}$ & $ \begin{pmatrix} {\bf v}^s\\ \phi( v_3^f- v_3^s) \end{pmatrix}$ \\
\hline
\ref{secPE}: Piezoelectric & $\begin{pmatrix}- \bftau_3 \\ {\bf E}_0 \end{pmatrix}$ & $\begin{pmatrix} {\bf v}\\  {\bf H}_0 \end{pmatrix}$ \\
\hline
\ref{secSE}: Seismoelectric & $\begin{pmatrix}- \bftau_3^b \\ p^f \\ {\bf E}_0 \end{pmatrix}$ & $\begin{pmatrix} {\bf v}^s\\ \phi( v_3^f- v_3^s) \\ {\bf H}_0\end{pmatrix}$ \\
\hline
\hline
\end{tabular}
\end{center}\end{table}

The unified matrix-vector wave equation has the form
\begin{eqnarray}\label{eq2.1}
\partial_3\bq =\bA\,\bq +\bd.
\end{eqnarray}
Here $\bq$ is the wave field vector,  $\bd$ the source vector and  $\bfAhat$ the operator matrix. 
All quantities are defined in the space-frequency domain, hence $\bq=\bq(\bx,\omega)$, etc., where $\bx$ denotes the
Cartesian coordinate vector $(x_1,x_2,x_3)$ and $\omega$ the angular frequency. The positive $x_3$-axis is pointing downward.
Operator $\partial_3$ stands for the spatial differential operator $\partial/\partial x_3$.
The vectors and matrix in equation (\ref{eq2.1}) are partitioned as follows
\begin{eqnarray}\label{Aeq7mvbbprff}
\bq=\begin{pmatrix} \bqa \\ \bqb \end{pmatrix},\quad
\bd=\begin{pmatrix} \bd_1 \\ \bd_2 \end{pmatrix},\quad
\bfAhat= \begin{pmatrix}\bfAhat_{11}      & \bfAhat_{12} \\
                \bfAhat_{21} & \bfAhat_{22}    \end{pmatrix},
\end{eqnarray}
hence,
\begin{eqnarray}
\partial_3\bqa&=&\bfAhat_{11}\bqa+\bfAhat_{12}\bqb+\bd_1,\label{eqH3}\\
\partial_3\bqb&=&\bfAhat_{21}\bqa+\bfAhat_{22}\bqb+\bd_2.\label{eqH4}
\end{eqnarray}
The vectors $\bqa$ and $\bqb$ are specified in rows \ref{secAC} to \ref{secSE}
of Table \ref{table1} for the different wave phenomena considered in this paper. 
The wavefield quantities making up these vectors are defined in Appendices \ref{secAC} to \ref{secSE}. 
For acoustic and quantum mechanical waves  (rows \ref{secAC}  and \ref{secQM}),  $\bqa$ and $\bqb$ are scalars. 
For electromagnetic, elastodynamic and poroelastodynamic waves (rows \ref{secEM}  to \ref{secPOE}) they are 
$2\times 1$, $3\times 1$ and $4\times 1$ vectors, respectively
(superscripts $b$, $f$ and $s$ in row \ref{secPOE} stand for bulk, fluid and solid, respectively).
Rows \ref{secPE}  and \ref{secSE} represent coupled electromagnetic and (poro)elastodynamic waves. 
For piezoelectric waves (row \ref{secPE}), constitutive equations (\ref{eqdijk}) and (\ref{eqdklm}) account for the coupling.
For this situation the vectors $\bqa$ and $\bqb$ are combinations of those for electromagnetic and elastodynamic waves (rows \ref{secEM} and \ref{secED}).
For seismoelectric waves (row \ref{secSE}), constitutive equations (\ref{eqL1}) and (\ref{eqL2}) account for the coupling.
In this case the vectors $\bqa$ and $\bqb$ are combinations of those for electromagnetic and poroelastodynamic waves (rows \ref{secEM} and \ref{secPOE}).

In all cases, except for quantum mechanical waves, the vectors $\bqa$ and $\bqb$ are defined such that they consitute the power-flux density $\J$ in the $x_3$-direction via
\begin{eqnarray}\label{eq50ab}
\J=\quarter(\bqa^\dagger\bqb+\bqb^\dagger\bqa),
\end{eqnarray}
where the superscript $\dagger$ denotes transposition and complex conjugation.
For quantum mechanical waves, $\J$ represents the probability current density in the $x_3$-direction.
Vectors $\bd_1$ and $\bd_2$ and operator matrices $\bfAhat_{11}$, $\bfAhat_{12}$, $\bfAhat_{21}$ and $\bfAhat_{22}$ 
are defined in Appendices \ref{secAC} to \ref{secSE} for the different wave phenomena.
The operator matrices contain specific combinations of space-dependent medium parameters (or, for quantum mechanics, the potential $V$ and mass $m$) 
and spatial differential operators $\partial_1$ and $\partial_2$
(standing for $\partial/\partial x_1$ and $\partial/\partial x_2$, respectively).

\revv{Equation (\ref{eq2.1}), with the operator matrix specified in the appendices, 
may be used as a starting point for generalising many of the applications mentioned in the introduction
(analysis of surface waves, derivation of long-wave effective medium parameters,
derivation of propagator matrices, decomposition into downgoing and upgoing waves, modelling wide-angle propagation in laterally variant media, 
etc.). A discussion of these applications is beyond the scope of this paper. 
Here we focus on the symmetry of the operator matrix and its use in unified reciprocity theorems and representations.}

\subsection{Symmetry properties of the operator matrix}\label{secsymx}

We discuss the symmetry properties of the operator matrix.
First, consider a scalar operator ${\cal U}$, containing space-dependent parameters  and differential operators $\partial_1$ and $\partial_2$.
We introduce its \revv{transpose} ${\cal U}^t$ and adjoint ${\cal U}^\dagger$ via their integral properties
\begin{eqnarray}\label{eq90c}
\int_\setA ({\cal U}f)g\,{\rm d}^2\bxh=\int_\setA f({\cal U}^tg)\,{\rm d}^2\bxh
\end{eqnarray}
and
\begin{eqnarray}\label{eq90d}
\int_\setA ({\cal U}f)^*g\,{\rm d}^2\bxh=\int_\setA f^*({\cal U}^\dagger g)\,{\rm d}^2\bxh.
\end{eqnarray}
Here $\bxh$ is the horizontal coordinate vector $(x_1,x_2)$, superscript $*$ denotes complex conjugation,
$\setA$ denotes an infinite horizontal integration boundary at arbitrary depth $x_3$, and
$f=f(\bx)$ and $g=g(\bx)$ are space-dependent functions with sufficient decay along $\setA$ towards infinity.
Equation (\ref{eq90c}) implies 
\begin{eqnarray}
({\cal U}{\cal V}{\cal W})^t={\cal W}^t{\cal V}^t{\cal U}^t,\label{eqH6}
\end{eqnarray}
where also ${\cal V}$ and ${\cal W}$ are scalar operators.
Equations (\ref{eq90c}) and (\ref{eq90d})  imply  
\begin{eqnarray}
{\cal U}^\dagger=({\cal U}^t)^*.\label{eqH8}
\end{eqnarray}
For the special case that ${\cal U}=\partial_1$, equation  (\ref{eq90c}) implies
(via integration by parts) $\partial_1^t=-\partial_1$.  Similarly,  $\partial_2^t=-\partial_2$. Hence,
\begin{eqnarray}
\partial_\alpha^t=-\partial_\alpha,\label{eqH7}
\end{eqnarray}
where Greek subscripts take on the values 1 and 2. 
\revv{Using this property and equation (\ref{eqH6}), we find for example for the operator in equation (\ref{eq21}), 
$(\partial_\alpha b_{\alpha\beta}\partial_\beta)^t=\partial_\beta b_{\alpha\beta}\partial_\alpha$
(Einstein's summation convention applies to repeated subscripts). Since $b_{\alpha\beta}=b_{\beta\alpha}$, this implies
$(\partial_\alpha b_{\alpha\beta}\partial_\beta)^t=\partial_\beta b_{\beta\alpha}\partial_\alpha=\partial_\alpha b_{\alpha\beta}\partial_\beta$ and, using equation (\ref{eqH8}), 
$(\partial_\alpha b_{\alpha\beta}\partial_\beta)^\dagger=(\partial_\alpha b_{\alpha\beta}\partial_\beta)^*=\partial_\alpha b_{\alpha\beta}^*\partial_\beta$. 
}

Next, we consider an operator matrix $\bU$, \revv{of which the entries are operators} containing space-dependent parameters and  differential operators $\partial_1$ and $\partial_2$.
Analogous to equations (\ref{eq90c}) and (\ref{eq90d}), we introduce its \revv{transpose} $\bU^t$ and its adjoint $\bU^\dagger$ via
\begin{eqnarray}\label{eq90}
\int_\setA (\bU{\bf f})^t{\bf g}\,{\rm d}^2\bxh=\int_\setA {\bf f}^t(\bU^t{\bf g})\,{\rm d}^2\bxh\
\end{eqnarray}
and
\begin{eqnarray}\label{eq90b}
\int_\setA (\bU{\bf f})^\dagger{\bf g}\,{\rm d}^2\bxh=\int_\setA {\bf f}^\dagger(\bU^\dagger{\bf g})\,{\rm d}^2\bxh,
\end{eqnarray}
where ${\bf f}={\bf f}(\bx)$ and ${\bf g}={\bf g}(\bx)$ are space-dependent vector functions with sufficient decay  along $\setA$ towards infinity.
Equation (\ref{eq90}) implies that  $\bU^t$ involves transposition of the matrix and transposition of the operators contained in the matrix. 
\revv{For example, for a $2\times 2$ operator matrix $\bU$, we have
\begin{eqnarray}
\begin{pmatrix} {\cal U}_{11} & {\cal U}_{12} \\ {\cal U}_{21} & {\cal U}_{22}\end{pmatrix}^t=\begin{pmatrix} {\cal U}_{11}^t & {\cal U}_{21}^t \\ {\cal U}_{12}^t & {\cal U}_{22}^t\end{pmatrix}.
\end{eqnarray}
Equation (\ref{eq90}) implies}
\begin{eqnarray}
({\bU}{\bV}{\bW})^t={\bW}^t{\bV}^t{\bU}^t,\label{eqH11}
\end{eqnarray}
where also ${\bV}$ and ${\bW}$ are operator matrices.
Equations (\ref{eq90}) and (\ref{eq90b}) imply 
\begin{eqnarray}
\bU^\dagger=(\bU^t)^*.\label{eqH12}
\end{eqnarray}
Using equations (\ref{eqH6}), (\ref{eqH7}) and (\ref{eqH11}), it follows that operator matrices $\bfAhat_{11}$, $\bfAhat_{12}$, $\bfAhat_{21}$ and $\bfAhat_{22}$, 
defined in Appendices \ref{secAC} to \ref{secSE}, obey the following symmetry relations
\begin{eqnarray}
{\bA}_{11}^t &=& -{\bA}_{22},\label{eq31}  \\ 
{\bA}_{12}^t &=& {\bA}_{12}, \label{eq32}\\ 
{\bA}_{21}^t &=& {\bA}_{21}.\label{eq33}
\end{eqnarray}

In Appendices  \ref{secAC} to \ref{secSE} we define adjoint medium parameters (or, for quantum mechanics, an adjoint potential).
When a medium is dissipative, its adjoint is effectual, and vice versa \citep{Hoop87RS, Hoop88JASA, Wapenaar2001RS}.
A wave propagating through an effectual medium gains energy. 
Effectual media play a role in the reciprocity theorems and representations, discussed in sections \ref{secRR} and \ref{secRRR}. 
\revv{An adjoint medium parameter is denoted by an overbar. An operator with an overbar means that the medium parameters  contained in that operator are replaced by their adjoints.
For example, for the operator in equation (\ref{eq21}) we have 
$\overline{\frac{1}{\i\omega}\partial_\alpha b_{\alpha\beta}\partial_\beta}=\frac{1}{\i\omega}\partial_\alpha \bar b_{\alpha\beta}\partial_\beta$.
Since $\bar b_{\alpha\beta}=b_{\alpha\beta}^*$ this becomes 
$\overline{\frac{1}{\i\omega}\partial_\alpha b_{\alpha\beta}\partial_\beta}=\frac{1}{\i\omega}\partial_\alpha b_{\alpha\beta}^*\partial_\beta=-(\frac{1}{\i\omega}\partial_\alpha b_{\alpha\beta}\partial_\beta)^*$.}

For the operator matrices in Appendices  \ref{secAC} to \ref{secSE} in an adjoint medium we have
\begin{eqnarray}
\bbA_{11}&=&{\bA}_{11}^*,\label{eq941}\\
\bbA_{12}&=&-{\bA}_{12}^*,\label{eq942}\\
\bbA_{21}&=&-{\bA}_{21}^*,\label{eq943}\\
\bbA_{22}&=&{\bA}_{22}^*.\label{eq944}
\end{eqnarray}
Using equation (\ref{eqH12}), we find from equations (\ref{eq31})  $-$ (\ref{eq944})
\begin{eqnarray}
{\bA}_{11}^\dagger &=& -\bbA_{22},\label{eq41}  \\ 
{\bA}_{12}^\dagger &=&-\bbA_{12},\label{eq42}\\ 
{\bA}_{21}^\dagger &=& -\bbA_{21},\label{eq43a}\\
{\bA}_{22}^\dagger &=& -\bbA_{11}.\label{eq43}
\end{eqnarray}
From equations (\ref{eq31})  $-$ (\ref{eq43}), we find for the 
operator matrix  $\bA$ defined in equation (\ref{Aeq7mvbbprff})
\begin{eqnarray}
\bA^t\bN&=&-\bN\bA,\label{eqsym}\\
\bA^*\bJ&=&\bJ\bbA,\label{eqsymcon}\\
\bA^\dagger\bK&=&-\bK\bbA,\label{eqsymad}
\end{eqnarray}
with
\begin{eqnarray}\label{eq4.3}
{\bN}=\begin{pmatrix} {\bf O} & {\bf I} \\ -{\bf I} & {\bf O} \end{pmatrix},
\quad {\bJ}=\begin{pmatrix} {\bf I} & {\bf O} \\ {\bf O} & -{\bf I} \end{pmatrix},
\quad {\bK}=\begin{pmatrix} {\bf O} & {\bf I} \\ {\bf I} & {\bf O} \end{pmatrix},
\end{eqnarray}
where ${\bf O}$ and ${\bf I}$ are zero and identity matrices of appropriate size.

Symmetry relations in the wavenumber-frequency domain for the special case of a laterally invariant medium (or potential) are given in Appendix \ref{secsymkx}.

\section{Matrix-vector wave field reciprocity theorems}\label{secRR}

In wave theory, a reciprocity theorem formulates a mathematical relation between two 
states (wave fields, sources and medium parameters) 
in the same spatial domain.
An early reference for the acoustic reciprocity theorem is \cite{Rayleigh78Book}, who referred to it as Helmholtz's theorem.
\cite{Lorentz1895KNAW} formulated a reciprocity theorem for electromagnetic fields.
Early references for elastodynamic reciprocity theorems are \cite{Knopoff59GEO} and \cite{Hoop66ASR}. 
\cite{Auld79WM} and \cite{Pride96JASA} formulated reciprocity theorems for piezoelectric and seismoelectric waves, respectively. 
Comprehensive overviews of the history
of reciprocity theorems  and their applications are given by \cite{Fokkema93Book}, \cite{Hoop95Book} and \cite{Achenbach2003Book}.

\begin{figure}
\vspace{0cm}
\centerline{\epsfysize=9 cm \epsfbox{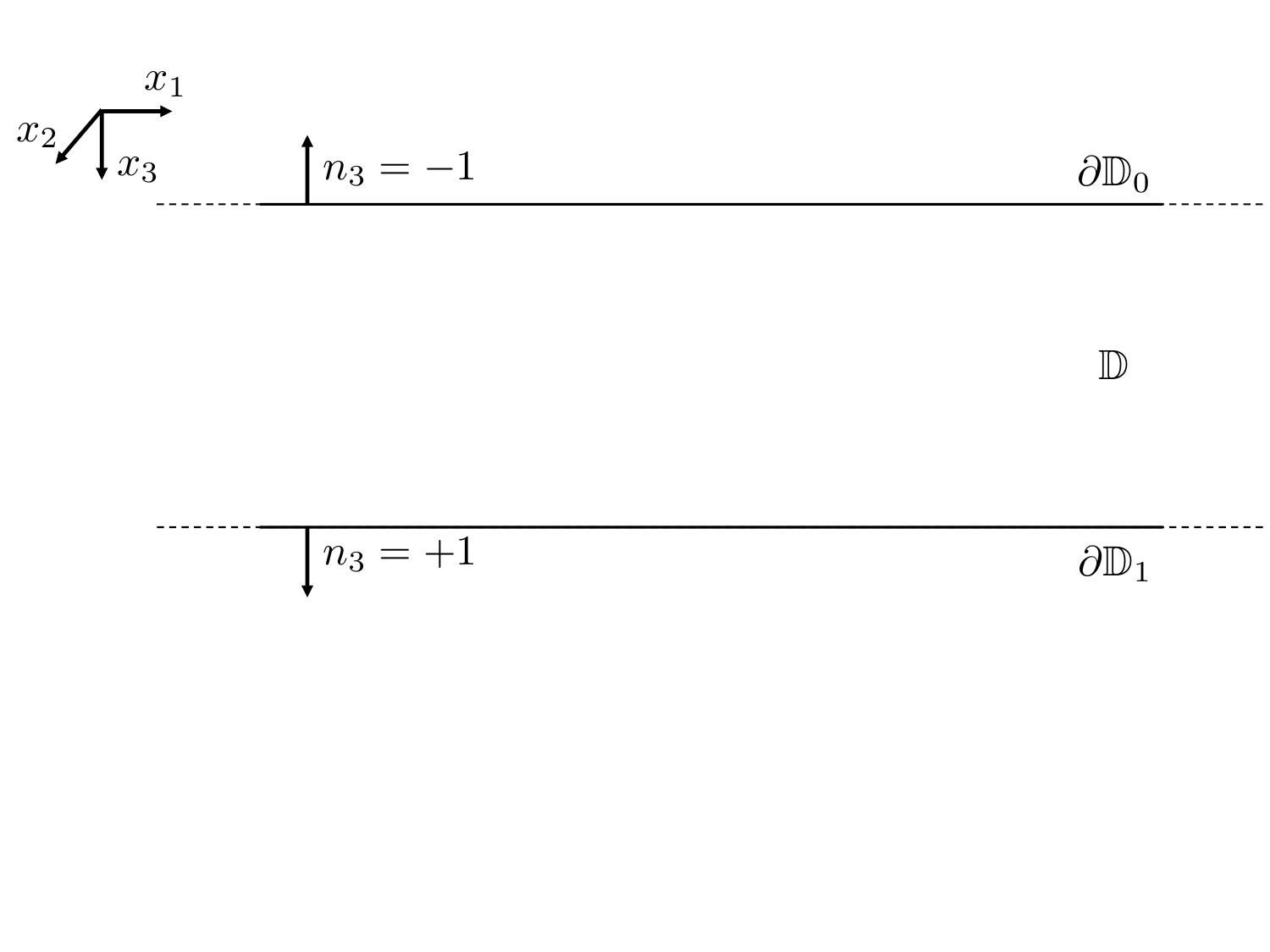}}
\vspace{-2.8cm}
\caption{\small Configuration for the matrix-vector reciprocity theorems, equations (\ref{eq4.1}) and (\ref{eq4.2}). 
The combination of boundaries $\setdD_0$ and $\setdD_1$ is called $\setdD$ in these equations.
}\label{Fig1}
\end{figure}

Matrix-vector wave equation (\ref{eq2.1}) and symmetry relations (\ref{eqsym}) and (\ref{eqsymad}) underly unified matrix-vector reciprocity theorems 
in an inhomogeneous medium (or potential). We consider two 
states $A$ and $B$, characterised by independent wave  vectors $\bq_A(\bx,\omega)$ and $\bq_B(\bx,\omega)$, obeying matrix-vector wave equation (\ref{eq2.1}),
with source vectors $\bd_A(\bx,\omega)$ and $\bd_B(\bx,\omega)$, and operator matrices $\bA_A(\bx,\omega)$ and $\bA_B(\bx,\omega)$.
The subscripts $A$ and $B$ of these operator matrices refer to the, possibly different, medium parameters in states $A$ and $B$.
We assume that outside a finite domain, the medium (or potential) and its adjoint are lossless in both states.
We consider a spatial domain $\setD$ enclosed by two infinite horizontal boundaries $\setdD_0$ and $\setdD_1$ (with $\setdD_1$ below $\setdD_0$), together denoted by $\setdD$, see Figure \ref{Fig1}.
The starting point for deriving reciprocity theorems for the wave fields in states $A$ and $B$ is given by
the quantities $\partial_3\{\bq_A^t\bN\bq_B\}$ and  $\partial_3\{\bq_A^\dagger\bK\bq_B\}$ in domain $\setD$. 
\revv{Applying the product rule for differentiation gives
\begin{eqnarray}
\partial_3\{\bq_A^t\bN\bq_B\} &=& (\partial_3\bq_A^t)\bN\bq_B + \bq_A^t\bN(\partial_3\bq_B),\label{eq431}\\
\partial_3\{\bq_A^\dagger\bK\bq_B\} &=& (\partial_3\bq_A^\dagger)\bK\bq_B + \bq_A^\dagger\bK(\partial_3\bq_B).\label{eq432}
\end{eqnarray}
Note that equation (\ref{eqH7}), which defines the transpose of the horizontal differential operator $\partial_\alpha$, 
does not apply to the vertical differential operator $\partial_3$. 
Hence, we may replace $(\partial_3\bq_A^t)$ by $(\partial_3\bq_A)^t$ in equation (\ref{eq431}), and $(\partial_3\bq_A^\dagger)$ by $(\partial_3\bq_A)^\dagger$ in equation (\ref{eq432}).
Using wave equation (\ref{eq2.1}) for both states in the right-hand sides of equations (\ref{eq431})  and (\ref{eq432}),
integrating both sides of these equations over domain $\setD$ and applying the theorem of Gauss to the left-hand sides}, we obtain 
\begin{eqnarray}\label{eq4.10}
\int_\setdD\bq_A^t\bN\bq_B n_3{\rm d}^2\bxh=\int_\setD\Bigl[\bigl((\bA_A\bq_A)^t+\bd_A^t\bigr)\bN\bq_B +\bq_A^t\bN\bigl(\bA_B\bq_B+\bd_B\bigr) \Bigr]{\rm d}^3\bx
\end{eqnarray}
and
\begin{eqnarray}\label{eq4.20}
\int_\setdD\bq_A^\dagger\bK\bq_B n_3{\rm d}^2\bxh=\int_\setD\Bigl[\bigl((\bA_A\bq_A)^\dagger+\bd_A^\dagger\bigr)\bK\bq_B +\bq_A^\dagger\bK\bigl(\bA_B\bq_B+\bd_B\bigr) \Bigr]{\rm d}^3\bx.
\end{eqnarray}
Here  $n_3$ is the vertical component of the outward pointing normal vector on $\setdD$, with $n_3=-1$ at the upper boundary $\setdD_0$
and $n_3=+1$ at the lower boundary $\setdD_1$. 
The integrals on the \revv{right}-hand sides can be written as 
\begin{eqnarray}
\int_\setD\{\cdots\}{\rm d}^3\bx=\int_{x_{3,0}}^{x_{3,1}}{\rm d}x_3\int_\setA\{\cdots\}{\rm d}^2\bxh,
\end{eqnarray}
where $x_{3,0}$ and $x_{3,1}$ denote the depths of $\setdD_0$ and $\setdD_1$, respectively.
\revv{Hence, at each depth level between $\setdD_0$ and $\setdD_1$ we can use the integral properties of transpose and adjoint operators, as formulated by}
 equations (\ref{eq90}) and (\ref{eq90b}). Together with the 
symmetry relations  (\ref{eqsym}) and (\ref{eqsymad}) for operator $\bA_A$, we thus obtain the following matrix-vector reciprocity theorems 
\begin{eqnarray}\label{eq4.1}
\int_\setD\bigl(\bd_A^t\bN\bq_B +\bq_A^t\bN\bd_B \bigr){\rm d}^3\bx=
\int_\setdD\bq_A^t\bN\bq_B n_3{\rm d}^2\bxh+\int_\setD\bq_A^t\bN(\bA_A-\bA_B)\bq_B {\rm d}^3\bx
\end{eqnarray}
and
\begin{eqnarray}\label{eq4.2}
\int_\setD\bigl(\bd_A^\dagger\bK\bq_B + \bq_A^\dagger\bK\bd_B\bigr){\rm d}^3\bx=
\int_\setdD\bq_A^\dagger\bK\bq_B n_3{\rm d}^2\bxh+\int_\setD\bq_A^\dagger\bK(\bbA_A-\bA_B)\bq_B {\rm d}^3\bx.
\end{eqnarray}
Equation (\ref{eq4.1}) is a convolution-type reciprocity theorem \citep{Fokkema93Book, Hoop95Book} because products like $\bq_A^t\bN\bq_B$
in the frequency domain correspond to convolutions in the time domain. Equation (\ref{eq4.2}) is a correlation-type reciprocity theorem \citep{Bojarski83JASA} 
because products like $\bq_A^\dagger\bK\bq_B$ in the frequency domain correspond to correlations in the time domain. 
These matrix-vector reciprocity  theorems have been previously derived for acoustic waves \citep{Haines96JMP, Wapenaar96JASA}.
Because these theorems follow from the unified matrix-vector wave equation (\ref{eq2.1}), with
unified symmetry relations (\ref{eqsym}) and (\ref{eqsymad}), they hold for all wave phenomena listed in Table \ref{table1}.
In the next section we use these theorems as the basis for matrix-vector wave field representations. Here we consider some special cases
of these theorems.\\

\noindent
$-$ {\it Power balance}

\noindent
When the sources, medium parameters and wave fields are identical in both states, we may drop the subscripts $A$ and $B$. In this case equation (\ref{eq4.2}) simplifies to
\begin{eqnarray}\label{eq4.3pp}
\int_\setD\quarter\bigl(\bd^\dagger\bK\bq + \bq^\dagger\bK\bd\bigr){\rm d}^3\bx=
\int_\setdD\quarter\bq^\dagger\bK\bq n_3{\rm d}^2\bxh+\int_\setD\quarter\bq^\dagger\bK(\bbA-\bA)\bq {\rm d}^3\bx.
\end{eqnarray}
Because $\quarter\bq^\dagger\bK\bq=\quarter(\bqa^\dagger\bqb+\bqb^\dagger\bqa)=\J$,
the first term on the right-hand side is the power flux (or probability current) through the  boundary \revv{$\setdD=\setdD_0\cup\setdD_1$ (i.e., the power leaving the domain $\setD$).} 
Hence, equation (\ref{eq4.3pp}) formulates the unified power  balance.
The term on the left-hand side is the power generated by the sources in $\setD$ and
the second term on the right-hand side is the dissipated power in $\setD$.\\

\noindent
$-$ {\it Propagation invariants}

\noindent
When there are no sources in $\setD$ and the medium parameters in $\setD$ are equal in the two states, \revv{the domain integrals in equation (\ref{eq4.1}) vanish, hence
%
\begin{eqnarray}
\int_{\setdD_0\cup\setdD_1}\bq_A^t\bN\bq_B n_3{\rm d}^2\bxh=0,
\end{eqnarray}
or, since $n_3=-1$ at  $\setdD_0$ and $n_3=+1$ at  $\setdD_1$,
\begin{eqnarray}
\int_{\setdD_0}\bq_A^t\bN\bq_B{\rm d}^2\bxh = \int_{\setdD_1}\bq_A^t\bN\bq_B{\rm d}^2\bxh.
\end{eqnarray}
Since this holds for any choice of the domain $\setD$, we infer that the quantity} 
\begin{eqnarray}\label{eq49}
\int_\setA\bq_A^t\bN\bq_B {\rm d}^2\bxh,
 \end{eqnarray}
\revv{with $\setA$ denoting a horizontal plane at arbitrary depth $x_3$,}
is a unified propagation invariant (i.e., it is independent of the depth $x_3$ of $\setA$).
On the other hand, when the medium parameters 
are each other's adjoints in the two states, 
\revv{we find in a similar way from equation (\ref{eq4.2}) that the quantity}
%
\begin{eqnarray}\label{eq50}
\int_\setA\bq_A^\dagger\bK\bq_B {\rm d}^2\bxh
 \end{eqnarray}
is a unified propagation invariant. 
Propagation invariants have been extensively used in the 
analysis of symmetry properties of reflection and transmission responses and for the design of efficient 
numerical modelling schemes for acoustic and elastodynamic wave fields \citep{Haines88GJI, Kennett90GJI, Koketsu91GJI, Takenaka93WM}.

\section{Matrix-vector wave field representations}\label{secRRR}

\subsection{Representation of the convolution type}

A wave field representation is obtained by replacing one of the states in a reciprocity theorem by a Green's state 
\citep{Knopoff56JASA, Hoop58PHD, Gangi70JGR, Pao76JASA}. Here we derive a unified matrix-vector wave field representation  
\revv{from} the matrix-vector reciprocity theorem of  the convolution type (equation \ref{eq4.1}).

We introduce the Green's matrix $\bG(\bx,\bx_A,\omega)$ 
(with the same dimensions as matrix $\bA$)
as the solution of the unified matrix-vector wave equation (\ref{eq2.1}), with the source vector 
$\bd$ replaced by a diagonal point-source matrix. Hence
\begin{eqnarray}\label{eq2.1g}
\partial_3\bG =\bA\bG +{\bf I}\delta(\bx-\bx_A),
\end{eqnarray}
where ${\bf I}$ is an identity matrix and $\bx_A$ defines the position of the point source.
We let $\bG$ represent the forward propagating solution of equation (\ref{eq2.1g}), 
which corresponds to imposing causality in the time domain, i.e., $\bG(\bx,\bx_A,t)={\bf O}$ for $t<0$, where ${\bf O}$ is a zero matrix
(the relation between functions in the time- and frequency domain is defined by  the Fourier transform, equation (\ref{eqFT})).
Before we derive the unified wave field representation, we first derive a reciprocity relation \revv{for the Green's matrix. To this end
we define a second forward propagating Green's matrix $\bG(\bx,\bx_B,\omega)$, with its point source at $\bx_B$.}
We assume that $\bx_A$ and $\bx_B$ are both situated in $\setD$. 
We replace   $\bq_A$ and $\bq_B$ in reciprocity theorem (\ref{eq4.1}) by $\bG(\bx,\bx_A,\omega)$ and
$\bG(\bx,\bx_B,\omega)$, respectively. Accordingly, we replace  $\bd_A$ and $\bd_B$ by ${\bf I}\delta(\bx-\bx_A)$ and ${\bf I}\delta(\bx-\bx_B)$, respectively.
Both Green's matrices are defined in the same medium, \revv{hence, $\bA_A=\bA_B$. This implies that} 
the second integral on the right-hand side of equation (\ref{eq4.1}) vanishes.
When Neumann or Dirichlet boundary conditions apply on $\setdD$, 
or when the medium outside $\setdD$ is homogeneous, the first integral on the right-hand side of equation (\ref{eq4.1}) vanishes as well.
We thus obtain
\begin{eqnarray}
\bN\bG(\bx_A,\bx_B,\omega) +\bG^t(\bx_B,\bx_A,\omega)\bN={\bf O}.
\end{eqnarray}
Using $\bN^{-1}=-\bN$ this gives
%
\begin{eqnarray}\label{eq65a}
\bG(\bx_A,\bx_B,\omega)=\bN\bG^t(\bx_B,\bx_A,\omega)\bN,
\end{eqnarray}
\revv{which is} the unified source-receiver reciprocity relation for the Green's matrix. 
 
Next, we use the \revv{same} reciprocity theorem to derive a representation for the \revv{actual} wave field vector $\bq$. 
\revv{We let state $B$ be the 
actual state (i.e., actual wave field, source and medium parameters). For convenience we drop the subscript $B$ from $\bq_B$, $\bd_B$ and $\bA_B$.} 
For state $A$ we choose again the Green's \revv{state.
Hence,} 
 we replace $\bq_A$ by $\bG(\bx,\bx_A,\omega)$ and  
$\bd_A$ by ${\bf I}\delta(\bx-\bx_A)$.
\revv{Operator $\bA_A$ may be defined in a reference medium or in the actual medium.}
Making these substitutions in equation (\ref{eq4.1}), pre-multiplying all terms by $-\bN$, using  equation  (\ref{eq65a}) and $-\bN\bN={\bf I}$, we obtain
\begin{eqnarray}
\chia(\bx_A)\bq(\bx_A,\omega)&=&\int_\setD \bG(\bx_A,\bx,\omega)\bd(\bx,\omega){\rm d}^3\bx
-\int_\setdD \bG(\bx_A,\bx,\omega)\bq(\bx,\omega)n_3{\rm d}^2\bxh\nonumber\\
&+&\int_\setD \bG(\bx_A,\bx,\omega)\{\bA-\bA_A\}\bq(\bx,\omega){\rm d}^3\bx,\label{eqrepgen}
\end{eqnarray}
where $\chia(\bx_A)$ is the characteristic function, defined as
\begin{eqnarray}\label{eqC3.2}
\chia(\bx_A)=
\begin{cases}
1,   &\text{for } \bx_A\text{ inside }{\setD}, \\
\half, &\text{for } \bx_A\text{ on }{\setdD},\\
0,  &\text{for } \bx_A\text{ outside }{\setD}.
\end{cases}
\end{eqnarray}
The left-hand side of equation (\ref{eqrepgen})  is the \revv{actual} wave field vector $\bq$, observed at $\bx_A$ (assuming $\bx_A$ is inside $\setD$). 
The right-hand side contains, respectively, a contribution from the source distribution $\bd(\bx,\omega)$ inside $\setD$, 
a contribution from the wave field $\bq(\bx,\omega)$ at the boundary $\setdD$, and
a contribution from the contrast operator $\bA-\bA_A$, applied to the wave field $\bq(\bx,\omega)$ inside $\setD$. 
This unified matrix-vector wave field representation holds for all wave phenomena listed in Table \ref{table1}.

This representation can often be simplified, which leads to different applications.
For example, when the medium outside the domain $\setD$  is homogeneous, source free and identical in both states, 
the boundary integral on the right-hand side vanishes. 
The remaining representation \revv{(with $\bA_A$ defined in a reference medium)} forms a basis for the analysis of forward scattering problems.
On the other hand, when \revv{$\bA_A$ is defined in the actual medium} (i.e., $\bA_A=\bA$) and the domain  $\setD$ is source free, only the boundary integral on the right-hand side remains.
In this case, equation (\ref{eqrepgen})
is a generalisation of the Kirchhoff-Helmholtz integral \citep{Morse53Book, Born65Book, Pao76JASA, Frazer85GJRAS, Berkhout85Book}, which finds applications 
in forward wave field extrapolation problems.

\subsection{Representation of the correlation type}

Representations of the correlation type find their application in inverse source problems \citep{Porter82JOSA, Hoop95Book}, inverse scattering problems
\citep{Devaney82UI, Bojarski83JASA, Bleistein84Book, Oristaglio89IP}, 
imaging \citep{Porter70JOSA, Schneider78GEO, Berkhout82Book,  Maynard85JASA, Esmersoy88GEO, Lindsey2004AJSS}, time-reversal acoustics
\citep{Fink2001IP}, and Green's function retrieval from ambient noise \citep{Derode2003JASA, Wapenaar2003GEO, Weaver2004JASA}.
There are several ways to approach the representation of the correlation type.
The homogeneous Green's function representation \citep{Porter70JOSA, Oristaglio89IP} elegantly covers most of the aforementioned applications for 
scalar wave fields. It is obtained by replacing both states in the reciprocity theorem of the correlation type by Green's states.
Here we derive a unified representation for the homogeneous Green's matrix by substituting
two Green's matrices into the matrix-vector reciprocity theorem of the correlation type (equation \ref{eq4.2}).

Before we discuss the homogeneous Green's matrix, we introduce the Green's matrix of the adjoint medium, $\bar\bG(\bx,\bx_A,\omega)$, 
as the forward propagating solution of the  following matrix-vector wave equation
\begin{eqnarray}\label{eq2.1gadj}
\partial_3\bar\bG =\bbA\bar\bG +{\bf I}\delta(\bx-\bx_A).
\end{eqnarray}
Pre- and post multiplying all terms by $\bJ$ and subsequently using equation (\ref{eqsymcon}) gives 
\begin{eqnarray}\label{eq2.1gadjb}
\partial_3\bJ\bar\bG\bJ =\bA^*\bJ\bar\bG\bJ +\bJ\bJ\delta(\bx-\bx_A).
\end{eqnarray}
Taking the complex conjugate of all terms and using $\bJ\bJ={\bf I}$ gives
\begin{eqnarray}\label{eq2.1gadjc}
\partial_3\bJ\bar\bG^*\bJ =\bA\bJ\bar\bG^*\bJ +{\bf I}\delta(\bx-\bx_A).
\end{eqnarray}
Subtracting all terms in this equation from the corresponding terms in equation (\ref{eq2.1g}) we obtain 
\begin{eqnarray}\label{eq2.1gadjd}
\partial_3\bG_{\rm h}(\bx,\bx_A,\omega) =\bA\bG_{\rm h}(\bx,\bx_A,\omega),
\end{eqnarray}
with
\begin{eqnarray}\label{eq750}
\bG_{\rm h}(\bx,\bx_A,\omega)=\bG(\bx,\bx_A,\omega)-\bJ\bar\bG^*\bJ(\bx,\bx_A,\omega).
\end{eqnarray}
Because $\bG_{\rm h}(\bx,\bx_A,\omega)$ obeys a matrix-vector wave equation without a source term, we call it the homogeneous Green's matrix.
The second term on the right-hand side represents a backward propagating wave field in the adjoint medium.

Next, we use the correlation-type reciprocity theorem (equation (\ref{eq4.2})) to derive a representation for the  homogeneous Green's matrix $\bG_{\rm h}$.
For state $A$ we choose the Green's matrix in the adjoint medium medium, hence, we replace $\bq_A$ by $\bar\bG(\bx,\bx_A,\omega)$, $\bd_A$ by ${\bf I}\delta(\bx-\bx_A)$, and $\bA_A$ by $\bbA$.
For state $B$ we choose the Green's matrix in the actual medium, hence, we replace $\bq_B$ by $\bG(\bx,\bx_B,\omega)$, $\bd_B$ by ${\bf I}\delta(\bx-\bx_B)$, and $\bA_B$ by $\bA$.
With these choices the contrast operator $\bbA_A-\bA_B=\,\,\,\,\bar{\bar{\!\!\!\!\bA}}-\bA$ vanishes.
Making these substitutions in equation (\ref{eq4.2}), taking $\bx_A$ and $\bx_B$ both inside $\setD$,
pre-multiplying all terms by $\bK$, using  equations  (\ref{eq65a}) and (\ref{eq750}), $\bK\bK={\bf I}$ and $\bK=\bJ\bN=-\bN\bJ$, we obtain
\begin{eqnarray}
\bG_{\rm h}(\bx_A,\bx_B,\omega)=\int_\setdD \bK\bar\bG^\dagger(\bx,\bx_A,\omega)\bK\bG(\bx,\bx_B,\omega)n_3{\rm d}^2\bxh.
\end{eqnarray}
This unified homogeneous Green's matrix representation holds for all wave phenomena listed in Table \ref{table1}.
It forms the basis for  generalising the applications mentioned at the beginning of this section.

\section{Conclusions}\label{secCO}

A unified  matrix-vector wave equation is presented for acoustic, quantum mechanical, electromagnetic, elastodynamic, 
poroelastodynamic, piezoelectric and seismoelectric waves. For most cases a 3D inhomogeneous, anisotropic, dissipative medium is considered. 
\revv{The unified equation may be used as a basis for generalising various applications of the elastodynamic matrix-vector wave equation, such as
the analysis of surface waves, the derivation of long-wave effective medium parameters,
the derivation of propagator matrices, decomposition into downgoing and upgoing waves, modelling wide-angle propagation in laterally variant media, etc.}

The operator matrix in the matrix-vector wave equation obeys unified  symmetry relations.
These symmetry relations underly unified reciprocity theorems of the convolution and correlation type, which, in turn, 
form the basis for representations of the wave vector and the homogeneous Green's matrix.
Reciprocity theorems and representations find applications in
forward modelling problems, inverse source and inverse scattering problems, imaging, time-reversal methods and Green's function retrieval from ambient noise.
The unified treatment in this paper provides a starting point for generalising these applications to a broad range of wave phenomena.

\section*{Acknowledgments}

The author thanks Evert Slob and Niels Grobbe for fruitful discussions. The constructive comments of the  reviewers and editors
\revv{(including Richard Gibson, Rune Mittet and J\"org Renner)}
are very much appreciated and helped a lot to improve the paper. 
This research  has received funding from the European Research Council (ERC) under the European Union's Horizon 2020 research 
and innovation programme (grant agreement No: 742703).

\bibliographystyle{gji}

\appendix

\section{Acoustic waves}\label{secAC}

\rev{The basic equations for acoustic wave propagation 
are  the linearised  equation of motion
\begin{eqnarray}
\partial_t \m_i+\partial_i \sigmaa=f_i\label{eqofmotrt}
\end{eqnarray}
and the linearised deformation equation
\begin{eqnarray}
-\partial_t \Theta + \partial_i v_i=q\label{eqststrt}
\end{eqnarray}
\citep{Hoop95Book, Willis2012CRM}. Here $\m_i=\m_i(\bx,t)$ is the momentum density as a function of spatial position $\bx$ and time $t$,
 $\sigmaa=\sigmaa(\bx,t)$ is the acoustic pressure,
  $\Theta=\Theta(\bx,t)$ the cubic dilatation,
 $v_i=v_i(\bx,t)$ the particle velocity, and 
 $f_i=f_i(\bx,t)$ and $q=q(\bx,t)$  represent the sources in terms of external force density and volume-injection rate density, respectively (the function $q$ should not be confused with vector $\bq$ and its components $\bqa$ and $\bqb$  in equations (\ref{eq2.1}) and (\ref{Aeq7mvbbprff})).
Operator $\partial_i$ stands for differentiation in the $x_i$-direction. 
Lower-case Latin subscripts (except $t$) take on the values 1, 2 and 3, and Einstein's summation convention applies to repeated subscripts.
Operator $\partial_t$  stands for the temporal differential operator $\partial/\partial t$.
The constitutive relations for an inhomogeneous, anisotropic fluid are given by
\begin{eqnarray}
\m_i&=&\rho_{ij}v_j,\label{eqconstac1}\\
\Theta&=&-\kappa\sigmaa,\label{eqconstac2}
\end{eqnarray}
where $\rho_{ij}=\rho_{ij}(\bx)$ and $\kappa=\kappa(\bx)$  are the mass density and  compressibility, respectively.
To account for anisotropy, the mass density is defined as a tensor. Although ideal fluids are by definition isotropic, inhomogeneities at the micro 
scale can often be represented by effective anisotropic parameters at the macro scale.
For example, a periodic stratified fluid can, in the long wavelength limit, be represented by a homogeneous fluid with an effective transverse isotropic mass density tensor 
and an effective isotropic compressibility \citep{Schoenberg83JASA}. The mass density tensor is symmetric, i.e., $\rho_{ij}=\rho_{ji}$.
Substituting the constitutive relations (\ref{eqconstac1}) and (\ref{eqconstac2}) into equations (\ref{eqofmotrt}) and (\ref{eqststrt}) yields
\begin{eqnarray}
\rho_{ij}\partial_t v_j+\partial_i \sigmaa&=&f_i,\label{eqofmotr}\\
\kappa\partial_t \sigmaa + \partial_i v_i&=&q.\label{eqststr}
\end{eqnarray}
}

We define the temporal Fourier transform of a space- and time-dependent function $h(\bx,t)$ as
\begin{eqnarray}\label{eqFT}
h(\bx,\omega)=\int_{-\infty}^\infty h(\bx,t){\rm exp}(\i\omega t){\rm d}t,
\end{eqnarray}
where $\i$ is the imaginary unit.
For notational convenience, we use the same symbol (here $h$) for quantities in the time domain and in the frequency domain.
We use equation  (\ref{eqFT}) to transform equations  (\ref{eqofmotr}) and (\ref{eqststr})
to the frequency domain. The time derivatives are thus replaced by $-\i\omega$,
hence
\begin{eqnarray}
-\i\omega\rho_{ij} v_j + \partial_i p&=&f_i,\label{eqofmotrf}\\
-\i\omega\kappa p + \partial_i v_i&=&q,\label{eqststrf}
\end{eqnarray}
with  $p=p(\bx,\omega)$, $v_i=v_i(\bx,\omega)$,   $f_i=f_i(\bx,\omega)$ and $q=q(\bx,\omega)$. 
In a lossless medium, the parameters  $\rho_{ij}(\bx)$ and $\kappa(\bx)$ are real-valued and frequency independent.
To account for losses, we replace them by complex-valued, frequency-dependent parameters $\rho_{ij}=\rho_{ij}(\bx,\omega)$ and
$\kappa=\kappa(\bx,\omega)$ \citep{Hoop95Book, Carcione2007Book}.

The quantities $p$ and $v_3$ constitute the power-flux density $\J$ in the $x_3$-direction, via
\begin{eqnarray}\label{eq9}
\J=\quarter(p^*v_3+v_3^*p).
\end{eqnarray}
We choose these quantities for the $1\times 1$ vectors $\bqa$ and $\bqb$ in equations (\ref{eqH3}) and (\ref{eqH4}), hence
\begin{eqnarray}
\bqa=p,\quad \bqb=v_3.\label{eqA6}
\end{eqnarray}
To arrive at a set of equations for these quantities, 
we need to eliminate the remaining wave field quantities, $v_1$ and $v_2$, from equations (\ref{eqofmotrf}) and (\ref{eqststrf}).
To this end, we first introduce the inverse of the mass density tensor, the so-called specific volume tensor $\s_{hi}=\s_{hi}(\bx,\omega)$, via
\begin{eqnarray}\label{eq10}
\s_{hi}\rho_{ij}=\delta_{hj},
\end{eqnarray}
with $\delta_{hj}$ denoting the Kronecker delta. On account of the symmetry of the mass density tensor and equation (\ref{eq10}), 
the specific volume tensor is symmetric as well,  hence $\s_{hi}=\s_{ih}$.
Applying $\s_{hi}$ to both sides of equation (\ref{eqofmotrf}), using equation (\ref{eq10}),  gives
\begin{eqnarray}\label{eqofmotrfaa}
-\i\omega v_h + \s_{hi}\partial_i p=\s_{hi}f_i.
\end{eqnarray}
We separate the derivatives in the $x_3$-direction from the lateral derivatives in equations (\ref{eqofmotrfaa}) and (\ref{eqststrf}), according to
\begin{eqnarray}
\partial_3 p&=&\s_{33}^{-1}\bigl( -\s_{3 \beta}\partial_\beta p  +  \i\omega v_3 + \s_{3 i}f_i \bigr),\label{eqA14}\\
\partial_3 v_3&=&  \i\omega\kappa p - \partial_\alpha v_\alpha + q.\label{eqA15}
\end{eqnarray}
Einstein's summation convention applies also to repeated Greek subscripts (which take on the values 1 and 2).
\revv{The particle velocity $v_\alpha$ needs to be eliminated from equation (\ref{eqA15}).}
From equation  (\ref{eqofmotrfaa}) we obtain
\begin{eqnarray}\label{eqA16}
v_\alpha=\frac{1}{\i\omega}\bigl(\s_{\alpha \beta}\partial_\beta p + \s_{\alpha 3}\partial_3 p - \s_{\alpha i}f_i \bigr).
\end{eqnarray}
Substituting equation (\ref{eqA16})  into equation  (\ref{eqA15}), using equation (\ref{eqA14}), we obtain
\begin{eqnarray}
\partial_3 v_3&=&  \i\omega\kappa p - \frac{1}{\i\omega}\partial_\alpha \bigl(\s_{\alpha \beta}\partial_\beta p + \s_{\alpha 3}\partial_3 p - \s_{\alpha i}f_i \bigr) + q\nonumber\\
&=&  \i\omega\kappa p - \frac{1}{\i\omega}\partial_\alpha \biggl(\s_{\alpha \beta}\partial_\beta p + 
\s_{\alpha 3}\s_{33}^{-1}\bigl( -\s_{3 \beta}\partial_\beta p  +  \i\omega v_3 + \s_{3 i}f_i \bigr) - \s_{\alpha i}f_i \biggr) + q.\label{eqA23}
\end{eqnarray}
\rev{We define
\begin{eqnarray}
b_{\alpha i}=\s_{\alpha i}-  \s_{\alpha 3}\s_{33}^{-1} \s_{3i},
\end{eqnarray}
with $b_{\alpha 3}=0$ and $b_{\alpha \beta}=b_{\beta \alpha}$ on account of $\s_{hi}=\s_{ih}$.}
Equations (\ref{eqA14}) and (\ref{eqA23}) have the form of equations (\ref{eqH3}) and (\ref{eqH4}), with $\bqa$ and $\bqb$
defined in equation (\ref{eqA6}), $1 \times 1$ vectors $\bd_1$ and $\bd_2$ defined as
\begin{eqnarray}
\bd_1= \s_{33}^{-1}\s_{3i}f_i,\quad \bd_2=\frac{1}{\i\omega}\partial_\alpha(\rev{ b_{\alpha \beta} f_\beta}) + q,\label{eqA6ag}
\end{eqnarray}
and $1\times 1$  operator matrices $\bfAhat_{11}$, $\bfAhat_{12}$, $\bfAhat_{21}$ and $\bfAhat_{22}$ defined as
\begin{eqnarray}
\bfAhat_{11} &=& -\s_{33}^{-1}\s_{3 \beta}\partial_\beta,\label{eq19}  \\ 
\bfAhat_{12} &=&  \i\omega \s_{33}^{-1}, \label{eq20}\\ 
\bfAhat_{21} &=&  \i\omega\kappa-\frac{1}{\i\omega}\partial_\alpha\rev{b_{\alpha\beta}}\partial_\beta, \label{eq21}\\ 
\bfAhat_{22} &=& -\partial_\alpha \s_{\alpha 3}\s_{33}^{-1}.  \label{eq22}
\end{eqnarray}
The notation in the right-hand side of these equations  should be understood in the sense that differential operators 
act on \revv{all factors} to the right of it.  \rev{For example, operator $\partial_\alpha b_{\alpha\beta}\partial_\beta$, applied via equation (\ref{eqH4}) to $p$, stands for $\partial_\alpha (b_{\alpha\beta} \partial_\beta p)$}, etc.
Operators $\bfAhat_{11}$, $\bfAhat_{12}$, $\bfAhat_{21}$ and $\bfAhat_{22}$ 
obey the symmetry relations (\ref{eq31}) $-$ (\ref{eq33}).

We define adjoint acoustic medium parameters as  $\bar \kappa = \kappa^*$, $\bar \s_{hi} = \s_{hi}^*$ \rev{and hence $\bar b_{\alpha\beta}=b_{\alpha\beta}^*$}.
Operators $\bbA_{11}$, $\bbA_{12}$, $\bbA_{21}$ and $\bbA_{22}$ in the adjoint medium
are defined as in equations (\ref{eq19}) $-$ (\ref{eq22}), but with  $\kappa$, $\s_{hi}$ and $b_{\alpha\beta}$ replaced by  $\bar\kappa$, $\bar \s_{hi}$  and $\bar b_{\alpha\beta}$, respectively.
These operators obey relations (\ref{eq941}) $-$ (\ref{eq944}).

For the special case of an isotropic fluid we have $\s_{hi}=\frac{1}{\rho}\delta_{hi}$, with $\rho$ denoting the mass density of the isotropic fluid. 
For this situation  equations (\ref{eq19}) $-$ (\ref{eq22}) reduce to the well-known expressions
\begin{eqnarray}
{\bA}_{11} &=&{\bA}_{22}=0, \\ 
{\bA}_{12} &=&  \i\omega \rho, \\ 
{\bA}_{21} &=& \i\omega\kappa-\frac{1}{\i\omega}\partial_\alpha\frac{1}{\rho}\partial_\alpha,
\end{eqnarray}
\citep{Corones75JMAA, Ursin83GEO, Fishman84JMP, Wapenaar89Book, Hoop96JMP}.

\section{Quantum mechanical waves}\label{secQM}

Schr\"odinger's wave equation for a particle with mass $m$ in a potential $V=V(\bx)$ is given by \citep{Messiah61Book, Merzbacher61Book}
\begin{eqnarray}
\i \hbar  \partial_t \psi  = - \frac {\hbar^2}{2m} \partial_i\partial_i \psi + V\psi,
\end{eqnarray}
where $\psi=\psi(\bx, t)$ is the wave function and $\hbar=h/2\pi$, with $h$ Planck's constant.
We use equation (\ref{eqFT}) to transform this equation to the space-frequency domain, which means we can replace $\partial_t$ by $-\i\omega$. 
This gives
\begin{eqnarray}\label{eq59}
\hbar  \omega \psi  = - \frac {\hbar^2}{2m} \partial_i\partial_i \psi + V\psi,
\end{eqnarray}
with $\psi=\psi(\bx,\omega)$. To account for losses, we replace $V(\bx)$ by a complex-valued, frequency-dependent function $V(\bx,\omega)$.

The quantities $\psi$ and  $\frac{2\hbar}{m\i}\partial_3\psi$ constitute the probability current density $\J$ in the $x_3$-direction, via
\begin{eqnarray}\label{eq940}
\J=\quarter\frac{2\hbar}{m\i}\bigl(\psi^*\partial_3\psi-\psi\partial_3\psi^*\bigr).
\end{eqnarray}
We choose these quantities for the $1\times 1$ vectors $\bqa$ and $\bqb$ in equations (\ref{eqH3}) and (\ref{eqH4}), hence
\begin{eqnarray}
\bqa=\psi,\quad \bqb=\frac{2\hbar}{m\i}\partial_3\psi .\label{eq61}
\end{eqnarray}
To arrive at a set of equations for these quantities, 
we first recast equation (\ref{eq59}) (using the fact that $\hbar$ and $m$ are constants) as
\begin{eqnarray}
\partial_3\Bigl(\frac{2\hbar}{m\i}\partial_3\psi\Bigr) = 4\i\Bigl(\omega-\frac{V}{\hbar}\Bigr)\psi-\frac{2\hbar}{m\i}\partial_\alpha\partial_\alpha\psi.
\end{eqnarray}
This equation, together with the trivial equation 
\begin{eqnarray}
\partial_3\psi= \frac{m\i}{2\hbar}\Bigl(\frac{2\hbar}{m\i}\partial_3\psi\Bigr),
\end{eqnarray}
have the form of equations (\ref{eqH4}) and (\ref{eqH3}), with $\bqa$ and $\bqb$
defined in equation (\ref{eq61}), $\bd_1=\bd_2=0$, 
and $1\times 1$  operator matrices $\bfAhat_{11}$, $\bfAhat_{12}$, $\bfAhat_{21}$ and $\bfAhat_{22}$ defined as
\begin{eqnarray}
{\bA}_{11} &=&{\bA}_{22}=0, \\ 
{\bA}_{12} &=&  \frac{m\i}{2\hbar}, \\ 
{\bA}_{21} &=&4\i\Bigl(\omega-\frac{V}{\hbar}\Bigr)-\frac{2\hbar}{m\i}\partial_\alpha\partial_\alpha.
\end{eqnarray}
Operators $\bfAhat_{12}$ and $\bfAhat_{21}$ 
obey the symmetry relations (\ref{eq32}) and (\ref{eq33}).

We define the adjoint  potential as $\bar V=V^*$. Operators $\bbA_{12}$ and $\bbA_{21}$ for the adjoint potential
obey relations (\ref{eq942}) and (\ref{eq943}).

\section{Electromagnetic waves}\label{secEM}

In the space-frequency domain, the Maxwell equations for electromagnetic wave propagation read
\citep{Landau60Book, Hoop95Book}
\begin{eqnarray}
-\i\omega D_i + J_i - \epsilon_{ijk}\partial_j  H_k&=& -  J_i^e,\label{gmax3b}\\
-\i\omega B_k + \epsilon_{klm}\partial_l  E_m&=& -  J_k^m,\label{gmax4b}
\end{eqnarray}
where $E_m= E_m({\bf x},\omega)$ is the electric field strength, $H_k=H_k({\bf x},\omega)$ the magnetic field strength, 
$D_i= D_i({\bf x},\omega)$ the electric flux density, $B_k=B_k({\bf x},\omega)$ the magnetic flux density, 
$J_i= J_i({\bf x},\omega)$ the induced electric current density,
$ J_i^e= J_i^e({\bf x},\omega)$ and $ J_k^m= J_k^m({\bf x},\omega)$ are source functions in terms of
external electric and magnetic current densities and, finally, $\epsilon_{ijk}$ is the alternating tensor (or Levi-Civita
tensor), with $\epsilon_{123}=\epsilon_{312}=\epsilon_{231}=-\epsilon_{213}=-\epsilon_{321}=-\epsilon_{132}=1$ and all other elements being equal to $0$.
The constitutive relations for an inhomogeneous, anisotropic, dissipative medium are given by
\begin{eqnarray}
D_i &=& \varepsilon_{ik} E_k= \varepsilon_0\varepsilon_{{\rm r},ik} E_k,\label{ieqA.21a}\\
B_k &=& \mu_{km} H_m = \mu_0\mu_{{\rm r},km}H_m,\label{ieqA.22a}\\
J_i &=& \sigma_{ik} E_k,\label{ieqA.23a}
\end{eqnarray}
where $\varepsilon_{ik}=\varepsilon_{ik}({\bf x},\omega)$, $\mu_{km}=\mu_{km}({\bf x},\omega)$ and
$\sigma_{ik}=\sigma_{ik}({\bf x},\omega)$ are the permittivity, permeability and
conductivity tensors, respectively. The subscripts $0$ refer to the
parameters in vacuum and the subscripts ${\rm r}$ denote relative
parameters for the anisotropic medium. 
These tensors obey the symmetry relations $\varepsilon_{ik}=\varepsilon_{ki}$, $\mu_{km}=\mu_{mk}$ and
$\sigma_{ik}=\sigma_{ki}$, respectively.
Substituting the constitutive relations (\ref{ieqA.21a}) $-$ (\ref{ieqA.23a}) into Maxwell's
electromagnetic field equations (\ref{gmax3b}) and (\ref{gmax4b}) yields
\begin{eqnarray}
-\i\omega{\cal E}_{ik} E_k - \epsilon_{ijk}\partial_j  H_k&=& -  J_i^e,\label{max3b}\\
-\i\omega\mu_{km} H_m + \epsilon_{klm}\partial_l  E_m&=& -  J_k^m,\label{max4b}
\end{eqnarray}
with
\begin{eqnarray}\label{max7b}
{\cal E}_{ik}=\varepsilon_{ik}- \frac{\sigma_{ik}}{\i\omega}.
\end{eqnarray}

A matrix-vector wave equation for electromagnetic waves in an isotropic stratified medium is given by \cite{Ursin83GEO} and \cite{Stralen97PHD}.
This has been extended for an anisotropic stratified medium by \cite{Loseth2007GJI}. Here we derive the matrix-vector wave equation for electromagnetic waves
in a  3D inhomogeneous, anisotropic, dissipative medium.

The quantities 
\begin{eqnarray}\label{eq80}
{\bf E}_0=\begin{pmatrix}  E_1 \\  E_2\end{pmatrix}
\quad \mbox{and}\quad {\bf H}_0=\begin{pmatrix}  H_2 \\- H_1  \end{pmatrix}
\end{eqnarray}
constitute the power-flux density $\J$ in the $x_3$-direction, via
\begin{eqnarray}
\J=\quarter({\bf E}_0^\dagger{\bf H}_0+{\bf H}_0^\dagger{\bf E}_0)= \quarter(E_1^*H_2-E_2^*H_1+H_2^*E_1-H_1^*E_2).
\end{eqnarray}
We choose these quantities for the $2\times 1$ vectors $\bqa$ and $\bqb$ in equations (\ref{eqH3}) and (\ref{eqH4}), hence
\begin{eqnarray}
\bqa={\bf E}_0,\quad \bqb={\bf H}_0.\label{eq79}
\end{eqnarray}
To arrive at a set of equations for these quantities, 
we need to eliminate the remaining wave field quantities, $E_3$ and $H_3$, from equations (\ref{max3b}) and (\ref{max4b}).
We start by rewriting these equations as
\begin{eqnarray}
-\i\omega \bfcalE_1  {\bf E}_0 -\i\omega \bfcalE_3E_3  + \partial_3 {\bf H}_0-
\bfpartialb H_3&=&- {\bf J}_0^e,\label{AEmax1a''}\\
-\i\omega \bfcalE_3^t {\bf E}_0 -\i\omega{\cal E}_{33} E_3 +\bfpartiala^t {\bf H}_0
&=&- J_3^e,\label{AEmax3a''}\\
-\i\omega \bfmu_1  {\bf H}_0 -\i\omega \bfmu_3H_3  + \partial_3 {\bf E}_0-
\bfpartiala E_3&=&- {\bf J}_0^m,\label{AEmax1am''}\\
-\i\omega \bfmu_3^t {\bf H}_0 -\i\omega\mu_{33} H_3 +\bfpartialb^t {\bf E}_0
&=&- J_3^m,\label{AEmax3am''}
\end{eqnarray}
with
\begin{eqnarray}
 \bfcalE_1=\begin{pmatrix} 
{\cal E}_{11} &  {\cal E}_{12} \\
{\cal E}_{12} &  {\cal E}_{22} 
\end{pmatrix},\quad
\bfcalE_3=\begin{pmatrix} 
{\cal E}_{13}  \\
{\cal E}_{23} 
\end{pmatrix},\quad
\bfmu_1=\begin{pmatrix} 
\mu_{22} &  -\mu_{12} \\
-\mu_{12} &  \mu_{11} 
\end{pmatrix},\quad
\bfmu_3=\begin{pmatrix} 
\mu_{23}  \\
-\mu_{13} 
\end{pmatrix},
\end{eqnarray}
\begin{eqnarray}\label{eq88}
\bfpartiala=\begin{pmatrix}\partial_1\\\partial_2\end{pmatrix},\quad
\bfpartialb=\begin{pmatrix}\partial_2\\-\partial_1\end{pmatrix},\quad
{\bf J}_0^e=\begin{pmatrix}  J_1^e \\  J_2^e\\
\end{pmatrix},\quad {\bf J}_0^m=\begin{pmatrix}  J_2^m \\- J_1^m\\ \end{pmatrix}.
\end{eqnarray}
Equation (\ref{eqH7}) implies 
\begin{eqnarray}
\bfpartiala^t=\begin{pmatrix}-\partial_1& - \partial_2\end{pmatrix},\quad
\bfpartialb^t=\begin{pmatrix}-\partial_2&  \partial_1\end{pmatrix}.
\end{eqnarray}
We separate the derivatives in the $x_3$-direction from the lateral derivatives in equations (\ref{AEmax1am''}) and (\ref{AEmax1a''}), according to
\begin{eqnarray}
\partial_3 {\bf E}_0 &=&\i\omega
\bfmu_1 {\bf H}_0 +\i\omega \bfmu_3H_3+
\bfpartiala E_3-  {\bf J}_0^m,\label{Aeqhh00a}\\
\partial_3 {\bf H}_0&=&
\i\omega \bfcalE_1  {\bf E}_0+\i\omega \bfcalE_3 E_3 +\bfpartialb  H_3- {\bf J}_0^e.\label{Aeqee00a}
\end{eqnarray}
The field components $E_3$ and $H_3$ need to be eliminated. 
From equations (\ref{AEmax3a''}) and (\ref{AEmax3am''}) we obtain
\begin{eqnarray}
E_3&=&{\cal E}_{33}^{-1}\Bigl(- \bfcalE_3^t {\bf E}_0  +\frac{1}{\i\omega}\bfpartiala^t {\bf H}_0+\frac{1}{\i\omega} J_3^e\Bigr),\label{eq93}\\
H_3&=&\mu_{33}^{-1}\Bigl(- \bfmu_3^t {\bf H}_0  +\frac{1}{\i\omega}\bfpartialb^t {\bf E}_0+\frac{1}{\i\omega} J_3^m\Bigr).\label{eq94}
\end{eqnarray}
Substituting equations (\ref{eq93}) and (\ref{eq94}) into equations (\ref{Aeqhh00a}) and (\ref{Aeqee00a}) we obtain
\begin{eqnarray}
\partial_3 {\bf E}_0 &=&
\Bigl(\bfmu_3\mu_{33}^{-1}\bfpartialb^t-\bfpartiala{\cal E}_{33}^{-1}\bfcalE_3^t \Bigr) {\bf E}_0
+\Bigl(\i\omega\bfmu_1 - \i\omega\bfmu_3\mu_{33}^{-1}\bfmu_3^t +\frac{1}{\i\omega}\bfpartiala{\cal E}_{33}^{-1}\bfpartiala^t\Bigr) {\bf H}_0
\nonumber\\&& 
+\frac{1}{\i\omega}\bfpartiala({\cal E}_{33}^{-1}J_3^e)
-{\bf J}_0^m+\bfmu_3\mu_{33}^{-1}J_3^m,\label{Aeqhh00aaa}\\
\partial_3 {\bf H}_0&=&
\Bigl(\i\omega \bfcalE_1 - \i\omega \bfcalE_3{\cal E}_{33}^{-1} \bfcalE_3^t +\frac{1}{\i\omega}\bfpartialb\mu_{33}^{-1}\bfpartialb^t\Bigr) {\bf E}_0
+\Bigl(\bfcalE_3{\cal E}_{33}^{-1}\bfpartiala^t - \bfpartialb \mu_{33}^{-1}\bfmu_3^t\Bigr) {\bf H}_0\nonumber\\
&&- {\bf J}_0^e+ \bfcalE_3{\cal E}_{33}^{-1}J_3^e + \frac{1}{\i\omega}\bfpartialb (\mu_{33}^{-1}J_3^m).\label{Aeqee00aaa}
\end{eqnarray}
Equations (\ref{Aeqhh00aaa}) and (\ref{Aeqee00aaa}) have the form of equations (\ref{eqH3}) and (\ref{eqH4}), with $\bqa$ and $\bqb$
defined in equation (\ref{eq79}), $2 \times 1$ vectors $\bd_1$ and $\bd_2$ defined as
\begin{eqnarray}
\bd_1&=&\frac{1}{\i\omega}\bfpartiala({\cal E}_{33}^{-1}J_3^e) -  {\bf J}_0^m+\bfmu_3\mu_{33}^{-1}J_3^m,\\
\bd_2&=& - {\bf J}_0^e+ \bfcalE_3{\cal E}_{33}^{-1}J_3^e + \frac{1}{\i\omega}\bfpartialb (\mu_{33}^{-1}J_3^m),
\end{eqnarray}
and $2\times 2$  operator matrices $\bfAhat_{11}$, $\bfAhat_{12}$, $\bfAhat_{21}$ and $\bfAhat_{22}$ defined as
\begin{eqnarray}
\bfAhat_{11}&=& \bfmu_3\mu_{33}^{-1}\bfpartialb^t-\bfpartiala{\cal E}_{33}^{-1}\bfcalE_3^t,\label{eq101} \\
\bfAhat_{12}&=& \i\omega(\bfmu_1 - \bfmu_3\mu_{33}^{-1}\bfmu_3^t) +\frac{1}{\i\omega}\bfpartiala{\cal E}_{33}^{-1}\bfpartiala^t,\\
\bfAhat_{21}&=& \i\omega( \bfcalE_1 - \bfcalE_3{\cal E}_{33}^{-1} \bfcalE_3^t) +\frac{1}{\i\omega}\bfpartialb\mu_{33}^{-1}\bfpartialb^t,\\
\bfAhat_{22}&=& \bfcalE_3{\cal E}_{33}^{-1}\bfpartiala^t - \bfpartialb\mu_{33}^{-1}\bfmu_3^t.\label{eq104}
\end{eqnarray}
These operators obey the symmetry relations (\ref{eq31}) $-$ (\ref{eq33}).

We define adjoint electromagnetic medium parameters as $\bar\varepsilon_{ik}=\varepsilon_{ik}^*$,
$\bar\mu_{km}=\mu_{km}^*$ and $\bar\sigma_{ik}=-\sigma_{ik}^*$.
Using equation (\ref{max7b}) it follows that $\bar{\cal E}_{ik}={\cal E}_{ik}^*$.
Similar relations hold for $\bfcalE_1$, $\bfcalE_3$, $\bfmu_1$ and $\bfmu_3$, which contain the parameters ${\cal E}_{ik}$ and $\mu_{km}$. 
Operators $\bbA_{11}$, $\bbA_{12}$, $\bbA_{21}$ and $\bbA_{22}$ in the adjoint medium
obey relations (\ref{eq941}) $-$ (\ref{eq944}).

Explicit expressions for the operator matrices in an isotropic medium are given in the supplemental material, Appendix I.

\section{Elastodynamic waves}\label{secED}

\rev{In the space-frequency domain, the elastodynamic equations of motion and deformation read  \citep{Achenbach73Book, Aki80Book, Hoop95Book, Willis2012CRM}
\begin{eqnarray}
-\i\omega \m_i - \partial_j\tau_{ij} &=& f_i,\label{eqmom}\\
\i\omega e_{kl}+ \half(\partial_k v_l+\partial_l v_k) &=& h_{kl},\label{eqstrain}
\end{eqnarray}
%
%
%
%
where $\m_i=\m_i(\bx,\omega)$ is the momentum density,
$\tau_{ij}=\tau_{ij}(\bx,\omega)$ the stress tensor,
$e_{kl}=e_{kl}(\bx,\omega)$ the strain tensor,
$v_k=v_k(\bx,\omega)$ the particle velocity,
and $f_i=f_i(\bx,\omega)$ and $h_{kl}=h_{kl}(\bx,\omega)$ are source functions in terms of external force density and deformation-rate density, respectively. 
The stress, strain and  deformation-rate tensors obey the  symmetry relations  $\tau_{ij}=\tau_{ji}$, $e_{kl}=e_{lk}$ and $h_{kl}=h_{lk}$.
The constitutive relations for an inhomogeneous, anisotropic, dissipative solid are given by
\begin{eqnarray}
\m_i &=& \rho_{ij}v_j,\label{eqconst1}\\
e_{kl} &=&s_{klmn}\tau_{mn},\label{eqconst2}
\end{eqnarray}
where $\rho_{ij}=\rho_{ij}(\bx,\omega)$ and  $s_{klmn}=s_{klmn}(\bx,\omega)$ are the mass density and compliance tensors, respectively.
These tensors obey the symmetry relations  $\rho_{ij}=\rho_{ji}$ and
$s_{klmn}=s_{lkmn}=s_{klnm}=s_{mnkl}$, respectively \citep{Aki80Book, Dahlen98Book}.
Substituting the constitutive relations (\ref{eqconst1}) and  (\ref{eqconst2}) into equations (\ref{eqmom}) and (\ref{eqstrain}) yields 
\begin{eqnarray}
-\i\omega\rho_{ij}v_j - \partial_j\tau_{ij} &=& f_i,\label{eq6.2aarf}\\
\i\omega s_{klmn}\tau_{mn}+ \half(\partial_k v_l+\partial_l v_k) &=&h_{kl}.\label{eq6.12bsc}
\end{eqnarray}
%
}
We introduce the stiffness tensor
$c_{ijkl}=c_{ijkl}({\bf x},\omega)$ as the inverse of the compliance tensor $s_{klmn}$, according to
\begin{eqnarray}\label{invcom}
c_{ijkl}s_{klmn}=s_{ijkl}c_{klmn}=
\half(\delta_{im}\delta_{jn}+\delta_{in}\delta_{jm}).
\end{eqnarray}
The stiffness tensor obeys the symmetry relation 
$c_{ijkl}=c_{jikl}=c_{ijlk}=c_{klij}$.
Multiplying all terms in equation (\ref{eq6.12bsc}) by $c_{ijkl}$, using the symmetry relations
$\tau_{ij}=\tau_{ji}$ and  $c_{ijkl}=c_{ijlk}$, we obtain an alternative form of 
equation (\ref{eq6.12bsc}), according to
\begin{eqnarray}
\i\omega\tau_{ij}+ c_{ijkl}\partial_l v_k&=&c_{ijkl}h_{kl}.\label{eq6.12b}
\end{eqnarray}

A matrix-vector wave equation for elastodynamic  waves in an inhomogeneous anisotropic medium is given by \cite{Woodhouse74GJR}.
Here we review this derivation, which also serves as a starting point for the
derivation of the matrix-vector wave equations for  poroelastodynamic waves (Appendix \ref{secPOE}), piezoelectric waves (Appendix \ref{secPE}) and 
seismoelectric waves (Appendix \ref{secSE}).

The quantities $-\bftau_3$ and ${\bf v}$ (which are $3\times 1$ vectors, with $(\bftau_3)_i=\tau_{i3}$ and $({\bf v})_i=v_i$)
constitute the power-flux density $\J$ in the $x_3$-direction, via
\begin{eqnarray}
\J=\quarter(-\bftau_3^\dagger{\bf v}-{\bf v}^\dagger\bftau_3)= \quarter (-\tau_{i3}^*v_i-v_i^*\tau_{i3}).
\end{eqnarray}
We choose these quantities for the $3\times 1$ vectors $\bqa$ and $\bqb$ in equations (\ref{eqH3}) and (\ref{eqH4}), hence
\begin{eqnarray}
\bqa=-\bftau_3,\quad \bqb={\bf v}.\label{eq120}
\end{eqnarray}
To arrive at a set of equations for these quantities, 
we need to eliminate the remaining wave field quantities, $3\times 1$ vectors $\bftau_1$ and $\bftau_2$ (with $(\bftau_j)_i=\tau_{ij}$), from equations (\ref{eq6.2aarf}) and (\ref{eq6.12b}).
We start by rewriting these equations as
\begin{eqnarray}
-\i\omega\R{\bf v} - \partial_j\bftau_j &=& {\bf f},\label{eq6.8b}\\
\i\omega\bftau_j+{\bfC}_{jl} \partial_l {\bf v}&=& {\bfC}_{jl}{\bf h}_l,\label{eq6.19b}
\end{eqnarray}
where $\R$ and ${\bfC}_{jl}$ are $3\times 3$ matrices, with  $({\R})_{ij}=\rho_{ij}$, $\R^t=\R$, $({\bfC}_{jl})_{ik}=c_{ijkl}$,  ${\bfC}_{jl}^t={\bfC}_{lj}$,
and where ${\bf f}$ and ${\bf h}_l$ are $3\times 1$ vectors, with $({\bf f})_i=f_i$ and $({\bf h}_l)_k=h_{kl}$.
We separate the derivatives in the $x_3$-direction from the lateral derivatives in equations (\ref{eq6.8b}) and (\ref{eq6.19b}), according to
\begin{eqnarray}
-\partial_3\bftau_3 &=& \i\omega\R{\bf v} + \partial_\alpha\bftau_\alpha+ {\bf f},\label{eq6.8bb}\\
\partial_3 {\bf v}&=& {\bfC}_{33}^{-1}\Bigl(-\i\omega\bftau_3 -{\bfC}_{3\beta}\partial_\beta {\bf v} +{\bfC}_{3l}{\bf h}_l\Bigr).\label{eq6.19bb}
\end{eqnarray}
The field components $\bftau_1$ and $\bftau_2$ need to be eliminated. 
From equation (\ref{eq6.19b}) we obtain
\begin{eqnarray}\label{eqC13}
\bftau_\alpha=-\frac{1}{\i\omega}\bigl({\bfC}_{\alpha \beta} \partial_\beta {\bf v} +{\bfC}_{\alpha 3} \partial_3 {\bf v} - {\bfC}_{\alpha l}{\bf h}_l\bigr).
\end{eqnarray}
Substituting equation (\ref{eq6.19bb}) into (\ref{eqC13}) and the result into equation  (\ref{eq6.8bb}), we obtain
\begin{eqnarray}\label{eq6.8bd}
-\partial_3\bftau_3 &=& \partial_\alpha\bigl({\bfC}_{\alpha 3}{\bfC}_{33}^{-1}\bftau_3\bigr)
+\i\omega\R{\bf v}-\frac{1}{\i\omega}\partial_\alpha 
\Bigl({\bfU}_{\alpha\beta}\partial_\beta {\bf v} -{\bfU}_{\alpha l}{\bf h}_l\Bigr) + {\bf f},
\end{eqnarray}
with
\begin{eqnarray}
{\bfU}_{\alpha l}={\bfC}_{\alpha l}-
{\bfC}_{\alpha 3}{\bfC}_{33}^{-1}{\bfC}_{3l},
\end{eqnarray}
where \rev{${\bfU}_{\alpha 3}={\bf O}$ and} ${\bfU}_{\alpha\beta}^t={\bfU}_{\beta\alpha}$ on account of $ {\bfC}_{jl}^t= {\bfC}_{lj}$.
Equations  (\ref{eq6.8bd}) and (\ref{eq6.19bb}) have the form of equations (\ref{eqH3}) and (\ref{eqH4}), with $\bqa$ and $\bqb$
defined in equation (\ref{eq120}), $3 \times 1$ vectors $\bd_1$ and $\bd_2$ defined as
\begin{eqnarray}
\bd_1&=&{\bf f}+\frac{1}{\i\omega}\partial_\alpha\bigl(\rev{{\bfU}_{\alpha \beta}{\bf h}_\beta}\bigr),\\
\bd_2&=& {\bfC}_{33}^{-1}{\bfC}_{3l}{\bf h}_l,
\end{eqnarray}
and $3\times 3$  operator matrices $\bfAhat_{11}$, $\bfAhat_{12}$, $\bfAhat_{21}$ and $\bfAhat_{22}$ defined as
\begin{eqnarray}
\bA_{11}&=& -\partial_\alpha{\bfC}_{\alpha 3} {\bfC}_{33}^{-1},\label{eq1.81}\\
\bA_{12}&=& \i\omega\R-\frac{1}{\i\omega}\partial_\alpha{\bfU}_{\alpha\beta}\partial_\beta,\label{eq1.82}\\
\bA_{21}&=&\i\omega{\bfC}_{33}^{-1},\label{eq1.83}\\
\bA_{22}&=&-{\bfC}_{33}^{-1} {\bfC}_{3\beta}\partial_\beta.\label{eq1.84}
\end{eqnarray}
These operators obey the symmetry relations (\ref{eq31}) $-$ (\ref{eq33}).

We define adjoint elastodynamic medium parameters as $\bar c_{ijkl}=c_{ijkl}^*$ and $\bar\rho_{ij}=\rho_{ij}^*$.
Similar relations hold for ${\bfC}_{jl}$, \rev{${\bfU}_{\alpha \beta}$} and $\R$, which contain the parameters $c_{ijkl}$ and $\rho_{ij}$. 
Operators $\bbA_{11}$, $\bbA_{12}$, $\bbA_{21}$ and $\bbA_{22}$ in the adjoint medium
obey relations (\ref{eq941}) $-$ (\ref{eq944}).

Explicit expressions for the operator matrices in an isotropic medium are given in the supplemental material, Appendix J.

\section{Poroelastodynamic waves}\label{secPOE}

In the space-frequency domain, the basic equations for poroelastodynamic wave propagation in an inhomogeneous, anisotropic, dissipative,
fluid-saturated porous solid read  \citep{Biot56JASA, Biot56JASA2, Pride92JASA,  Pride96JASA}
\begin{eqnarray}
-\i\omega\rho_{ij}^b{v}_j^s -\i\omega\rho_{ij}^f{w}_j - \partial_j{\tau}_{ij}^b&=& f_i^b,\label{eqg19prpe}\\
-\frac{\i\omega}{\eta} k_{ij}\rho_{jl}^f{v}_l^s +w_i + \frac{1}{\eta} k_{ij}\partial_j{p^f}&=& \frac{1}{\eta} k_{ij} f_j^f,\label{eqg23prpe}\\
\i\omega{\tau}_{ij}^b+ \c_{ijkl}\partial_l{v}_k^s +  C_{ij}\partial_k{w}_k&=&\c_{ijkl}\h_{kl}+C_{ij}\qf,\label{eqg24prpe}\\
-\i\omega{p^f}+ C_{kl}\partial_l{v}_k^s+ M\partial_k{w}_k&=&C_{kl}\h_{kl}+M\qf,\label{eqg26prpe}
\end{eqnarray}
with
\begin{eqnarray}
w_j&=&\phi( v_j^f- v_j^s),\label{eqg17prpe}\\
v_j^b&=&\phi v_j^f+(1-\phi)v_j^s=v_j^s+w_j,\\
\tau_{ij}^b&=&\phi \tau_{ij}^f+(1-\phi)\tau_{ij}^s=-\phi \delta_{ij}p^f+(1-\phi)\tau_{ij}^s,\\
f_i^b&=&\phi f_i^f+(1-\phi)f_i^s,\\
\rho_{ij}^b&=&\phi \rho_{ij}^f+(1-\phi)\rho_{ij}^s.
\end{eqnarray}
Superscripts $b$, $f$ and $s$ stand for bulk, fluid and solid, respectively. The wave field quantity
$ v_j= v_j({\bf x},\omega)$ is the averaged particle velocity in the bulk, fluid or solid (depending on the superscript),
$ w_j= w_j({\bf x},\omega)$ is the filtration velocity, $\tau_{ij}=\tau_{ij}({\bf x},\omega)$  the averaged stress in the bulk, fluid or solid
and $ p^f= p^f({\bf x},\omega)$ the averaged fluid pressure.
The stress tensors are symmetric, i.e., $\tau_{ij}=\tau_{ji}$.
The medium parameter $\rho_{ij}=\rho_{ij}({\bf x},\omega)$
is the mass density of the bulk, fluid or solid (depending on the superscript). 
Furthermore,  $k_{ij}=k_{ij}({\bf x},\omega)$ is the dynamic permeability tensor,
$\eta=\eta({\bf x},\omega)$ is the fluid viscosity parameter and $\phi=\phi({\bf x})$ the porosity.
Moreover, $ \c_{ijkl}= \c_{ijkl}({\bf x},\omega)$, $ C_{ij}= C_{ij}({\bf x},\omega)$ and $ M= M({\bf x},\omega)$ are
stiffness parameters of the porous solid. The medium parameters obey the following symmetry relations
$\rho_{ij}=\rho_{ji}$, $k_{ij}=k_{ji}$, $ \c_{ijkl}= \c_{jikl}= \c_{ijlk}= \c_{klij}$ and  $ C_{ij}= C_{ji}$.
The source function $ f_i= f_i({\bf x},\omega)$ is the volume density of external force on the bulk, fluid or solid.
For many source types the forces on the bulk and fluid are equal but in the following they will be treated distinctly.
The source functions $\h_{kl}=\h_{kl}({\bf x},\omega)$ and $\qf=\qf({\bf x},\omega)$ are the volume densities of external deformation rate on the bulk and volume-injection rate in the fluid
\citep{Wapenaar89Book, Pride94PRB, Hoop95Book, Grobbe2016PHD}. The deformation rate tensor is symmetric, i.e., $\h_{kl}=\h_{lk}$.
For later convenience, we eliminate $\partial_k{w}_k$ from equation
(\ref{eqg24prpe}), using equation (\ref{eqg26prpe}). This yields
\begin{eqnarray}\label{Eeqg24wanpes}
\i\omega{\tau}_{ij}^b+ \e_{ijkl}\partial_l{v}_k^s +\frac{\i\omega}{ M}C_{ij}{p^f}=\e_{ijkl}\h_{kl},
\end{eqnarray}
with  $ \e_{ijkl}= \e_{ijkl}({\bf x},\omega)$ defined as 
\begin{eqnarray}
\e_{ijkl}= \c_{ijkl}-\frac{1}{ M}C_{ij}C_{kl}.
\end{eqnarray}

A matrix-vector wave equation for normal-incidence poroelastodynamic waves in a stratified isotropic medium is given by \cite{Norris93JASA} and \cite{Gurevich95GJI}.
This has been extended for oblique-incidence poroelastodynamic waves in a stratified anisotropic medium, separately for P-SV and SH propagation, by \cite{Gelinsky97GEO}.
Here we derive the matrix-vector wave equation for poroelastodynamic waves in a 3D inhomogeneous, anisotropic, dissipative, fluid-saturated porous solid. 
The quantities $-\bftau_3^b$, ${p^f}$, ${\bf v}^s$ and $w_3$ (with $(\bftau_j^b)_i=\tau_{ij}^b$ and $({\bf v}^s)_i=v_i^s$)
constitute the power-flux density $\J$ in the $x_3$-direction, via
\begin{eqnarray}
\J&=&\quarter(-(\bftau_3^b)^\dagger{\bf v}^s+ p^{f*}w_3 - ({\bf v}^s)^\dagger\bftau_3^b + w_3^*p^f)\nonumber\\
&=&\quarter(-\tau_{i3}^{b*}v_i^s + p^{f*}w_3 - v_i^{s*}\tau_{i3}^b + w_3^*p^f).
\end{eqnarray}
We choose these quantities for the $4\times 1$ vectors $\bqa$ and $\bqb$ in equations (\ref{eqH3}) and (\ref{eqH4}), hence
\begin{eqnarray}
\bqa=\begin{pmatrix}- \bftau_3^b \\p^f  \end{pmatrix},
\quad \bqb=\begin{pmatrix} {\bf v}^s\\ w_3 \end{pmatrix}.\label{eq120pe}
\end{eqnarray}
To arrive at a set of equations for these quantities, 
we need to eliminate the remaining wave field quantities  from equations  (\ref{eqg19prpe}), (\ref{eqg23prpe}), (\ref{Eeqg24wanpes}) and (\ref{eqg26prpe}).
We start by rewriting these equations as
\begin{eqnarray}
-\i\omega{\R}^b{\bf v}^s -\i\omega{\R}^f\bfdelta_j w_j -\partial_j\bftau_j^b &=& {\bf f}^b,\label{Eeqg27anpes}\\
-\frac{\i\omega}{\eta}{\bfk}{\R}^f{\bf v}^s +\bfdelta_j w_j + \frac{1}{\eta}{\bfk}\bfdelta_j\partial_j{p^f}&=&\frac{1}{\eta}{\bfk}\bfdelta_j f_j^f,\label{Eeqg28anpes'}\\
\i\omega\bftau_j^b+  {\bfC}_{jl}\partial_l {\bf v}^s+ \frac{\i\omega}{ M}\bfc_j{p^f} &=&{\bfC}_{jl}\hh_l,\label{Eeqg29anpes}\\
-\i\omega{p^f}+ \bfc_l^t \partial_l{\bf v}^s+  M \partial_k{w}_k &=& \bfc_l^t\hh_l +M\qf,\label{Eeqg30anpes}
\end{eqnarray}
where $\R$, ${\bfk}$ and ${\bfC}_{jl}$ are $3\times 3$ matrices, with  $({\R})_{ij}=\rho_{ij}$, $\R^t=\R$,  $({\bfk})_{ij}=k_{ij}$, ${\bfk}^t={\bfk}$, $({\bfC}_{jl})_{ik}=\e_{ijkl}$,  
${\bfC}_{jl}^t={\bfC}_{lj}$, and where $\bfc_j$, ${\bf f}^b$, $\hh_l$ and $\bfdelta_j$ are $3\times 1$ vectors, with $(\bfc_j)_i=C_{ij}$, $({\bf f}^b)_i=f_i^b$, $(\hh_l)_k=\h_{kl}$, 
and $(\bfdelta_j)_i=\delta_{ij}$.
Equations (\ref{Eeqg27anpes}) $-$ (\ref{Eeqg30anpes}) form the starting point for deriving matrix-vector equations in the form of
equations (\ref{eqH3}) and (\ref{eqH4}), with $\bqa$ and $\bqb$ defined in equation  (\ref{eq120pe}).
The other quantities ($\bftau_1^b$, $\bftau_2^b$, $w_1$ and $w_2$) need to be eliminated. 
The detailed derivation can be found in the supplemental material, Appendix K. 
The $4\times 1$ vectors $\bd_1$ and $\bd_2$ are defined as
\begin{eqnarray}
\bd_1  \!\!\!\!&=&\!\!\!\! \begin{pmatrix}
\frac{\i\omega}{\eta}{\R}^f\bfdelta_\alpha\bfdelta_\alpha^t{\bfk}\bigl({\bf I}+\frac{1}{\b}\bfdelta_3\bfdelta_3^t{\bfk}^{-1}\bfdelta_\gamma\bfdelta_\gamma^t{\bfk}\bigr)\bfdelta_jf_j^f
-\frac{\i\omega}{\eta\b}{\R}^f\bfdelta_\alpha\bfdelta_\alpha^t{\bfk}\bfdelta_3f_3^f
+{\bf f}^b+\frac{1}{\i\omega}\partial_\alpha \bigl(\rev{{\bfU}_{\alpha \beta}\hh_\beta}\bigr)\\
\frac{1}{\b}\bigl(-\bfdelta_3^t{\bfk}^{-1}\bfdelta_\alpha\bfdelta_\alpha^t{\bfk}\bfdelta_jf_j^f+f_3^f\bigr)
\end{pmatrix},\\
\bd_2 \!\!\!\!  &=&\!\!\!\!  \begin{pmatrix}
{\bfC}_{33}^{-1}
{\bfC}_{3l}\hh_l\\
-\partial_\alpha\Bigl(\frac{1}{\eta}\bfdelta_\alpha^t{\bfk}\bigl({\bf I}+
\frac{1}{\b}\bfdelta_3\bfdelta_3^t{\bfk}^{-1}\bfdelta_\gamma\bfdelta_\gamma^t{\bfk}\bigr)\bfdelta_jf_j^f-\frac{1}{\eta\b}\bfdelta_\alpha^t{\bfk}\bfdelta_3f_3^f \Bigr)+
\frac{1}{M}\rev{{\bfu}_\alpha^t\hh_\alpha}+\qf
\end{pmatrix}
\end{eqnarray}
and the $4\times 4$ operator matrices $\bfAhat_{11}$, $\bfAhat_{12}$, $\bfAhat_{21}$ and $\bfAhat_{22}$ as
\begin{eqnarray}
\bfAhat_{11} &=& \begin{pmatrix}
\bfAhat_{11}^{11} & \bfahat_{11}^{12}\\
\bfahat_{11}^{21} & \bfahat_{11}^{22}\end{pmatrix},\quad
\bfAhat_{12} = \begin{pmatrix}
\bfAhat_{12}^{11} & \bfahat_{12}^{12}\\
(\bfahat_{12}^{12})^t & \bfahat_{12}^{22}\end{pmatrix},\label{eqD15}\\
\bfAhat_{21} &=& \begin{pmatrix}
\bfAhat_{21}^{11} & \bfahat_{21}^{12}\\
(\bfahat_{21}^{12})^t & \bfahat_{21}^{22}\end{pmatrix},\quad
\bfAhat_{22} =-\bfAhat_{11}^t.\label{eqD16}
\end{eqnarray}
Here
\begin{eqnarray}
\bfAhat_{11}^{11}&=& -\partial_\alpha{\bfC}_{\alpha 3}{\bfC}_{33}^{-1},\\
\bfahat_{11}^{12}
&=&-\frac{\i\omega}{\eta}{\R}^f\Bigl({\bf I}-\frac{1}{\b}{\bfk}\bfdelta_3\bfdelta_3^t{\bfk}^{-1}\bfdelta_3\bfdelta_3^t\Bigl){\bfk}\bfdelta_\beta\partial_\beta 
-\partial_\alpha\frac{1}{M}{\bfu}_\alpha,\\
\bfahat_{11}^{21}&=&{\bf 0}^t,\\
\bfAhat_{11}^{22}&=&-\frac{1}{\b}\bfdelta_3^t{\bfk}^{-1}\bfdelta_3\bfdelta_3^t{\bfk}\bfdelta_\beta\partial_\beta,
\end{eqnarray}
\begin{eqnarray}
\bfAhat_{12}^{11}
&=&\i\omega{\R}^b-\frac{1}{\i\omega}\partial_\alpha {\bfU}_{\alpha\beta}\partial_\beta
-\frac{\omega^2}{\eta}{\R}^f\bfdelta_\alpha\bfdelta_\alpha^t\bigl({\bfk}-\frac{1}{\b}{\bfk}\bfdelta_3\bfdelta_3^t{\bfk}^{-1}\bfdelta_3\bfdelta_3^t{\bfk}\bigr)\bfdelta_\beta\bfdelta_\beta^t{\R}^f,\\
\bfahat_{12}^{12}
&=&\frac{\i\omega}{\b}{\R}^f{\bfk}\bfdelta_3\bfdelta_3^t{\bfk}^{-1}\bfdelta_3,\\
\bfAhat_{12}^{22}&=&-\frac{\eta}{\b}\bfdelta_3^t{\bfk}^{-1}\bfdelta_3,\\
\bfAhat_{21}^{11}&=&\i\omega{\bfC}_{33}^{-1},\\
\bfahat_{21}^{12}&=&-\frac{\i\omega}{M}{\bfC}_{33}^{-1}\bfc_3,\\
\bfAhat_{21}^{22}&=&\frac{ \i\omega}{M^2}\bfc_3^t{\bfC}_{33}^{-1}\bfc_3 +\frac{\i\omega}{M}
+\partial_\alpha\frac{1}{\eta}\bigl(\bfdelta_\alpha^t{\bfk}\bfdelta_\beta - \frac{1}{\b}\bfdelta_\alpha^t{\bfk}\bfdelta_3\bfdelta_3^t{\bfk}^{-1}\bfdelta_3\bfdelta_3^t{\bfk}\bfdelta_\beta\bigr)\partial_\beta,
\end{eqnarray}
with ${\bf 0}$ denoting a  zero vector and
\begin{eqnarray}
\rev{{\bfU}_{\alpha \beta}}&=&\rev{ {\bfC}_{\alpha \beta}-
{\bfC}_{\alpha 3} {\bfC}_{33}^{-1} {\bfC}_{3\beta}},\\
\rev{{\bfu}_{\alpha}}&=&\rev{\bfc_\alpha-  {\bfC}_{\alpha 3} {\bfC}_{33}^{-1}\bfc_3},\\
\b&=&1-\bfdelta_3^t{\bfk}^{-1}\bfdelta_\alpha\bfdelta_\alpha^t{\bfk}\bfdelta_3.
\end{eqnarray}

Operators $\bfAhat_{11}$, $\bfAhat_{12}$, $\bfAhat_{21}$ and $\bfAhat_{22}$ 
obey the symmetry relations formulated in equations (\ref{eq31}) $-$ (\ref{eq33}).

We defined the adjoints of the medium parameters  $c_{ijkl}$ and $\rho_{ij}$ in Appendix \ref{secED}
(where $\rho_{ij}$ now has superscript $b$ or $f$). Moreover, we define
$\bar k_{ij}=-k_{ij}^*$, $\bar \eta = \eta^*$, $\bar C_{ij}=C_{ij}^*$ and $\bar M= M^*$.
Similar relations hold for  ${\bfC}_{jl}$, $\R^b$, $\R^f$, ${\bfk}$ and $\bfc_j$, 
which contain the parameters $ \e_{ijkl}= \c_{ijkl}-\frac{1}{ M}C_{ij}C_{kl}$, $\rho_{ij}^b$,  $\rho_{ij}^f$, $k_{ij}$ and $C_{ij}$.
Operators $\bbA_{11}$, $\bbA_{12}$, $\bbA_{21}$ and $\bbA_{22}$ in the adjoint medium
obey relations (\ref{eq941}) $-$ (\ref{eq944}).

Explicit expressions for the operator matrices in an isotropic medium are given in the supplemental material, Appendix K.

\section{Piezoelectric waves}\label{secPE}

Piezoelectric waves are governed by the equations for electromagnetic waves (Appendix \ref{secEM}) and elastodynamic waves (Appendix \ref{secED}), in which two of the constitutive 
relations need to be modified to account for the coupling between the two wave types. For piezoelectric waves, the modified constitutive relations are \citep{Auld73Book}
\begin{eqnarray}
D_i &=& \varepsilon_{ik} E_k + d_{ijk}\tau_{jk}\label{eqdijk},\\
e_{kl}&=&d_{klm}E_m+s_{klmn}\tau_{mn}.\label{eqdklm}
\end{eqnarray}
The field quantities and medium parameters (except $d_{ijk}$) have been defined in Appendices \ref{secEM} and \ref{secED}.
Parameters $\varepsilon_{ik}$  in equation (\ref{eqdijk}) and $s_{klmn}$ in equation (\ref{eqdklm})
are \rev{defined} under constant stress and constant electric field, respectively.
The coupling tensor $d_{ijk}=d_{ijk}(\bx,\omega)$ obeys the symmetry relation $d_{ijk}=d_{jik}=d_{ikj}$.
Equation (\ref{eqdijk}) replaces constitutive relation (\ref{ieqA.21a}) and is substituted, together with constitutive relations (\ref{ieqA.22a}) and (\ref{ieqA.23a}),
into the Maxwell equations (\ref{gmax3b}) and (\ref{gmax4b}).
Equation (\ref{eqdklm}) replaces stress-strain relation (\ref{eqconst2})
\rev{and is substituted, together with constitutive relation (\ref{eqconst1}), into equations (\ref{eqstrain}) and (\ref{eqmom}).  
Subsequently, all terms in the latter equation are multiplied by $c_{ijkl}$, using equation (\ref{invcom}) as well as the symmetry relations
$\tau_{ij}=\tau_{ji}$ and  $c_{ijkl}=c_{ijlk}$.}
The basic equations for coupled electromagnetic and elastodynamic waves thus read
\begin{eqnarray}
-\i\omega{\cal E}_{ik}E_k -\epsilon_{ijk}\partial_j H_k-\i\omega d_{ijk}\tau_{jk}&=&-J_i^e,\label{eqA.1pef}\\
-\i\omega\mu_{km}H_m+\epsilon_{klm}\partial_l E_m&=&-J_k^m,\label{eqA.2pef}\\
-\i\omega\rho_{ij}v_j -\partial_j\tau_{ij}& =& f_i,\label{eqA.3pef}\\
\i\omega\tau_{ij}+ c_{ijkl}\partial_l v_k+ \i\omega c_{ijkl}d_{klm}E_m&=& c_{ijkl}h_{kl},\label{eqA.5pef}
\end{eqnarray}
with ${\cal E}_{ik}=\varepsilon_{ik}- \frac{\sigma_{ik}}{\i\omega}$.

A matrix-vector wave equation in the quasi-static approximation for  2D piezoelectric waves  in an anisotropic stratified medium  is given 
by \cite{Honein91JIMSS}, \cite{Wang2002APL} and \cite{Zhao2012JAP}. 
Here we derive the exact matrix-vector wave equation for piezoelectric waves in a 3D inhomogeneous, anisotropic, dissipative,
piezoelectric medium. 

The quantities $-\bftau_3$, ${\bf E}_0$, ${\bf v}$ and ${\bf H}_0$
(with ${\bf E}_0$ and ${\bf H}_0$ defined in Appendix \ref{secEM} and $\bftau_3$  and ${\bf v}$ defined in Appendix \ref{secED})
constitute the power-flux density $\J$ in the $x_3$-direction, via
\begin{eqnarray}
\J&=&\quarter(-\bftau_3^\dagger{\bf v}+{\bf E}_0^\dagger{\bf H}_0-{\bf v}^\dagger\bftau_3+{\bf H}_0^\dagger{\bf E}_0)\nonumber\\
&=&\rev{\quarter(-\tau_{i3}^*v_i+E_1^*H_2-E_2^*H_1-v_i^*\tau_{i3}+H_2^*E_1-H_1^*E_2).}
\end{eqnarray}
We choose these quantities for the $5\times 1$ vectors $\bqa$ and $\bqb$ in equations (\ref{eqH3}) and (\ref{eqH4}), hence
\begin{eqnarray}
\bqa=\begin{pmatrix}- \bftau_3 \\ {\bf E}_0  \end{pmatrix},
\quad \bqb=\begin{pmatrix} {\bf v}\\  {\bf H}_0 \end{pmatrix}.\label{eq163}
\end{eqnarray}
To arrive at a set of equations for these quantities, 
we need to eliminate the remaining wave field quantities  from equations (\ref{eqA.1pef}) $-$ (\ref{eqA.5pef}).
Using the notation introduced in Appendices \ref{secEM} and \ref{secED},
we rewrite equations (\ref{eqA.1pef}) $-$ (\ref{eqA.5pef}) as
\begin{eqnarray}
-\i\omega \bfcalE_1  {\bf E}_0 -\i\omega \bfcalE_3E_3  + \partial_3 {\bf H}_0-\bfpartialb H_3-\i\omega{\bf D}_{1k}^t\bftau_k&=&- {\bf J}_0^e,\label{AEmax1ape}\\
-\i\omega \bfcalE_3^t {\bf E}_0 -\i\omega{\cal E}_{33} E_3 +\bfpartiala^t {\bf H}_0-\i\omega{\bf D}_{3k}^t\bftau_k&=&- J_3^e,\label{AEmax3ape}\\
-\i\omega \bfmu_1  {\bf H}_0 -\i\omega \bfmu_3H_3  + \partial_3 {\bf E}_0-\bfpartiala E_3&=&- {\bf J}_0^m,\label{AEmax1ampe}\\
-\i\omega \bfmu_3^t {\bf H}_0 -\i\omega\mu_{33} H_3 +\bfpartialb^t {\bf E}_0&=&- J_3^m,\label{AEmax3ampe}\\
-\i\omega\R{\bf v} - \partial_j\bftau_j &=& {\bf f},\label{eq6.8bpe}\\
\i\omega\bftau_j+{\bfC}_{jl} \partial_l {\bf v}+\i\omega{\bfC}_{jl}\bigl({\bf D}_{1l}{\bf E}_0+{\bf D}_{3l}E_3\bigr)&=& {\bfC}_{jl}{\bf h}_l,\label{eq6.19bpe}
\end{eqnarray}
with
\begin{eqnarray}
{\bf D}_{1k}=\begin{pmatrix}
d_{11k} & d_{12k} \\
d_{21k} & d_{22k} \\
d_{31k} & d_{32k} 
\end{pmatrix},\quad
{\bf D}_{3k}=\begin{pmatrix}
d_{13k} \\ d_{23k} \\ d_{33k}
\end{pmatrix}.
\end{eqnarray}
Equations (\ref{AEmax1ape}) $-$ (\ref{eq6.19bpe}) form the starting point for deriving 
matrix-vector equations in the form of equations (\ref{eqH3}) and (\ref{eqH4}), with $\bqa$ and $\bqb$ defined in equation (\ref{eq163}).
The other quantities ($\bftau_1$,  $\bftau_2$, $E_3$ and $H_3$) need to be eliminated.
The detailed derivation can be found in the supplemental material, Appendix L. 
The $5 \times 1$ vectors $\bd_1$ and $\bd_2$ are defined as
\begin{eqnarray}
\bd_1&=&\begin{pmatrix} 
{\bf f}+\frac{1}{\i\omega}\partial_\alpha( {\bfU}_{\alpha \beta}'{\bf h}_\beta)  - \frac{1}{\i\omega}\partial_\alpha \bigl(({\cal E}_{33}')^{-1}{\bfU}_{\alpha \beta}{\bf D}_{3\beta} J_3^e\bigr)\\
\frac{1}{\i\omega}\bfpartiala\bigl(({\cal E}_{33}')^{-1}(J_3^e-{\bf D}_{3\alpha}^t{\bfU}_{\alpha \beta} {\bf h}_\beta)\bigr)-  {\bf J}_0^m+\bfmu_3\mu_{33}^{-1} J_3^m\\
\end{pmatrix},\\
\bd_2&=& \begin{pmatrix} 
{\bfC}_{33}^{-1}\bigl({\bfC}_{3m}'{\bf h}_m-({\cal E}_{33}')^{-1}{\bfC}_{3l}{\bf D}_{3l} J_3^e\bigr)\\
- {\bf J}_0^e  + ({\cal E}_{33}')^{-1} \bfcalE_3'J_3^e+\frac{1}{\i\omega}\bfpartialb(\mu_{33}^{-1}J_3^m)+ ({\bf D}_{1\alpha}')^t{\bfU}_{\alpha \beta}{\bf h}_\beta
\end{pmatrix}
\end{eqnarray}
and the $5\times 5$  operator matrices $\bfAhat_{11}$, $\bfAhat_{12}$, $\bfAhat_{21}$ and $\bfAhat_{22}$ 
having the form defined in equations (\ref{eqD15}) and (\ref{eqD16}), where 
\begin{eqnarray}
\bfAhat_{11}^{11}&=&-\partial_\alpha({\bfC}_{3\alpha}')^t{\bfC}_{33}^{-1},\label{eq217}\\
\bfahat_{11}^{12}&=&-\partial_\alpha{\bfU}_{\alpha \beta}{\bf D}_{1\beta}',\\
\bfahat_{11}^{21}&=&\bfpartiala({\cal E}_{33}')^{-1}{\bf D}_{3k}^t{\bfC}_{k 3}{\bfC}_{33}^{-1},\\
\bfAhat_{11}^{22}&=&\bfmu_3\mu_{33}^{-1}\bfpartialb^t -\bfpartiala ({\cal E}_{33}')^{-1}(\bfcalE_3')^t,\\
\bfAhat_{12}^{11}&=&\i\omega\R-\frac{1}{\i\omega}\partial_\alpha{\bfU}_{\alpha \beta}' \partial_\beta, \\
\bfahat_{12}^{12}&=&-\frac{1}{\i\omega}\partial_\alpha({\cal E}_{33}')^{-1}{\bfU}_{\alpha \beta}{\bf D}_{3\beta}\bfpartiala^t,\\
\bfAhat_{12}^{22}&=&\i\omega\bigl(\bfmu_1- \bfmu_3\mu_{33}^{-1}\bfmu_3^t\bigr)+\frac{1}{\i\omega}\bfpartiala({\cal E}_{33}')^{-1}\bfpartiala^t,\\
\bfAhat_{21}^{11}&=&\i\omega \bigl({\bfC}_{33}^{-1} - ({\cal E}_{33}')^{-1} {\bfC}_{33}^{-1}{\bfC}_{3l}{\bf D}_{3l}{\bf D}_{3k}^t{\bfC}_{k 3}{\bfC}_{33}^{-1} \bigr),\\
\bfahat_{21}^{12}&=&-\i\omega {\bfC}_{33}^{-1}{\bfC}_{3l}{\bf D}_{1l}',
\end{eqnarray}
\begin{eqnarray}
\bfAhat_{21}^{22}&=&\i\omega\bigl(\bfcalE_1' -\bfcalE_3'({\cal E}_{33}')^{-1}(\bfcalE_3')^t\bigr)+\frac{1}{\i\omega}\bfpartialb \mu_{33}^{-1}\bfpartialb^t,
\end{eqnarray}
with
\begin{eqnarray}
\rev{{\bfU}_{\alpha \beta}}&=&\rev{{\bfC}_{\alpha \beta}-{\bfC}_{\alpha 3}{\bfC}_{33}^{-1}{\bfC}_{3\beta}},\\
\bfcalE_1'&=&\bfcalE_1-{\bf D}_{1\alpha}^t{\bfU}_{\alpha\beta} {\bf D}_{1\beta},\\
\bfcalE_3'&=&\bfcalE_3-{\bf D}_{1\alpha}^t{\bfU}_{\alpha\beta}{\bf D}_{3\beta},\\
{\cal E}_{33}'&=&{\cal E}_{33}-{\bf D}_{3\alpha}^t{\bfU}_{\alpha\beta}{\bf D}_{3\beta},\\
{\bfU}_{\alpha \beta}'&=&{\bfU}_{\alpha \beta}+({\cal E}_{33}')^{-1}{\bfU}_{\alpha \gamma} {\bf D}_{3\gamma}{\bf D}_{3\delta}^t {\bfU}_{\delta \beta},\\
{\bfC}_{3m}'&=&{\bfC}_{3m}+ ({\cal E}_{33}')^{-1}{\bfC}_{3l}{\bf D}_{3l}{\bf D}_{3\alpha}^t{\bfU}_{\alpha m},\\
{\bf D}_{1l}' &=&{\bf D}_{1l}  - ({\cal E}_{33}')^{-1}{\bf D}_{3l} (\bfcalE_3')^t,
\end{eqnarray}
\rev{with ${\bfU}_{\alpha 3}={\bf O}$}. Operators $\bfAhat_{11}$, $\bfAhat_{12}$, $\bfAhat_{21}$ and $\bfAhat_{22}$ 
obey the symmetry relations formulated in equations (\ref{eq31}) $-$ (\ref{eq33}).

We defined the adjoints of the medium parameters ${\cal E}_{ik}$,  $\mu_{km}$, $c_{ijkl}$ and $\rho_{ij}$ in Appendices \ref{secEM} and \ref{secED}.
Moreover, we define $\bar d_{ijk}=d_{ijk}^*$.
Similar relations hold for $\bfcalE_1$, $\bfcalE_3$, $\bfmu_1$, $\bfmu_3$,  ${\bfC}_{jl}$, $\R$, ${\bf D}_{1k}$ and ${\bf D}_{3k}$, 
which contain the parameters ${\cal E}_{ik}$, $\mu_{km}$, $c_{ijkl}$, $\rho_{ij}$ and $d_{ijk}$.
Operators $\bbA_{11}$, $\bbA_{12}$, $\bbA_{21}$ and $\bbA_{22}$ in the adjoint medium obey relations (\ref{eq941}) $-$ (\ref{eq944}).

\section{Seismoelectric waves}\label{secSE}

Seismoelectric waves are governed by the equations for electromagnetic waves (Appendix \ref{secEM}) and poroelastodynamic waves (Appendix \ref{secPOE}), in which two of 
the constitutive relations need to be modified to account for the coupling  between the two wave types. In this appendix we consider an isotropic medium
(the derivation for the anisotropic situation is disproportionally long). For seismoelectric waves, the modified constitutive relations are  \citep{Pride94PRB, Pride96JASA}
\begin{eqnarray}
J_i&=&\sigma E_i + L(-\partial_ip^f+\i\omega\rho^fv_i^s+f_i^f),\label{eqL1}\\
w_i&=&L E_i + \frac{k}{\eta}(-\partial_ip^f+\i\omega\rho^fv_i^s+f_i^f).\label{eqL2}
\end{eqnarray}
The field quantities, sources and medium parameters (except $L$)  have been defined in Appendices
\ref{secEM} and \ref{secPOE} (except that tensors are now replaced by scalars).
Here  ${L}={L}({\bf x},\omega)$ accounts for the coupling  between the elastodynamic and
electromagnetic waves and vice versa.
Equations (\ref{eqL1}) and (\ref{eqL2}) contain the same  coupling coefficient $L$  (due to Onsager's reciprocity relation, \cite{Pride94PRB}). 
Equation (\ref{eqL1}) replaces the isotropic version of 
constitutive relation (\ref{ieqA.23a}) and is substituted, together with the isotropic versions of constitutive relations (\ref{ieqA.21a}) and (\ref{ieqA.22a}),
into the Maxwell equations (\ref{gmax3b}) and (\ref{gmax4b}). Equation (\ref{eqL2}) replaces the isotropic version of equation (\ref{eqg23prpe}).
The basic equations for coupled electromagnetic and poroelastodynamic waves thus read
\begin{eqnarray}
-\i\omega\rho^b{v}_i^s -\i\omega\rho^f{w}_i - \partial_j{\tau}_{ij}^b&=& f_i^b,\label{eqg19prse}\\
-\i\omega\rho^f{v}_i^s +\frac{\eta}{k}(w_i - LE_i) + \partial_i{p^f}&=&  f_i^f,\label{eqg23prse}\\
\i\omega{\tau}_{ij}^b+ \c_{ijkl}\partial_l{v}_k^s +  C\delta_{ij}\partial_k{w}_k&=&\c_{ijkl}\h_{kl}+C\delta_{ij}\qf,\label{eqg24prse}\\
-\i\omega{p^f}+C\delta_{kl}\partial_l{v}_k^s+ M\partial_k{w}_k&=&C\delta_{kl}\h_{kl}+M\qf,\label{eqg26prse}\\
-\i\omega \varepsilon E_i+\sigma E_i +L(-\partial_ip^f+\i\omega\rho^fv_i^s) - \epsilon_{ijk}\partial_j H_k&=&- J_i^e - Lf_i^f,\label{max10}\\
-\i\omega \mu H_k + \epsilon_{klm}\partial_l E_m&=& - J_k^m.\label{max20}
\end{eqnarray}
For the isotropic medium we have
\begin{eqnarray}\label{eeq6.16bpe}
\c_{ijkl}&=&( K_G-\frac{2}{3} G_{\rm fr})\delta_{ij}\delta_{kl}+ G_{\rm fr}(\delta_{ik}\delta_{jl}+\delta_{il}\delta_{jk}),
\end{eqnarray}
where $ G_{\rm fr}$ is the
shear modulus of the framework of the grains when the fluid is absent and $K_G$ is the Gassmann modulus \citep{Pride92JASA}. 
The permittivity and permeability are defined as $\varepsilon=\varepsilon_0\varepsilon_{\rm r}$ and $\mu=\mu_0\mu_{\rm r}$. 
The subscripts $0$ refer to the parameters in vacuum and the subscripts ${\rm r}$ denote
relative parameters. For $\varepsilon_{\rm r}$ and $\mu_{\rm r}$ we have  \citep{Pride94PRB} 
\begin{eqnarray}
\varepsilon_{\rm r}&=&\frac{\phi}{\alpha_\infty}(\kappa^f-\kappa^s)+\kappa^s,\\
\mu_{\rm r}&\approx&1,
\end{eqnarray}
where $\kappa^f$ and $\kappa^s$ are the dielectric parameters of the fluid and solid, respectively, and $\alpha_\infty$ is the tortuosity at infinite frequency.
For later convenience, we eliminate $\partial_k{w}_k$ from equation
(\ref{eqg24prse}), using equation (\ref{eqg26prse}). This yields
\begin{eqnarray}\label{Eeqg24w}
\i\omega{\tau}_{ij}^b+ \e_{ijkl}\partial_l{v}_k^s +\i\omega\frac{ C}{ M}\delta_{ij}{p^f}= \e_{ijkl}\h_{kl},
\end{eqnarray}
with $ \e_{ijkl}= \e_{ijkl}({\bf x},\omega)$ defined  as
\begin{eqnarray}\label{Eeqg25w}
\e_{ijkl}= \c_{ijkl}-\frac{ C^2}{ M}\delta_{ij}\delta_{kl}.
\end{eqnarray}
Also for later convenience, we add ${L}$ times equation (\ref{eqg23prse}) to
equation (\ref{max10}) in order to compensate for the term $L(-\partial_ip^f+\i\omega\rho^fv_i^s)$. This yields
\begin{eqnarray}\label{eqF19}
-\i\omega {\cal E}  E_i +\frac{\eta}{k}{L} w_i- \epsilon_{ijk}\partial_j H_k&=&- J_i^e,\label{max1'}
\end{eqnarray}
with 
\begin{eqnarray}\label{max3'}
{\cal E}= \varepsilon-\frac{1}{\i\omega}\bigl(\sigma -\frac{\eta}{k} L^2\bigr).
\end{eqnarray}

A matrix-vector wave equation for oblique-incidence seismoelectric waves in a stratified isotropic medium, separately for  P-SV-TM and SH-TE propagation, is given by
\cite{Haartsen97JGR},  \cite{White2006SIAM} and \cite{Grobbe2016PHD}.
Here we derive the matrix-vector wave equation for a 3D inhomogeneous, isotropic, dissipative, fluid-saturated porous solid.
The quantities $-\bftau_3^b$, ${p^f}$, ${\bf E}_0$, ${\bf v}^s$, $w_3$ and ${\bf H}_0$ 
constitute the power-flux density $\J$ in the $x_3$-direction, via
\begin{eqnarray}
\J&=&\quarter(-(\bftau_3^b)^\dagger{\bf v}^s+ p^{f*}w_3 +{\bf E}_0^\dagger{\bf H}_0 - ({\bf v}^s)^\dagger\bftau_3^b + w_3^*p^f +{\bf H}_0^\dagger{\bf E}_0)\nonumber\\
&=&\quarter(-\tau_{i3}^{b*}v_i^s + p^{f*}w_3 + E_1^*H_2-E_2^*H_1 - v_i^{s*}\tau_{i3}^b + w_3^*p^f +H_2^*E_1-H_1^*E_2).
\end{eqnarray}
We choose these quantities for the $6\times 1$ vectors $\bqa$ and $\bqb$ in equations (\ref{eqH3}) and (\ref{eqH4}), hence
\begin{eqnarray}
\bqa=\begin{pmatrix}- \bftau_3^b \\p^f  \\ {\bf E}_0\end{pmatrix},
\quad \bqb=\begin{pmatrix} {\bf v}^s\\ w_3 \\ {\bf H}_0\end{pmatrix}.\label{eq120pese}
\end{eqnarray}
To arrive at a set of equations for these quantities, we need to eliminate the remaining wave field quantities from
equations (\ref{eqg19prse}), (\ref{eqg23prse}), (\ref{Eeqg24w}), (\ref{eqg26prse}), (\ref{eqF19}) and (\ref{max20}).
We start by rewriting these equations  as
\begin{eqnarray}
-\i\omega\rho^b{\bf v}^s -\i\omega\rho^f\bfdelta_j w_j -\partial_j\bftau_j^b &=& {\bf f}^b,\label{Eeqg27}\\
-\i\omega\rho^f\bfdelta_i^t{\bf v}^s +\frac{\eta}{k}\bigl( w_i -{L}(\bfgamma_i^t{\bf E}_0+\delta_{3i} E_3)\bigr)+ \partial_i{p^f}&=&f_i^f,\label{Eeqg28}\\
\i\omega\bftau_j^b+  {\bfC}_{jl}\partial_l {\bf v}^s+\i\omega\frac{ C}{ M}\bfdelta_j{p^f} &=& {\bfC}_{jl}\hh_l,\label{Eeqg29}\\
-\i\omega{p^f}+  C\bfdelta_l^t \partial_l{\bf v}^s+ M \partial_k{w}_k &=&C\bfdelta_l^t \hh_l+M\qf,\label{Eeqg30}\\
-\i\omega {\cal E}  {\bf E}_0 +\frac{\eta}{ k}{L}\bfgamma_\alpha w_\alpha+\partial_3 {\bf H}_0-\bfpartialb H_3&=&- {\bf J}_0^e,\label{Emax1''}\\
-\i\omega {\cal E}  E_3 +\frac{\eta}{ k}{L} w_3+\bfpartiala^t {\bf H}_0&=&- J_3^e,\label{Emax3''}\\
-\i\omega  \mu {\bf H}_0 + \partial_3 {\bf E}_0 -\bfpartiala E_3&=&-{\bf J}_0^m,\label{Emax2''}\\
-\i\omega  \mu H_3 +\bfpartialb^t {\bf E}_0&=& - J_3^m,\label{Emax4'}
\end{eqnarray}
with most of the vectors and matrices defined in Appendices \ref{secEM} and  \ref{secPOE}. In addition, $\bfgamma_i$ is a $2\times 1$ unit vector, with $(\bfgamma_i)_\beta=\delta_{\beta i}$.
Equations (\ref{Eeqg27}) $-$ (\ref{Emax4'}) form the starting point for deriving matrix-vector equations in the form of
equations (\ref{eqH3}) and (\ref{eqH4}), with $\bqa$ and $\bqb$ defined in equation  (\ref{eq120pese}).
The other quantities ($\bftau_1^b$, $\bftau_2^b$, $E_3$, $w_1$, $w_2$ and $H_3$)  need to be eliminated. 
The detailed derivation can be found in the supplemental material, Appendix M. 
The $6\times 1$ vectors $\bd_1$ and $\bd_2$ are defined as
\begin{eqnarray}
\bd_1 & = & \begin{pmatrix}
{\bf f}^b+\i\omega \rho^f\frac{k}{\eta}\bfdelta_\alpha f_\alpha^f+\frac{1}{\i\omega}\partial_\alpha\bigl({\bfU}_{\alpha \beta}\hh_\beta\bigr)\\
\frac{1}{\i\omega{\cal E}}\frac{\eta}{ k}{L}J_3^e+ f_3^f\\
-{\bf J}_0^m+\bfpartiala\bigl(\frac{1}{\i\omega{\cal E}}J_3^e\bigr) 
\end{pmatrix},
\end{eqnarray}
\begin{eqnarray}
\bd_2 & = & \begin{pmatrix}
{\bfC}_{33}^{-1}{\bfC}_{3l}\hh_l\\
\qf+\frac{1}{M}{\bfu}_\alpha^t\hh_\alpha-\partial_\beta\bigl( \frac{ k}{\eta} f_\beta^f\bigr)\\
-{\bf J}_0^e +\bfpartialb \bigl(\frac{1}{\i\omega \mu} J_3^m\bigr)-{L}\bfgamma_\alpha f_\alpha^f
\end{pmatrix}
\end{eqnarray}
and the $6\times 6$  operator matrices $\bfAhat_{11}$, $\bfAhat_{12}$, $\bfAhat_{21}$ and $\bfAhat_{22}$ as
\begin{eqnarray}
\bfAhat_{11} &=& \begin{pmatrix}
\bfAhat_{11}^{11} & \bfahat_{11}^{12} & \bfAhat_{11}^{13}\\
{\bf 0}^t& 0 &{\bf 0}^t\\
{\bf O} &{\bf 0}& {\bf O}
\end{pmatrix},\quad
\bfAhat_{12} = \begin{pmatrix}
\bfAhat_{12}^{11} & \bfahat_{12}^{12} &{\bf O}\\
(\bfahat_{12}^{12})^t &\bfAhat_{12}^{22} & \bfahat_{12}^{23}\\
{\bf O} & (\bfahat_{12}^{23})^t & \bfAhat_{12}^{33}
\end{pmatrix},\\
\bfAhat_{21} &=& \begin{pmatrix}
\bfAhat_{21}^{11} & \bfahat_{21}^{12} &{\bf O}\\
(\bfahat_{21}^{12})^t &\bfAhat_{21}^{22} & \bfahat_{21}^{23}\\
{\bf O} & (\bfahat_{21}^{23})^t & \bfAhat_{21}^{33}
\end{pmatrix},\quad
\bfAhat_{22} = -\bfAhat_{11}^t.
\end{eqnarray}
Here
\begin{eqnarray}
\bfAhat_{11}^{11}&=& -\partial_\alpha {\bfC}_{\alpha 3}{\bfC}_{33}^{-1},\\
\bfahat_{11}^{12}&=&-\i\omega \rho^f\frac{k}{\eta}\bfdelta_\alpha \partial_\alpha-\partial_\alpha\frac{1}{ M}{\bfu}_\alpha,\\
\bfAhat_{11}^{13}&=&\i\omega \rho^f{L}\bfdelta_\alpha\bfgamma_\alpha^t,\\
\bfAhat_{12}^{11}&=&-\frac{1}{\i\omega}\partial_\alpha  {\bfU}_{\alpha\beta}\partial_\beta+\i\omega\Bigl( \rho^b{\bf I}_3+\i\omega( \rho^f)^2\frac{k}{\eta}\bfdelta_\alpha\bfdelta_\alpha^t\Bigr),\\
\bfahat_{12}^{12}&=&\i\omega \rho^f\bfdelta_3,\\
\bfAhat_{12}^{22}&=&-\frac{\eta}{ k}\Bigl(1-\frac{1}{\i\omega{\cal E}}\frac{\eta}{ k}{L}^2\Bigr),\\
\bfahat_{12}^{23}&=&\frac{1}{\i\omega{\cal E}}\frac{\eta}{ k}{L} \bfpartiala^t,\\
\bfAhat_{12}^{33}&=&\i\omega  \mu{\bf I}_2 +\bfpartiala\frac{1}{\i\omega{\cal E}}\bfpartiala^t,\\
\bfAhat_{21}^{11}&=&\i\omega {\bfC}_{33}^{-1},\\
\bfahat_{21}^{12}&=&-\i\omega\frac{ C}{ M} {\bfC}_{33}^{-1}\bfdelta_3,\\
\bfAhat_{21}^{22}&=& \i\omega\frac{ C^2}{ M^2}\bfdelta_3^t {\bfC}_{33}^{-1}\bfdelta_3 +\frac{\i\omega}{ M}+\partial_\beta\frac{ k}{\eta}\partial_\beta,\\
\bfahat_{21}^{23}&=&-\partial_\beta{L}\bfgamma_\beta^t,\\
\bfAhat_{21}^{33}&=&\Bigl(\i\omega {\cal E}-\frac{\eta}{ k}{L}^2\Bigr){\bf I}_2+\bfpartialb\frac{1}{\i\omega \mu}\bfpartialb^t,
\end{eqnarray}
where ${\bf I}_3$ is a $3\times 3$ identity matrix, ${\bf I}_2$  a $2\times 2$ identity matrix and
\begin{eqnarray}
\rev{{\bfU}_{\alpha \beta}}&=& \rev{{\bfC}_{\alpha \beta}- {\bfC}_{\alpha 3} {\bfC}_{33}^{-1} {\bfC}_{3\beta}},\\
\rev{{\bfu}_{\alpha}}&=&\rev{C(\bfdelta_\alpha-  {\bfC}_{\alpha3} {\bfC}_{33}^{-1}\bfdelta_3)}.
\end{eqnarray}

Operators $\bfAhat_{11}$, $\bfAhat_{12}$, $\bfAhat_{21}$ and $\bfAhat_{22}$ 
obey the symmetry relations formulated in equations (\ref{eq31}) $-$ (\ref{eq33}).

We define the adjoints of the medium parameters  $\epsilon$, $\mu$, $\sigma$, $c_{ijkl}$, $\rho$ (with superscript $b$ or $f$), $k$, $\eta$, $C$ and $M$ 
similar as in Appendices \ref{secEM}, \ref{secED} and \ref{secPOE} but for the isotropic situation.
Moreover, we define $\bar L = -L^*$.
Operators $\bbA_{11}$, $\bbA_{12}$, $\bbA_{21}$ and $\bbA_{22}$ in the adjoint medium
obey relations (\ref{eq941}) $-$ (\ref{eq944}).

Explicit expressions for the operator matrices are given in the supplemental material, Appendix M.

\section{Symmetry properties of the operator matrix in the wavenumber-frequency domain}\label{secsymkx}

We derive symmetry properties of the operator matrix for the special case of a laterally invariant medium (or potential) in the wavenumber-frequency domain.
We define the spatial Fourier transform of a space- and frequency-dependent quantity $h(\bx,\omega)$ as
\begin{eqnarray}\label{eqFTX}
\tilde h(k_\alpha,x_3,\omega)=\int_\setA h(\bx,\omega){\rm exp}(-\i k_\alpha x_\alpha){\rm d}^2\bxh,
\end{eqnarray}
with $k_\alpha$ for $\alpha=1,2$ representing the horizontal wavenumbers.
Lateral derivatives $\partial_\alpha h(\bx,\omega)$ in the space-frequency domain are replaced by products
$\i k_\alpha \tilde h(k_\alpha,x_3,\omega)$ in the wavenumber-frequency domain. We denote the Fourier transform
of $\partial_\alpha$ as $\partial_\alpha \Rightarrow \i k_\alpha$.
Similarly, for an operator matrix $\bU$ in a laterally invariant medium, containing the differential operator $\partial_\alpha$, we denote the Fourier transform as 
$\bU(\partial_\alpha) \Rightarrow \tilde\bU(k_\alpha)$.
Using equations (\ref{eqH7}) and (\ref{eqH12}), we find
\begin{eqnarray}
\{\bU(\partial_\alpha)\}^t &\Rightarrow& \{\tilde\bU(-k_\alpha)\}^t,\label{eqap197}\\
\{\bU(\partial_\alpha)\}^* &\Rightarrow& \{\tilde\bU(-k_\alpha)\}^*,\label{eqap198}\\
\{\bU(\partial_\alpha)\}^\dagger &\Rightarrow& \{\tilde\bU(k_\alpha)\}^\dagger.\label{eqap199}
\end{eqnarray}
We use equation (\ref{eqFTX}) to transform equation  (\ref{eq2.1})  to the wavenumber-frequency domain, according to 
$\partial_3\tilde\bq =\tbA\tilde\bq +\,\,\tilde{\!\!\bd}$.
We find the symmetry properties of $\tbA (k_\alpha,x_3,\omega) $ by applying equations (\ref{eqap197}) $-$ (\ref{eqap199}) to
 the left-hand sides of equations (\ref{eqsym}) $-$ (\ref{eqsymad}). This gives
\begin{eqnarray}
\{\tbA(-k_\alpha,x_3,\omega)\}^t\bN&=&-\bN\tbA (k_\alpha,x_3,\omega),\label{eqsymkx}\\
\{\tbA(-k_\alpha,x_3,\omega)\}^*\bJ&=&\bJ\,\,\,\tilde{\bar{\!\!\!\bA}}(k_\alpha,x_3,\omega),\label{eqsymconkx}\\
\{\tbA(k_\alpha,x_3,\omega)\}^\dagger\bK&=&-\bK\,\,\,\tilde{\bar{\!\!\!\bA}} (k_\alpha,x_3,\omega).\label{eqsymadkx}
\end{eqnarray}

\newpage
\noindent
{\Huge Unified matrix-vector wave equation, reciprocity and representations: Supplemental material}

\begin{footnotesize}
\section{Electromagnetic waves}

The electromagnetic matrix-vector wave equation for an inhomogeneous, anisotropic, dissipative medium is derived in Appendix C in the main paper.
For the special case of an isotropic medium we have ${\cal E}_{ik}={\cal E}\delta_{ik}$ and ${\mu}_{km}={\mu}\delta_{km}$,  or
\begin{eqnarray}
 \bfcalE_1=\begin{pmatrix} 
{\cal E} & 0 \\
0 &  {\cal E} 
\end{pmatrix},\quad
 \bfcalE_3=\begin{pmatrix} 
0  \\
0
\end{pmatrix},\quad
{\cal E}_{33}={\cal E},\quad
 \bfmu_1=\begin{pmatrix} 
\mu & 0 \\
0 &  \mu 
\end{pmatrix},\quad
 \bfmu_3=\begin{pmatrix} 
0  \\
0
\end{pmatrix},\quad
\mu_{33}=\mu.
\end{eqnarray}
For this situation the wave and source vectors read 
\begin{eqnarray}\label{AEeq6mvffgg}
\bq  = \begin{pmatrix} {\bf E}_0\\
{\bf H}_0 \end{pmatrix},\quad
  \bd  = \begin{pmatrix}
\frac{1}{\i\omega}\bfpartiala({\cal E}^{-1}J_3^e) -  {\bf J}_0^m\\
 - {\bf J}_0^e+ \frac{1}{\i\omega}\bfpartialb (\mu^{-1}J_3^m)\end{pmatrix},
\end{eqnarray}
and the operator matrices reduce to
\begin{eqnarray}
\bfAhat_{12}&=& \i\omega\bfmu_1  +\frac{1}{\i\omega}\bfpartiala{\cal E}^{-1}\bfpartiala^t=
\begin{pmatrix}
\i\omega\mu-\frac{1}{\i\omega}\partial_1
\frac{1}{{\cal E}}\partial_1& -\frac{1}{\i\omega}\partial_1
\frac{1}{{\cal E}}\partial_2\\
-\frac{1}{\i\omega}\partial_2
\frac{1}{{\cal E}}\partial_1
&\i\omega\mu-\frac{1}{\i\omega}\partial_2
\frac{1}{{\cal E}}\partial_2
\end{pmatrix},\label{Aeqa12pr}\\
\bfAhat_{21}&=&\i\omega \bfcalE_1 +\frac{1}{\i\omega}\bfpartialb\mu^{-1}\bfpartialb^t=
\begin{pmatrix} \i\omega {\cal E}
-\frac{1}{\i\omega}\partial_2
\frac{1}{\mu}\partial_2&
\frac{1}{\i\omega}\partial_2
\frac{1}{\mu}\partial_1\\
\frac{1}{\i\omega}\partial_1
\frac{1}{\mu}\partial_2
&\i\omega  {\cal E} -\frac{1}{\i\omega}\partial_1 \frac{1}{\mu}\partial_1
\end{pmatrix}\label{Aeqa21pr}
\end{eqnarray}
and $\bfAhat_{11}=\bfAhat_{22}=\bfO$, where $\bfO$ is a $2\times 2$ zero matrix.

\section{Elastodynamic waves}

The elastodynamic matrix-vector wave equation for an inhomogeneous, anisotropic, dissipative solid is derived in Appendix D in the main paper.
For the special case of an isotropic medium we have
\begin{eqnarray}
c_{ijkl}&=&\lambda\delta_{ij}\delta_{kl}+ \mu(\delta_{ik}\delta_{jl}+\delta_{il}\delta_{jk}),\label{eq143a}\\
\rho_{ij}&=&\rho\delta_{ij},\label{eq144a}
\end{eqnarray}
with Lam\'e parameters $\lambda=\lambda({\bf x},\omega)$ and $\mu=\mu({\bf x},\omega)$ and mass density  $\rho=\rho(\bx,\omega)$.
Hence,   the mass density matrix reduces to $\R=\rho{\bf I}$. For the stiffness matrices ${\bfC}_{jl}$ we have $({\bfC}_{jl})_{ik}=c_{ijkl}=\lambda\delta_{ij}\delta_{kl}+ \mu(\delta_{ik}\delta_{jl}+\delta_{il}\delta_{jk})$, hence
\begin{eqnarray}
\begin{array}{llll}
{\bfC}_{11}=\begin{pmatrix} K_c&0&0 \\ 0&\mu&0 \\
0&0&\mu\end{pmatrix},& {\bfC}_{12}=
\begin{pmatrix}0&\lambda&0 \\ \mu&0&0 \\ 0&0&0\end{pmatrix},&
{\bfC}_{13}=
\begin{pmatrix}0&0&\lambda \\ 0&0&0 \\ \mu&0&0\end{pmatrix},&\\
&&&\\
{\bfC}_{21}=
\begin{pmatrix}0&\mu&0 \\ \lambda&0&0 \\ 0&0&0\end{pmatrix},&
{\bfC}_{22}=
\begin{pmatrix}\mu&0&0 \\ 0&K_c&0 \\ 0&0&\mu\end{pmatrix},&
{\bfC}_{23}=
\begin{pmatrix}0&0&0 \\ 0&0&\lambda \\ 0&\mu&0\end{pmatrix},&
\end{array}
\end{eqnarray}
\begin{eqnarray}
\begin{array}{llll}
{\bfC}_{31}=
\begin{pmatrix}0&0&\mu \\ 0&0&0 \\ \lambda&0&0\end{pmatrix},&
{\bfC}_{32}=
\begin{pmatrix}0&0&0 \\ 0&0&\mu \\ 0&\lambda&0\end{pmatrix},&
{\bfC}_{33}=
\begin{pmatrix}\mu&0&0 \\ 0&\mu&0 \\ 0&0&K_c\end{pmatrix},&\\
\end{array}
\end{eqnarray}
with $K_c=K_c({\bf x},\omega)=\lambda({\bf x},\omega) +2\mu({\bf x},\omega)$.
For this situation the operator matrices reduce to
\begin{eqnarray}
\bfAhat_{11}&=&\begin{pmatrix} 0 & 0 &
-\partial_1
\frac{\lambda}{\lambda+2\mu}\\
0 & 0 & -\partial_2
\frac{\lambda}{\lambda+2\mu}\\
-\partial_1 & -\partial_2&0
\end{pmatrix},\label{eqa11el}\\
&&\nonumber\\
\bfAhat_{12}&=&\begin{pmatrix}
\i\omega\rho-\frac{1}{\i\omega}\bigl(
\partial_1\nu_1\partial_1+
\partial_2\mu\partial_2\bigr)
&-\frac{1}{\i\omega}\bigl(\partial_2 \mu
\partial_1+ \partial_1 \nu_2\partial_2\bigr)&0\\
-\frac{1}{\i\omega}\bigl(\partial_2
\nu_2
\partial_1+ \partial_1 \mu\partial_2 \bigr)
&\i\omega\rho-\frac{1}{\i\omega}\bigl(
\partial_1\mu\partial_1+ \partial_2\nu_1\partial_2\bigr)&0\\
0&0&\i\omega\rho
\end{pmatrix},\label{eqa12el}\\
&&\nonumber\\
\bfAhat_{21}&=&\begin{pmatrix}
\frac{\i\omega}{\mu} & 0 & 0\\
0 & \frac{\i\omega}{\mu} & 0\\
0 & 0 & \frac{\i\omega}{\lambda+2\mu}
\end{pmatrix},\label{eqa21el}\\
&&\nonumber\\
\bfAhat_{22}&=&\begin{pmatrix} 0 & 0 &
-\partial_1
\\
0 & 0 & -\partial_2
\\
-\frac{\lambda}{\lambda+2\mu}\partial_1 &
-\frac{\lambda}{\lambda+2\mu}\partial_2&0
\end{pmatrix},\label{eqa22el}
\end{eqnarray}
where
\begin{eqnarray}
\nu_1&=&\nu_1({\bf x},\omega)=
4\mu\Bigl(\frac{\lambda+\mu}{\lambda+2\mu}\Bigr),\\
\nu_2&=&\nu_2({\bf x},\omega)=
2\mu\Bigl(\frac{\lambda}{\lambda+2\mu}\Bigr).
\end{eqnarray}

\section{Poroelastodynamic waves}\label{sec3}

Equations (E.14) $-$ (E.17) in Appendix E in the main paper form the starting point for deriving a poroelastodynamic
matrix-vector wave equation 
for the quantities $\bftau_3^b$, $p^f$,   ${\bf v}^s$ and $w_3$ in vector $\bq$. 
Pre-multiplying all terms in  equation (E.15) by $\eta\bfdelta_i^t{\bfk}^{-1}$, with $3\times 1$ unit vector $(\bfdelta_i)_j=\delta_{ij}$, gives
\begin{eqnarray}\label{Eeqg28anpes}
-\i\omega\bfdelta_i^t{\R}^f{\bf v}^s +\eta\bfdelta_i^t{\bfk}^{-1}\bfdelta_j w_j + \partial_i{p^f}= f_i^f.
\end{eqnarray}
We separate the derivatives in the $x_3$-direction from the lateral
derivatives in equations (E.14),
(\ref{Eeqg28anpes}),
(E.16) and (E.17), according to
\begin{eqnarray}
-\partial_3\bftau_3^b &=&
\i\omega{\R}^b{\bf v}^s
+\i\omega{\R}^f (\bfdelta_\alpha w_\alpha+\bfdelta_3 w_3 \bigr)
+\partial_\alpha\bftau_\alpha^b+
{\bf f}^b,
\label{Eeq6.8bbvanpes}\\
\partial_3  p^f &= &
\i\omega\bfdelta_3^t{\R}^f{\bf v}^s -\eta\bfdelta_3^t{\bfk}^{-1}\bigl(\bfdelta_\alpha w_\alpha+\bfdelta_3 w_3 \bigr)+ f_3^f,\label{Eeq5.8mvanpes}\\
\partial_3 {\bf v}^s&=&  {\bfC}_{33}^{-1}\Bigl(-\i\omega\bftau_3^b-\frac{\i\omega}{ M}\bfc_3 p^f - {\bfC}_{3\beta}\partial_\beta{\bf v}^s
+{\bfC}_{3l}\hh_l\Bigr),\label{Eeq6.19bbvanpes}\\
\partial_3  w_3& =&  \frac{\i\omega}{ M} p^f - \frac{ 1}{ M}\bigl( \bfc_\beta^t \partial_\beta{\bf v}^s+ \bfc_3^t \partial_3{\bf v}^s\bigr)
- \partial_\alpha w_\alpha+
\frac{1}{M}\bfc_l^t\hh_l+\qf.\label{Eeq5.14mvabpes}
\end{eqnarray}
The field components $\bftau_\alpha^b$ and $w_\alpha$ need to be eliminated.
From equation (E.16) we obtain
\begin{eqnarray}\label{eq264}
\bftau_\alpha^b=-\frac{1}{\i\omega}\Bigl(  {\bfC}_{\alpha\beta}\partial_\beta {\bf v}^s+  {\bfC}_{\alpha 3}\partial_3 {\bf v}^s+\frac{\i\omega}{ M}\bfc_\alpha p^f -{\bfC}_{\alpha l}\hh_l\Bigr).
\end{eqnarray}
Pre-multiplying all terms in  equation (E.15) by $\bfdelta_\alpha^t$ gives
\begin{eqnarray}\label{eqF169}
 w_\alpha =\bfdelta_\alpha^t\Bigl(\frac{\i\omega}{\eta}{\bfk}{\R}^f{\bf v}^s -  \frac{1}{\eta}{\bfk}\bigl(\bfdelta_\beta\partial_\beta{p^f}+\bfdelta_3\partial_3{p^f}\bigr)+\frac{1}{\eta}{\bfk}\bfdelta_j f_j^f\Bigr).
\end{eqnarray}
Using this in equation (\ref{Eeq5.8mvanpes}) gives
\begin{footnotesize}
\begin{eqnarray}
\partial_3  p^f &=& 
\i\omega\bfdelta_3^t{\R}^f{\bf v}^s -\eta\bfdelta_3^t{\bfk}^{-1}\bfdelta_3 w_3
-\bfdelta_3^t{\bfk}^{-1}\bfdelta_\alpha\bfdelta_\alpha^t\Bigl(\i\omega{\bfk}{\R}^f{\bf v}^s -  {\bfk}\bigl(\bfdelta_\beta\partial_\beta{p^f}+\bfdelta_3\partial_3{p^f}\bigr)+{\bfk}\bfdelta_j f_j^f\Bigr)+ f_3^f,\nonumber\\&&\\
\bigl(1-\bfdelta_3^t{\bfk}^{-1}\bfdelta_\alpha\bfdelta_\alpha^t{\bfk}\bfdelta_3\bigr)\partial_3  p^f &=&\i\omega\bfdelta_3^t{\R}^f{\bf v}^s -\eta\bfdelta_3^t{\bfk}^{-1}\bfdelta_3 w_3
-\bfdelta_3^t{\bfk}^{-1}\bfdelta_\alpha\bfdelta_\alpha^t\bigl(\i\omega{\bfk}{\R}^f{\bf v}^s - {\bfk}\bfdelta_\beta\partial_\beta{p^f}+{\bfk}\bfdelta_j f_j^f\bigr)+ f_3^f,\\
\partial_3  p^f &=&\frac{1}{\b}\Bigl(\i\omega\bfdelta_3^t{\R}^f{\bf v}^s -\eta\bfdelta_3^t{\bfk}^{-1}\bfdelta_3 w_3
-\bfdelta_3^t{\bfk}^{-1}\bfdelta_\alpha\bfdelta_\alpha^t\bigl(\i\omega{\bfk}{\R}^f{\bf v}^s - {\bfk}\bfdelta_\beta\partial_\beta{p^f}+{\bfk}\bfdelta_j f_j^f\bigr)+ f_3^f\Bigr)\label{eqF169b}\\
&=&\frac{1}{\b}\Bigl(\bfdelta_3^t{\bfk}^{-1}\bfdelta_\alpha\bfdelta_\alpha^t{\bfk}\bfdelta_\beta\partial_\beta{p^f}
+\i\omega\bfdelta_3^t({\bf I}-{\bfk}^{-1}\bfdelta_\alpha\bfdelta_\alpha^t{\bfk}){\R}^f{\bf v}^s-\eta\bfdelta_3^t{\bfk}^{-1}\bfdelta_3 w_3
-\bfdelta_3^t{\bfk}^{-1}\bfdelta_\alpha\bfdelta_\alpha^t{\bfk}\bfdelta_jf_j^f+f_3^f\Bigr),\nonumber
\end{eqnarray}
\end{footnotesize}
with
\begin{eqnarray}
\b=1-\bfdelta_3^t{\bfk}^{-1}\bfdelta_\alpha\bfdelta_\alpha^t{\bfk}\bfdelta_3.
\end{eqnarray}
Substituting  this into equation (\ref{eqF169}) gives
\begin{footnotesize}
\begin{eqnarray}
w_\alpha &=&\bfdelta_\alpha^t\Bigl(\frac{\i\omega}{\eta}{\bfk}{\R}^f{\bf v}^s -  \frac{1}{\eta}{\bfk}\bfdelta_\beta\partial_\beta{p^f}+\frac{1}{\eta}{\bfk}\bfdelta_j f_j^f\Bigr)\nonumber\\
&&
-\frac{1}{\eta\b}\bfdelta_\alpha^t{\bfk}\bfdelta_3\Bigl(\bfdelta_3^t{\bfk}^{-1}\bfdelta_\gamma\bfdelta_\gamma^t{\bfk}\bfdelta_\beta\partial_\beta{p^f}
+\i\omega\bfdelta_3^t({\bf I}-{\bfk}^{-1}\bfdelta_\gamma\bfdelta_\gamma^t{\bfk}){\R}^f{\bf v}^s-\eta\bfdelta_3^t{\bfk}^{-1}\bfdelta_3 w_3
-\bfdelta_3^t{\bfk}^{-1}\bfdelta_\gamma\bfdelta_\gamma^t{\bfk}\bfdelta_jf_j^f+f_3^f\Bigr),\\
w_\alpha &=&-\frac{1}{\eta}\bigl(\bfdelta_\alpha^t{\bfk}\bfdelta_\beta + \frac{1}{\b}\bfdelta_\alpha^t{\bfk}\bfdelta_3\bfdelta_3^t{\bfk}^{-1}\bfdelta_\gamma\bfdelta_\gamma^t{\bfk}\bfdelta_\beta\bigr)\partial_\beta p^f
+\frac{\i\omega}{\eta}\bfdelta_\alpha^t{\bfk}\bigl({\bf I}-\frac{1}{\b}\bfdelta_3\bfdelta_3^t({\bf I}-{\bfk}^{-1}\bfdelta_\gamma\bfdelta_\gamma^t{\bfk})\bigr){\R}^f{\bf v}^s\nonumber\\
&&+\frac{1}{\b}\bfdelta_\alpha^t{\bfk}\bfdelta_3\bfdelta_3^t{\bfk}^{-1}\bfdelta_3 w_3+\frac{1}{\eta}\bfdelta_\alpha^t{\bfk}\bigl({\bf I}+
\frac{1}{\b}\bfdelta_3\bfdelta_3^t{\bfk}^{-1}\bfdelta_\gamma\bfdelta_\gamma^t{\bfk}\bigr)\bfdelta_jf_j^f-\frac{1}{\eta\b}\bfdelta_\alpha^t{\bfk}\bfdelta_3f_3^f.
\label{eqF181sk}
\end{eqnarray}
\end{footnotesize}
We are now ready to eliminate $\bftau_\alpha^b$ and $w_\alpha$ from equations (\ref{Eeq6.8bbvanpes}) $-$ (\ref{Eeq5.14mvabpes}).
The expression for $\partial_3{\bf v}^s$, equation (\ref{Eeq6.19bbvanpes}), already has the desired form.
Substituting equations (\ref{eq264}) and (\ref{eqF181sk}) into equation (\ref{Eeq6.8bbvanpes}), we obtain
\begin{footnotesize}
\begin{eqnarray}\label{Eeq6.8bcvanpes}
-\partial_3\bftau_3^b &=&
-\frac{\i\omega}{\eta}{\R}^f\bfdelta_\alpha\bigl(\bfdelta_\alpha^t{\bfk}\bfdelta_\beta + \frac{1}{\b}\bfdelta_\alpha^t{\bfk}\bfdelta_3\bfdelta_3^t{\bfk}^{-1}\bfdelta_\gamma\bfdelta_\gamma^t{\bfk}\bfdelta_\beta\bigr)\partial_\beta p^f
+\i\omega{\R}^b{\bf v}^s-\frac{\omega^2}{\eta}{\R}^f\bfdelta_\alpha\bfdelta_\alpha^t{\bfk}\bigl({\bf I}-\frac{1}{\b}\bfdelta_3\bfdelta_3^t({\bf I}-{\bfk}^{-1}\bfdelta_\gamma\bfdelta_\gamma^t{\bfk})\bigr){\R}^f{\bf v}^s
\nonumber\\
&&+\i\omega{\R}^f\bigl({\bf I}+\frac{1}{\b} \bfdelta_\alpha\bfdelta_\alpha^t{\bfk}\bfdelta_3\bfdelta_3^t{\bfk}^{-1}  \bigr)\bfdelta_3 w_3
+\frac{\i\omega}{\eta}{\R}^f\bfdelta_\alpha\bfdelta_\alpha^t{\bfk}\bigl({\bf I}+\frac{1}{\b}\bfdelta_3\bfdelta_3^t{\bfk}^{-1}\bfdelta_\gamma\bfdelta_\gamma^t{\bfk}\bigr)\bfdelta_jf_j^f
-\frac{\i\omega}{\eta\b}{\R}^f\bfdelta_\alpha\bfdelta_\alpha^t{\bfk}\bfdelta_3f_3^f
\nonumber\\
&&-\frac{1}{\i\omega}\partial_\alpha \Bigl(
 {\bfC}_{\alpha\beta}\partial_\beta {\bf v}^s + {\bfC}_{\alpha 3}\partial_3 {\bf v}^s
+\frac{\i\omega}{ M}\bfc_\alpha p^f\Bigr) + {\bf f}^b
+\frac{1}{\i\omega}\partial_\alpha \bigl(
{\bfC}_{\alpha l}\hh_l\bigr).
\end{eqnarray}
\end{footnotesize}
Upon substitution of equation (\ref{Eeq6.19bbvanpes}) and using
\begin{eqnarray}
{\bfk}^{-1}\bfdelta_\gamma\bfdelta_\gamma^t{\bfk}&=&{\bfk}^{-1}\bigl({\bf I}-\bfdelta_3\bfdelta_3^t\bigr){\bfk}= {\bf I}-{\bfk}^{-1}\bfdelta_3\bfdelta_3^t{\bfk},\label{eqFF174}\\
\bfdelta_3^t{\bfk}^{-1}\bfdelta_\gamma\bfdelta_\gamma^t{\bfk}\bfdelta_\beta&=&
\bfdelta_3^t({\bf I}-{\bfk}^{-1}\bfdelta_3\bfdelta_3^t{\bfk})\bfdelta_\beta=
 -\bfdelta_3^t{\bfk}^{-1}\bfdelta_3\bfdelta_3^t{\bfk}\bfdelta_\beta,\label{eqFF175}
\end{eqnarray}
we obtain
\begin{footnotesize}
\begin{eqnarray}\label{Eeq6.8bdvanpes}
-\partial_3\bftau_3^b &=&\partial_\alpha\bigl( {\bfC}_{\alpha 3}
 {\bfC}_{33}^{-1}\bftau_3^b\bigr)
-\frac{\i\omega}{\eta}{\R}^f\bfdelta_\alpha\bigl(\bfdelta_\alpha^t{\bfk}\bfdelta_\beta - \frac{1}{\b}\bfdelta_\alpha^t{\bfk}\bfdelta_3\bfdelta_3^t{\bfk}^{-1}\bfdelta_3\bfdelta_3^t{\bfk}\bfdelta_\beta\bigr)\partial_\beta p^f
-\frac{1}{\i\omega}\partial_\alpha\Bigl(\frac{\i\omega}{ M} {\bfu}_\alpha p^f + {\bfU}_{\alpha\beta}\partial_\beta {\bf v}^s\Bigr)\nonumber\\
&&+\i\omega{\R}^b{\bf v}^s-\frac{\omega^2}{\eta}{\R}^f\bfdelta_\alpha\bfdelta_\alpha^t{\bfk}\bigl({\bf I}-\frac{1}{\b}\bfdelta_3\bfdelta_3^t{\bfk}^{-1}\bfdelta_3\bfdelta_3^t{\bfk}\bigr){\R}^f{\bf v}^s
+\i\omega{\R}^f\bigl({\bf I}+\frac{1}{\b} \bfdelta_\alpha\bfdelta_\alpha^t{\bfk}\bfdelta_3\bfdelta_3^t{\bfk}^{-1}  \bigr)\bfdelta_3 w_3
\nonumber\\&&
+\frac{\i\omega}{\eta}{\R}^f\bfdelta_\alpha\bfdelta_\alpha^t{\bfk}\bigl({\bf I}+\frac{1}{\b}\bfdelta_3\bfdelta_3^t{\bfk}^{-1}\bfdelta_\gamma\bfdelta_\gamma^t{\bfk}\bigr)\bfdelta_jf_j^f
-\frac{\i\omega}{\eta\b}{\R}^f\bfdelta_\alpha\bfdelta_\alpha^t{\bfk}\bfdelta_3f_3^f
+ {\bf f}^b
+\frac{1}{\i\omega}\partial_\alpha \bigl(
{\bfU}_{\alpha l}\hh_l\bigr),
\end{eqnarray}
\end{footnotesize}
with 
\begin{eqnarray}
 {\bfU}_{\alpha l}&=& {\bfC}_{\alpha l}-
 {\bfC}_{\alpha 3} {\bfC}_{33}^{-1} {\bfC}_{3l},\\
 {\bfu}_{l}&=&\bfc_l-  {\bfC}_{l3} {\bfC}_{33}^{-1}\bfc_3,
\end{eqnarray}
where ${\bfU}_{\alpha 3}={\bf O}$,  ${\bfu}_{3}={\bf 0}$ and
\begin{eqnarray}\label{Esymstif3vanpes}
 {\bfU}_{\alpha\beta}^t= {\bfU}_{\beta\alpha}
\end{eqnarray}
on account of ${\bfC}_{jl}^t={\bfC}_{lj}$.
Using equations (\ref{eqFF174}) and  (\ref{eqFF175})  in equation (\ref{eqF169b}), we obtain
\begin{eqnarray}
\partial_3 p^f
&=&\frac{1}{\b}\Bigl(-\bfdelta_3^t{\bfk}^{-1}\bfdelta_3\bfdelta_3^t{\bfk}\bfdelta_\beta\partial_\beta{p^f}
+\i\omega\bfdelta_3^t{\bfk}^{-1}\bfdelta_3\bfdelta_3^t{\bfk}{\R}^f{\bf v}^s-\eta\bfdelta_3^t{\bfk}^{-1}\bfdelta_3 w_3
-\bfdelta_3^t{\bfk}^{-1}\bfdelta_\alpha\bfdelta_\alpha^t{\bfk}\bfdelta_jf_j^f+f_3^f\Bigr),\label{eqF169bag}
\end{eqnarray}
with 
%
\begin{eqnarray}\label{eq280}
\b=1-\bfdelta_3^t{\bfk}^{-1}\bfdelta_\alpha\bfdelta_\alpha^t{\bfk}\bfdelta_3=
\bfdelta_3^t({\bf I}-{\bfk}^{-1}\bfdelta_\alpha\bfdelta_\alpha^t{\bfk})\bfdelta_3=
\bfdelta_3^t{\bfk}^{-1}\bfdelta_3\bfdelta_3^t{\bfk}\bfdelta_3=
(\bfdelta_3^t{\bfk}^{-1}\bfdelta_3)(\bfdelta_3^t{\bfk}\bfdelta_3)=
(\bfdelta_3^t{\bfk}\bfdelta_3)(\bfdelta_3^t{\bfk}^{-1}\bfdelta_3)=
\bfdelta_3^t{\bfk}\bfdelta_3\bfdelta_3^t{\bfk}^{-1}\bfdelta_3.\nonumber\\&&
\end{eqnarray}
Substituting equations (\ref{Eeq6.19bbvanpes}) and (\ref{eqF181sk})
into equation (\ref{Eeq5.14mvabpes}), we obtain
\begin{eqnarray}
\partial_3  w_3 &=&
 \frac{\i\omega}{ M} p^f
  - \frac{ 1}{ M}\bfc_\beta^t \partial_\beta{\bf v}^s  
  +\frac{ 1}{ M}  \bfc_3^t  {\bfC}_{33}^{-1}\Bigl(\i\omega\bftau_3^b
+\frac{\i\omega}{ M}\bfc_3 p^f + {\bfC}_{3\beta}\partial_\beta{\bf v}^s-{\bfC}_{3l}\hh_l
\Bigr)\nonumber\\
  && -\partial_\alpha\Bigl(-\frac{1}{\eta}\bigl(\bfdelta_\alpha^t{\bfk}\bfdelta_\beta - \frac{1}{\b}\bfdelta_\alpha^t{\bfk}\bfdelta_3\bfdelta_3^t{\bfk}^{-1}\bfdelta_3\bfdelta_3^t{\bfk}\bfdelta_\beta\bigr)\partial_\beta p^f
+\frac{\i\omega}{\eta}\bfdelta_\alpha^t{\bfk}\bigl({\bf I}-\frac{1}{\b}\bfdelta_3\bfdelta_3^t{\bfk}^{-1}\bfdelta_3\bfdelta_3^t{\bfk}\bigr){\R}^f{\bf v}^s\nonumber\\
&&+\frac{1}{\b}\bfdelta_\alpha^t{\bfk}\bfdelta_3\bfdelta_3^t{\bfk}^{-1}\bfdelta_3 w_3+\frac{1}{\eta}\bfdelta_\alpha^t{\bfk}\bigl({\bf I}+
\frac{1}{\b}\bfdelta_3\bfdelta_3^t{\bfk}^{-1}\bfdelta_\gamma\bfdelta_\gamma^t{\bfk}\bigr)\bfdelta_jf_j^f-\frac{1}{\eta\b}\bfdelta_\alpha^t{\bfk}\bfdelta_3f_3^f \Bigr)+
\frac{1}{M}\bfc_l^t\hh_l+\qf,
\end{eqnarray}
or
\begin{eqnarray}
\partial_3  w_3 &=&\frac{\i\omega}{ M}  \bfc_3^t  {\bfC}_{33}^{-1}\bftau_3^b+ \frac{\i\omega}{ M} p^f + \frac{\i\omega}{M^2}  \bfc_3^t  {\bfC}_{33}^{-1}\bfc_3 p^f
+\partial_\alpha\frac{1}{\eta}\bigl(\bfdelta_\alpha^t{\bfk}\bfdelta_\beta - \frac{1}{\b}\bfdelta_\alpha^t{\bfk}\bfdelta_3\bfdelta_3^t{\bfk}^{-1}\bfdelta_3\bfdelta_3^t{\bfk}\bfdelta_\beta\bigr)\partial_\beta p^f\nonumber\\
&&  - \frac{ 1}{ M}{\bfu}_\beta^t \partial_\beta{\bf v}^s
-\partial_\alpha\frac{\i\omega}{\eta}\bfdelta_\alpha^t{\bfk}\bigl({\bf I}-\frac{1}{\b}\bfdelta_3\bfdelta_3^t{\bfk}^{-1}\bfdelta_3\bfdelta_3^t{\bfk}\bigr){\R}^f{\bf v}^s\nonumber\\
&&-\partial_\alpha\Bigl(\frac{1}{\b}\bfdelta_\alpha^t{\bfk}\bfdelta_3\bfdelta_3^t{\bfk}^{-1}\bfdelta_3 w_3+\frac{1}{\eta}\bfdelta_\alpha^t{\bfk}\bigl({\bf I}+
\frac{1}{\b}\bfdelta_3\bfdelta_3^t{\bfk}^{-1}\bfdelta_\gamma\bfdelta_\gamma^t{\bfk}\bigr)\bfdelta_jf_j^f-\frac{1}{\eta\b}\bfdelta_\alpha^t{\bfk}\bfdelta_3f_3^f \Bigr)+
\frac{1}{M}{\bfu}_\alpha^t\hh_\alpha+\qf. \label{Eeq5.14mxanpes} 
\end{eqnarray}
Equations (\ref{Eeq6.8bdvanpes}), (\ref{eqF169bag}), (\ref{Eeq6.19bbvanpes}) and (\ref{Eeq5.14mxanpes}) can be cast  in the form of matrix-vector wave equation  (1) in the main paper,
with the  wave vector $\bq =\bq ({\bf x},\omega)$ and
 the source vector $\bd =\bd ({\bf x},\omega)$ defined as
\begin{eqnarray}\label{Eeq6mvanpes}
\bq  = \begin{pmatrix}- \bftau_3^b \\ p^f \\ 
{\bf v}^s\\ w_3 \\ 
\end{pmatrix},\quad
  \bd  = \begin{pmatrix}
 \frac{\i\omega}{\eta}{\R}^f\bfdelta_\alpha\bfdelta_\alpha^t{\bfk}\bigl({\bf I}+\frac{1}{\b}\bfdelta_3\bfdelta_3^t{\bfk}^{-1}\bfdelta_\gamma\bfdelta_\gamma^t{\bfk}\bigr)\bfdelta_jf_j^f
-\frac{\i\omega}{\eta\b}{\R}^f\bfdelta_\alpha\bfdelta_\alpha^t{\bfk}\bfdelta_3f_3^f
+ {\bf f}^b+\frac{1}{\i\omega}\partial_\alpha \bigl(
{\bfU}_{\alpha \beta}\hh_\beta\bigr)\\
\frac{1}{\b}\bigl(-\bfdelta_3^t{\bfk}^{-1}\bfdelta_\alpha\bfdelta_\alpha^t{\bfk}\bfdelta_jf_j^f+f_3^f\bigr)\\
{\bfC}_{33}^{-1}
{\bfC}_{3l}\hh_l\\
-\partial_\alpha\Bigl(\frac{1}{\eta}\bfdelta_\alpha^t{\bfk}\bigl({\bf I}+
\frac{1}{\b}\bfdelta_3\bfdelta_3^t{\bfk}^{-1}\bfdelta_\gamma\bfdelta_\gamma^t{\bfk}\bigr)\bfdelta_jf_j^f-\frac{1}{\eta\b}\bfdelta_\alpha^t{\bfk}\bfdelta_3f_3^f \Bigr)+
\frac{1}{M}{\bfu}_\alpha^t\hh_\alpha+\qf
   \end{pmatrix}
\end{eqnarray}
and the operator matrix $\bfAhat=\bfAhat({\bf x},\omega)$ having the form of equation  (2) in the main paper, with
\begin{eqnarray}
\bfAhat_{11} &=& \begin{pmatrix}
\bfAhat_{11}^{11} & \bfahat_{11}^{12}\\
\bfahat_{11}^{21} & \bfAhat_{11}^{22}\end{pmatrix},\quad
\bfAhat_{12} = \begin{pmatrix}
\bfAhat_{12}^{11} & \bfahat_{12}^{12}\\
\bfahat_{12}^{21} & \bfAhat_{12}^{22}\end{pmatrix},\label{eq221}\\
\bfAhat_{21} &=& \begin{pmatrix}
\bfAhat_{21}^{11} & \bfahat_{21}^{12}\\
\bfahat_{21}^{21} & \bfAhat_{21}^{22}\end{pmatrix},\quad
\bfAhat_{22} = \begin{pmatrix}
\bfAhat_{22}^{11} & \bfahat_{22}^{12}\\
\bfahat_{22}^{21} & \bfAhat_{22}^{22}\end{pmatrix},\label{eq222}
\end{eqnarray}
where, using equation (\ref{eq280}),
\begin{eqnarray}
\bfAhat_{11}^{11}&=& -\partial_\alpha{\bfC}_{\alpha 3}
{\bfC}_{33}^{-1},\\
\bfahat_{11}^{12}
&=&-\frac{\i\omega}{\eta\b}{\R}^f\Bigl(\b\bfdelta_\alpha\bfdelta_\alpha^t - \bfdelta_\alpha\bfdelta_\alpha^t{\bfk}\bfdelta_3\bfdelta_3^t{\bfk}^{-1}\bfdelta_3\bfdelta_3^t\Bigl){\bfk}\bfdelta_\beta\partial_\beta 
-\partial_\alpha\frac{1}{M}{\bfu}_\alpha
\nonumber\\&=&
-\frac{\i\omega}{\eta\b}{\R}^f\Bigl(\b{\bf I}-\bfdelta_3\b\bfdelta_3^t - \bfdelta_\alpha\bfdelta_\alpha^t{\bfk}\bfdelta_3\bfdelta_3^t{\bfk}^{-1}\bfdelta_3\bfdelta_3^t\Bigl){\bfk}\bfdelta_\beta\partial_\beta 
-\partial_\alpha\frac{1}{M}{\bfu}_\alpha\nonumber\\
&=&-\frac{\i\omega}{\eta\b}{\R}^f\Bigl(\b{\bf I}-(\bfdelta_3\bfdelta_3^t + \bfdelta_\alpha\bfdelta_\alpha^t){\bfk}\bfdelta_3\bfdelta_3^t{\bfk}^{-1}\bfdelta_3\bfdelta_3^t\Bigl){\bfk}\bfdelta_\beta\partial_\beta 
-\partial_\alpha\frac{1}{M}{\bfu}_\alpha\nonumber\\
&=&-\frac{\i\omega}{\eta}{\R}^f\Bigl({\bf I}-\frac{1}{\b}{\bfk}\bfdelta_3\bfdelta_3^t{\bfk}^{-1}\bfdelta_3\bfdelta_3^t\Bigl){\bfk}\bfdelta_\beta\partial_\beta 
-\partial_\alpha\frac{1}{M}{\bfu}_\alpha,\\
%
\bfahat_{11}^{21}&=&{\bf 0}^t,\\
\bfAhat_{11}^{22}&=&-\frac{1}{\b}\bfdelta_3^t{\bfk}^{-1}\bfdelta_3\bfdelta_3^t{\bfk}\bfdelta_\beta\partial_\beta,\\
%
\bfAhat_{12}^{11}&=& \i\omega{\R}^b-\frac{1}{\i\omega}\partial_\alpha {\bfU}_{\alpha\beta}\partial_\beta
-\frac{\omega^2}{\eta\b}{\R}^f\bfdelta_\alpha\bfdelta_\alpha^t{\bfk}\bigl(\b{\bf I}-\bfdelta_3\bfdelta_3^t{\bfk}^{-1}\bfdelta_3\bfdelta_3^t{\bfk}\bigr){\R}^f\nonumber\\
&=&\i\omega{\R}^b-\frac{1}{\i\omega}\partial_\alpha {\bfU}_{\alpha\beta}\partial_\beta
-\frac{\omega^2}{\eta\b}{\R}^f\bfdelta_\alpha\bfdelta_\alpha^t{\bfk}\bigl(\b\bfdelta_\beta\bfdelta_\beta^t+\bfdelta_3\b\bfdelta_3^t-\bfdelta_3\bfdelta_3^t{\bfk}^{-1}\bfdelta_3\bfdelta_3^t{\bfk}\bigr){\R}^f\nonumber\\
&=&\i\omega{\R}^b-\frac{1}{\i\omega}\partial_\alpha {\bfU}_{\alpha\beta}\partial_\beta
-\frac{\omega^2}{\eta\b}{\R}^f\bfdelta_\alpha\bfdelta_\alpha^t{\bfk}\bigl(\b\bfdelta_\beta\bfdelta_\beta^t-\bfdelta_3\bfdelta_3^t{\bfk}^{-1}\bfdelta_3\bfdelta_3^t{\bfk}({\bf I}-\bfdelta_3\bfdelta_3^t)\bigr){\R}^f\nonumber\\
&=&\i\omega{\R}^b-\frac{1}{\i\omega}\partial_\alpha {\bfU}_{\alpha\beta}\partial_\beta
-\frac{\omega^2}{\eta}{\R}^f\bfdelta_\alpha\bfdelta_\alpha^t\bigl({\bfk}-\frac{1}{\b}{\bfk}\bfdelta_3\bfdelta_3^t{\bfk}^{-1}\bfdelta_3\bfdelta_3^t{\bfk}\bigr)\bfdelta_\beta\bfdelta_\beta^t{\R}^f,\\
\bfahat_{12}^{12}&=&\frac{\i\omega}{\b}{\R}^f\bigl(\bfdelta_3\b+ \bfdelta_\alpha\bfdelta_\alpha^t{\bfk}\bfdelta_3\bfdelta_3^t{\bfk}^{-1} \bfdelta_3 \bigr)
=\frac{\i\omega}{\b}{\R}^f\bigl(\bfdelta_3\bfdelta_3^t+\bfdelta_\alpha\bfdelta_\alpha^t){\bfk}\bfdelta_3\bfdelta_3^t{\bfk}^{-1}\bfdelta_3\nonumber\\
&=&\frac{\i\omega}{\b}{\R}^f{\bfk}\bfdelta_3\bfdelta_3^t{\bfk}^{-1}\bfdelta_3,\\
\bfahat_{12}^{21}&=&\frac{\i\omega}{\b}\bfdelta_3^t{\bfk}^{-1}\bfdelta_3\bfdelta_3^t{\bfk}{\R}^f,\\
\bfAhat_{12}^{22}&=&-\frac{\eta}{\b}\bfdelta_3^t{\bfk}^{-1}\bfdelta_3,\\
%
\bfAhat_{21}^{11}&=&\i\omega{\bfC}_{33}^{-1},\\
\bfahat_{21}^{12}&=&-\frac{\i\omega}{M}{\bfC}_{33}^{-1}\bfc_3,\\
\bfahat_{21}^{21}&=&-\frac{\i\omega}{M}\bfc_3^t{\bfC}_{33}^{-1},\\
\bfAhat_{21}^{22}&=&\frac{ \i\omega}{M^2}\bfc_3^t{\bfC}_{33}^{-1}\bfc_3 +\frac{\i\omega}{M}
+\partial_\alpha\frac{1}{\eta}\bigl(\bfdelta_\alpha^t{\bfk}\bfdelta_\beta - \frac{1}{\b}\bfdelta_\alpha^t{\bfk}\bfdelta_3\bfdelta_3^t{\bfk}^{-1}\bfdelta_3\bfdelta_3^t{\bfk}\bfdelta_\beta\bigr)\partial_\beta,
\end{eqnarray}
\begin{eqnarray}
\bfAhat_{22}^{11}&=&-{\bfC}_{33}^{-1}
{\bfC}_{3\beta}\partial_\beta,\\
\bfahat_{22}^{12}&=&{\bf 0},\\
\bfahat_{22}^{21}&=& - \frac{ 1}{ M}{\bfu}_\beta^t \partial_\beta
-\partial_\alpha\frac{\i\omega}{\eta}\bfdelta_\alpha^t{\bfk}\bigl({\bf I}-\frac{1}{\b}\bfdelta_3\bfdelta_3^t{\bfk}^{-1}\bfdelta_3\bfdelta_3^t{\bfk}\bigr){\R}^f,\\
\bfAhat_{22}^{22}&=&-\partial_\alpha\frac{1}{\b}\bfdelta_\alpha^t{\bfk}\bfdelta_3\bfdelta_3^t{\bfk}^{-1}\bfdelta_3.\end{eqnarray}
For the special case of an isotropic medium we have
\begin{eqnarray}
 \c_{ijkl}&=&( K_{\rm G}-\frac{2}{3} G_{\rm fr})\delta_{ij}\delta_{kl}+ G_{\rm fr}(\delta_{ik}\delta_{jl}+\delta_{il}\delta_{jk}),\\
 C_{ij}&=& C\delta_{ij},\\
\rho_{ij}^b&=&\rho^b\delta_{ij},\\
\rho_{ij}^f&=&\rho^f\delta_{ij},\\
k_{ij}&=&k\delta_{ij}.
\end{eqnarray}
Here $ G_{\rm fr}$
is the complex frequency-dependent shear modulus of the framework of the grains when the fluid is absent. 
The elastic parameters $ K_{\rm G}$ (Gassmann modulus), $ C$ and $ M$ are given by \citep{Pride92JASA, Pride94PRB} 
\begin{eqnarray}
 K_{\rm G} &=& \frac{ K_{\rm fr}+\phi  K^f+(1+\phi) K^s \Delta}{1+ \Delta},\label{eqh9pe}\\
 C &=& \frac{ K^f+ K^s \Delta}{1+ \Delta},\label{eqh10pe}\\
%
 M &=& \frac{1}{\phi}\frac{ K^f}{1+ \Delta},\label{eqh11pe}\\
 \Delta &=& \frac{ K^f}{\phi ( K^s)^2}\bigl((1-\phi)
K^s- K_{\rm fr}\bigr),\label{eqh12pe}
\end{eqnarray}
where $ K^s$ and $ K^f$ are the solid and fluid  compression moduli
and $ K_{\rm fr}$ is the
compression modulus of the framework of the grains. These parameters can be
expressed in terms of Biot's parameters $ A$, $ N$, $Q$ and $ R$ \citep{Biot56JASA, Biot56JASA2}, according to
\begin{eqnarray}
 K_{\rm G}-\frac{2}{3} G_{\rm fr}= A+2 Q+ R,\quad
 G_{\rm fr}= N, \quad  C=\frac{ Q+ R}{\phi},
\quad  M=\frac{ R}{\phi^2}.
\end{eqnarray}
Hence,
\begin{eqnarray}
({\bfC}_{jl})_{ik}&=&\e_{ijkl}=\c_{ijkl}-\frac{C^2}{M}\delta_{ij}\delta_{kl}=S\delta_{ij}\delta_{kl}+N(\delta_{ik}\delta_{jl}+\delta_{il}\delta_{jk}),\label{eq138}\\
S&=&A-\frac{Q^2}{R},\\
\bfc_j&=&C\bfdelta_j,\\
\R^b&=&\rho^b{\bf I},\\
\R^f&=&\rho^f{\bf I},\\
{\bfk}&=&k{\bf I},\\
\b&=&1.
\end{eqnarray}
With these substitutions, we obtain for the  source vector
\begin{eqnarray}\label{Eeq6mvanpesis}
  \bd  = \begin{pmatrix}
\i\omega\rho^f \frac{k}{\eta}\bfdelta_\alpha f_\alpha^f
+ {\bf f}^b+\frac{1}{\i\omega}\partial_\alpha \bigl({\bfU}_{\alpha \beta}\hh_\beta\bigr)\\
f_3^f\\
{\bfC}_{33}^{-1}
{\bfC}_{3l}\hh_l\\
-\partial_\alpha\bigl(\frac{k}{\eta}f_\alpha^f\bigr)+
\frac{1}{M}{\bfu}_\alpha^t\hh_\alpha+\qf
\end{pmatrix}
\end{eqnarray}
and for the operator matrices
\begin{eqnarray}
\bfAhat_{11}^{11}&=& -\partial_\alpha{\bfC}_{\alpha 3}
{\bfC}_{33}^{-1},\\
\bfahat_{11}^{12}&=&-\i\omega\rho^f\frac{ k}{\eta}\bfdelta_\beta \partial_\beta-
\partial_\alpha\frac{1}{M}{\bfu}_\alpha,\\
\bfahat_{11}^{21}&=&{\bf 0}^t,\\
\bfAhat_{11}^{22}&=&0,\\
%
\bfAhat_{12}^{11}&=& -\frac{1}{\i\omega}\partial_\alpha  {\bfU}_{\alpha\beta}\partial_\beta+\i\omega\Bigl(\rho^b{\bf I}
+\i\omega(\rho^f)^2\frac{k}{\eta}\bfdelta_\alpha\bfdelta_\alpha^t\Bigr),\\
\bfahat_{12}^{12}&=&\i\omega\rho^f\bfdelta_3,\\
\bfahat_{12}^{21}&=&\i\omega\rho^f\bfdelta_3^t,\\
\bfAhat_{12}^{22}&=&-\frac{\eta}{ k},
\end{eqnarray}
\begin{eqnarray}
\bfAhat_{21}^{11}&=&\i\omega{\bfC}_{33}^{-1},\\
\bfahat_{21}^{12}&=&-\i\omega\frac{C}{M}{\bfC}_{33}^{-1}\bfdelta_3,\\
\bfahat_{21}^{21}&=&-\i\omega\frac{C}{M}\bfdelta_3^t{\bfC}_{33}^{-1},\\
\bfAhat_{21}^{22}&=&\i\omega\frac{C^2}{M^2}\bfdelta_3^t{\bfC}_{33}^{-1}\bfdelta_3 +\frac{\i\omega}{M}+\partial_\alpha\frac{ k}{\eta}\partial_\alpha,\\
%
\bfAhat_{22}^{11}&=&-{\bfC}_{33}^{-1}
{\bfC}_{3\beta}\partial_\beta,\\
\bfahat_{22}^{12}&=&{\bf 0},\\
\bfahat_{22}^{21}&=& -\i\omega\partial_\alpha\rho^f \frac{ k}{\eta}\bfdelta_\alpha^t-\frac{1}{M}{\bfu}_\beta^t\partial_\beta,\\
\bfAhat_{22}^{22}&=&0.\end{eqnarray}
Using this in equations (\ref{eq221}) and (\ref{eq222}) we obtain
\begin{eqnarray}
\bfAhat_{11}&=&\begin{pmatrix} 0 & 0 &
-\partial_1 \frac{S}{K_c}
&\frac{\rho^f}{\rho^E} \partial_1-\partial_1\frac{2CN}{MK_c}\\
0 & 0 & -\partial_2
\frac{S}{K_c}& \frac{\rho^f}{\rho^E}
\partial_2-\partial_2
\frac{2CN}{MK_c}\\
-\partial_1 & -\partial_2&0&0\\
0 & 0& 0 &0
\end{pmatrix},\label{eqa11biot}\\
\bfAhat_{12}&=&\begin{pmatrix}
\i\omega\bigl(\rho^b-\frac{(\rho^f)^2}{\rho^E}\bigr)
-\frac{1}{\i\omega}\bigl( \partial_1\nu_1\partial_1+
\partial_2N\partial_2\bigr)
&-\frac{1}{\i\omega}\bigl(\partial_2 N \partial_1+ \partial_1 \nu_2\partial_2 \bigr)
& 0 &0\\
-\frac{1}{\i\omega}\bigl(\partial_2 \nu_2\partial_1+ \partial_1 N\partial_2 \bigr)
&\i\omega\bigl(\rho^b-\frac{(\rho^f)^2}{\rho^E}\bigr)-\frac{1}{\i\omega}\bigl(
\partial_1N\partial_1+ \partial_2\nu_1\partial_2\bigr) & 0
& 0
\\ 0 & 0 & \i\omega\rho^b & \i\omega\rho^f  \\ 0 & 0 & \i\omega\rho^f & \i\omega\rho^E
\end{pmatrix},\nonumber\\
&&\label{eqa12biot}\\
\bfAhat_{21}&=&\begin{pmatrix}
\frac{\i\omega}{N} & 0 & 0 & 0\\
0 & \frac{\i\omega}{N} & 0 & 0\\
0 & 0 & \frac{\i\omega}{K_c} & -\frac{\i\omega C}{MK_c} \\
0 & 0 & -\frac{\i\omega C}{MK_c} &\i\omega\bigl(\frac{C^2}{M^2K_c} +\frac{1}{M}\bigr)-\partial_\alpha \frac{1}{\i\omega\rho^E}\partial_\alpha
\end{pmatrix},\label{eqa21biot}\\
\bfAhat_{22}&=&\begin{pmatrix} 0 & 0 &
-\partial_1 & 0
\\
0 & 0 & -\partial_2 & 0
\\
-\frac{S}{K_c}\partial_1 &
-\frac{S}{K_c}\partial_2&0 & 0\\
\partial_1\frac{\rho^f}{\rho^E}
-\frac{2CN}{MK_c}\partial_1 &
\partial_2\frac{\rho^f}{\rho^E}
-\frac{2CN}{MK_c}\partial_2  & 0 & 0
\end{pmatrix},\label{eqa22biot}
\end{eqnarray}
where
\begin{eqnarray}
\rho^E&=&\rho^E({\bf x},\omega)=-\frac{\eta}{\i\omega k},\\
\nu_1&=&\nu_1({\bf x},\omega)=
4N\Bigl(\frac{S+N}{K_c}\Bigr),\\
\nu_2&=&\nu_2({\bf x},\omega)=
2N\Bigl(\frac{S}{K_c}\Bigr),\\
 K_c&=& K_c({\bf x},\omega)= S+2 N.
\end{eqnarray}

\section{Piezoelectric waves}

Equations (F.9) $-$ (F.14) in Appendix F in the main paper form the starting point for deriving a piezoelectric
matrix-vector wave equation 
for the quantities $\bftau_3$, $ {\bf E}_0$, $ {\bf v}$ and $ {\bf H}_0$.
We separate the derivatives in the $x_3$-direction from the lateral derivatives, according to
\begin{eqnarray}
-\partial_3\bftau_3 &=& \i\omega\R{\bf v} + \partial_\alpha\bftau_\alpha+ {\bf f},\label{eq6.8bbpe}\\
\partial_3 {\bf E}_0 &=&\i\omega\bfmu_1 {\bf H}_0 +\i\omega \bfmu_3H_3+\bfpartiala E_3-  {\bf J}_0^m,\label{Aeqhh00ape}\\
\partial_3 {\bf v}&=& {\bfC}_{33}^{-1}\bigl(-\i\omega\bftau_3 -{\bfC}_{3\beta}\partial_\beta {\bf v} -\i\omega{\bfC}_{3l}({\bf D}_{1l}{\bf E}_0+{\bf D}_{3l}E_3)+{\bfC}_{3l}{\bf h}_l\bigr),\label{eq6.19bbpe}\\
 \partial_3 {\bf H}_0&=&\i\omega \bfcalE_1  {\bf E}_0+\i\omega \bfcalE_3 E_3+\bfpartialb  H_3 + \i\omega{\bf D}_{1k}^t\bftau_k  - {\bf J}_0^e.\label{Aeqee00ape}
\end{eqnarray}
The field components  $\bftau_\alpha$,  $E_3$ and $H_3$  need to be eliminated. We start by deriving an explicit expression for $E_3$. 
From equations (F.10) and (F.14) we obtain
\begin{eqnarray}
E_3&=&{\cal E}_{33}^{-1}\biggl(- \bfcalE_3^t {\bf E}_0  +\frac{1}{\i\omega}\bfpartiala^t {\bf H}_0
+{\bf D}_{3k}^t\Bigl(\frac{1}{\i\omega}{\bfC}_{k l} \partial_l {\bf v}  +{\bfC}_{k l}({\bf D}_{1l}{\bf E}_0+{\bf D}_{3l}E_3)- \frac{1}{\i\omega}{\bfC}_{k l}{\bf h}_l\Bigr)+\frac{1}{\i\omega} J_3^e\biggr),\nonumber\\&&\\
({\cal E}_{33}-{\bf D}_{3k}^t{\bfC}_{k l} {\bf D}_{3l})E_3&=&- \bfcalE_3^t {\bf E}_0  +\frac{1}{\i\omega}\bfpartiala^t {\bf H}_0
+{\bf D}_{3k}^t\Bigl(\frac{1}{\i\omega}{\bfC}_{k l} \partial_l {\bf v}  +{\bfC}_{k l}{\bf D}_{1l}{\bf E}_0- \frac{1}{\i\omega}{\bfC}_{k l}{\bf h}_l\Bigr)+\frac{1}{\i\omega} J_3^e,\\
E_3&=&(\eep)^{-1}\biggl(- (\beep)^t {\bf E}_0  +\frac{1}{\i\omega}\bfpartiala^t {\bf H}_0
+\frac{1}{\i\omega}{\bf D}_{3k}^t\Bigl({\bfC}_{k \beta} \partial_\beta {\bf v} +{\bfC}_{k 3} \partial_3 {\bf v} - {\bfC}_{k l}{\bf h}_l\Bigr)+\frac{1}{\i\omega} J_3^e\biggr),\label{eqe27}
\end{eqnarray}
with
\begin{eqnarray}
\eep&=&{\cal E}_{33}-{\bf D}_{3k}^t{\bfC}_{k l} {\bf D}_{3l},\\
(\beep)^t&=&\bfcalE_3^t -{\bf D}_{3k}^t{\bfC}_{k l} {\bf D}_{1l}.
\end{eqnarray}
The term $\partial_3 {\bf v}$ is eliminated from equation (\ref{eqe27}) by substituting
equation (\ref{eq6.19bbpe}), hence 
\begin{eqnarray}
E_3&=&(\eep)^{-1}\biggl(- (\beep)^t {\bf E}_0  +\frac{1}{\i\omega}\bfpartiala^t {\bf H}_0 +\frac{1}{\i\omega}{\bf D}_{3k}^t\Bigl({\bfC}_{k \beta} \partial_\beta {\bf v}  - {\bfC}_{k l}{\bf h}_l\Bigr)+\frac{1}{\i\omega} J_3^e\label{eqe35}\\
&&+\frac{1}{\i\omega}{\bf D}_{3k}^t{\bfC}_{k3}{\bfC}_{33}^{-1}\bigl(-\i\omega\bftau_3 -{\bfC}_{3\beta}\partial_\beta {\bf v} -\i\omega{\bfC}_{3l}({\bf D}_{1l}{\bf E}_0+{\bf D}_{3l}E_3)+{\bfC}_{3l}{\bf h}_l\bigr)\biggr),\nonumber\\
(\eep+{\bf D}_{3k}^t{\bfC}_{k3}{\bfC}_{33}^{-1}{\bfC}_{3l}{\bf D}_{3l})E_3&=&
- (\beep)^t {\bf E}_0  +\frac{1}{\i\omega}\bfpartiala^t {\bf H}_0 +\frac{1}{\i\omega}{\bf D}_{3k}^t\Bigl({\bfC}_{k \beta} \partial_\beta {\bf v}  - {\bfC}_{k l}{\bf h}_l\Bigr)+\frac{1}{\i\omega} J_3^e\label{eqe36}\\
&&+\frac{1}{\i\omega}{\bf D}_{3k}^t{\bfC}_{k3}{\bfC}_{33}^{-1}\bigl(-\i\omega\bftau_3 -{\bfC}_{3\beta}\partial_\beta {\bf v} -\i\omega{\bfC}_{3l}{\bf D}_{1l}{\bf E}_0+{\bfC}_{3l}{\bf h}_l\bigr),\nonumber\\
E_3&=&(\eepp)^{-1}\biggl(- (\beepp)^t {\bf E}_0  +\frac{1}{\i\omega}\bfpartiala^t {\bf H}_0+\frac{1}{\i\omega}{\bf D}_{3k}^t\Bigl({\bfU}_{k \beta} \partial_\beta {\bf v} - {\bfU}_{k l}{\bf h}_l\Bigr)
-{\bf D}_{3k}^t{\bfC}_{k 3}{\bfC}_{33}^{-1}\bftau_3+\frac{1}{\i\omega} J_3^e\biggr),\nonumber\\
&&\label{eqe37}
\end{eqnarray}
with
\begin{eqnarray}
\eepp&=&\eep+{\bf D}_{3k}^t{\bfC}_{k 3}{\bfC}_{33}^{-1}{\bfC}_{3l}{\bf D}_{3l}={\cal E}_{33}-{\bf D}_{3k}^t{\bfU}_{kl}{\bf D}_{3l},\\
(\beepp)^t&=&(\beep)^t+{\bf D}_{3k}^t{\bfC}_{k 3}{\bfC}_{33}^{-1}{\bfC}_{3l}{\bf D}_{1l}=\bfcalE_3^t-{\bf D}_{3k}^t{\bfU}_{kl}{\bf D}_{1l},\\
{\bfU}_{kl}&=&{\bfC}_{kl}-{\bfC}_{k 3}{\bfC}_{33}^{-1}{\bfC}_{3l},
\end{eqnarray}
where ${\bfU}_{k 3}={\bfU}_{3 l}={\bf O}$ and
\begin{eqnarray}\label{Esymstif3vanpess}
 {\bfU}_{\alpha\beta}^t= {\bfU}_{\beta\alpha}
\end{eqnarray}
on account of ${\bfC}_{jl}^t={\bfC}_{lj}$.
Next, we derive an  expression for $H_3$ from equation (F.12), according to
\begin{eqnarray}
H_3=\mu_{33}^{-1}\Bigl(- \bfmu_3^t {\bf H}_0  +\frac{1}{\i\omega}\bfpartialb^t {\bf E}_0+\frac{1}{\i\omega} J_3^m\Bigr).\label{eq94f}
\end{eqnarray}
From equation (F.14) we obtain the following expression for $\bftau_\alpha$ 
\begin{eqnarray}
-\bftau_\alpha=\frac{1}{\i\omega}{\bfC}_{\alpha\beta} \partial_\beta {\bf v}+\frac{1}{\i\omega}{\bfC}_{\alpha 3} \partial_3 {\bf v}+{\bfC}_{\alpha l}\bigl({\bf D}_{1l}{\bf E}_0+{\bf D}_{3l}E_3\bigr)-\frac{1}{\i\omega}{\bfC}_{\alpha l}{\bf h}_l,\label{eq6.19bpegg}
\end{eqnarray}
from which $ \partial_3 {\bf v}$ and $E_3$ need to be eliminated.
Substituting equation (\ref{eqe37}) into equation (\ref{eq6.19bbpe}) yields
\begin{eqnarray}
\partial_3 {\bf v}&=& {\bfC}_{33}^{-1}\Biggl(-\i\omega\bftau_3 -{\bfC}_{3\beta}\partial_\beta {\bf v} -\i\omega{\bfC}_{3l}{\bf D}_{1l}{\bf E}_0+{\bfC}_{3l}{\bf h}_l\label{eqe30}\\
&&-\i\omega(\eepp)^{-1}{\bfC}_{3l}{\bf D}_{3l}\biggl(- (\beepp)^t {\bf E}_0  +\frac{1}{\i\omega}\bfpartiala^t {\bf H}_0+\frac{1}{\i\omega}{\bf D}_{3k}^t\Bigl({\bfU}_{k \beta} \partial_\beta {\bf v} - {\bfU}_{k m}{\bf h}_m\Bigr)
-{\bf D}_{3k}^t{\bfC}_{k 3}{\bfC}_{33}^{-1}\bftau_3+\frac{1}{\i\omega} J_3^e\biggr)\Biggr),\nonumber\\
\partial_3 {\bf v}&=& {\bfC}_{33}^{-1}\Bigl(-\i\omega{\bf I}'\bftau_3
-\i\omega{\bfC}_{3l}{\bf D}_{1l}'{\bf E}_0 -{\bfC}_{3\beta}'\partial_\beta {\bf v} 
-(\eepp)^{-1}{\bfC}_{3l}{\bf D}_{3l}\bfpartiala^t {\bf H}_0
-(\eepp)^{-1}{\bfC}_{3l}{\bf D}_{3l} J_3^e+{\bfC}_{3m}'{\bf h}_m\Bigr),\label{eqe32}
\end{eqnarray}
with
\begin{eqnarray}
{\bf I}'&=&{\bf I}-(\eepp)^{-1}{\bfC}_{3l}{\bf D}_{3l}{\bf D}_{3k}^t{\bfC}_{k 3}{\bfC}_{33}^{-1}\\
{\bfC}_{3m}'&=&{\bfC}_{3m}+ (\eepp)^{-1}{\bfC}_{3l}{\bf D}_{3l}{\bf D}_{3k}^t{\bfU}_{k m},\quad\mbox{for}\quad m=\beta(=1,2), 3,\\
{\bf D}_{1l}' &=&{\bf D}_{1l}  - (\eepp)^{-1}{\bf D}_{3l} (\beepp)^t. 
\end{eqnarray}
Substituting equations (\ref{eqe37}) and (\ref{eqe32}) into equation (\ref{eq6.19bpegg}) gives
\begin{eqnarray}
-\bftau_\alpha&=&\frac{1}{\i\omega}{\bfC}_{\alpha \beta} \partial_\beta {\bf v}
+\frac{1}{\i\omega}{\bfC}_{\alpha 3} {\bfC}_{33}^{-1}\Bigl(-\i\omega{\bf I}'\bftau_3
-\i\omega{\bfC}_{3l}{\bf D}_{1l}'{\bf E}_0 -{\bfC}_{3\beta}'\partial_\beta {\bf v} 
-(\eepp)^{-1}{\bfC}_{3l}{\bf D}_{3l}\bfpartiala^t {\bf H}_0\label{eqe42}\\
&&-(\eepp)^{-1}{\bfC}_{3l}{\bf D}_{3l} J_3^e+{\bfC}_{3m}'{\bf h}_m\Bigr) +{\bfC}_{\alpha l}{\bf D}_{1l}{\bf E}_0\nonumber\\
&&+(\eepp)^{-1}{\bfC}_{\alpha l}{\bf D}_{3l}\biggl(- (\beepp)^t {\bf E}_0  +\frac{1}{\i\omega}\bfpartiala^t {\bf H}_0+\frac{1}{\i\omega}{\bf D}_{3k}^t\Bigl({\bfU}_{k \beta} \partial_\beta {\bf v} - {\bfU}_{k m}{\bf h}_m\Bigr)
-{\bf D}_{3k}^t{\bfC}_{k 3}{\bfC}_{33}^{-1}\bftau_3+\frac{1}{\i\omega} J_3^e\biggr)- \frac{1}{\i\omega}{\bfC}_{\alpha m}{\bf h}_m,\nonumber\\
-\bftau_\alpha&=& - {\bfS}_\alpha\bftau_3 - {\bfT}_\alpha{\bf E}_0 + \frac{1}{\i\omega}{\bfU}_{\alpha \beta}' \partial_\beta {\bf v}  - \frac{1}{\i\omega}{\bfV}_\alpha\bfpartiala^t {\bf H}_0
-\frac{1}{\i\omega}{\bfU}_{\alpha m}'{\bf h}_m -\frac{1}{\i\omega}{\bfV}_\alpha J_3^e,\label{eqe43}
\end{eqnarray}
with
\begin{eqnarray}
{\bfS}_\alpha&=& {\bfC}_{\alpha 3}{\bfC}_{33}^{-1}{\bf I}'+(\eepp)^{-1}{\bfC}_{\alpha l}{\bf D}_{3l}{\bf D}_{3k}^t{\bfC}_{k 3}{\bfC}_{33}^{-1}=\bigl({\bfC}_{\alpha 3}+(\eepp)^{-1}{\bfU}_{\alpha l}{\bf D}_{3l}{\bf D}_{3k}^t{\bfC}_{k 3}\bigr){\bfC}_{33}^{-1},\\
{\bfT}_\alpha&=&{\bfC}_{\alpha 3} {\bfC}_{33}^{-1}{\bfC}_{3l}{\bf D}_{1l}'-{\bfC}_{\alpha l}{\bf D}_{1l}+(\eepp)^{-1}{\bfC}_{\alpha l}{\bf D}_{3l} (\beepp)^t=
-{\bfU}_{\alpha l}\bigl({\bf D}_{1l}-(\eepp)^{-1}{\bf D}_{3l} (\beepp)^t\bigr), \\
{\bfU}_{\alpha m}'&=&{\bfC}_{\alpha m}-{\bfC}_{\alpha 3}{\bfC}_{33}^{-1}{\bfC}_{3m}'+(\eepp)^{-1}{\bfC}_{\alpha l}{\bf D}_{3l}{\bf D}_{3k}^t{\bfU}_{km} ={\bfU}_{\alpha m}+(\eepp)^{-1}{\bfU}_{\alpha l} {\bf D}_{3l}{\bf D}_{3k}^t {\bfU}_{k m},\\
{\bfV}_\alpha&=&(\eepp)^{-1}{\bfC}_{\alpha 3} {\bfC}_{33}^{-1}{\bfC}_{3l}{\bf D}_{3l}-(\eepp)^{-1}{\bfC}_{\alpha l}{\bf D}_{3l}=-(\eepp)^{-1}{\bfU}_{\alpha l}{\bf D}_{3l},
\end{eqnarray}
where ${\bfU}_{\alpha 3}'={\bf O}$ and
\begin{eqnarray}
({\bfU}_{\alpha\beta}')^t={\bfU}_{\beta\alpha}'.
\end{eqnarray}
We are now ready to eliminate  $\bftau_\alpha$,  $E_3$ and $H_3$ from equations (\ref{eq6.8bbpe}) $-$ (\ref{Aeqee00ape}).
The expression for $\partial_3{\bf v}$, equation (\ref{eqe32}), already has the desired form.
Substituting equation (\ref{eqe43}) into (\ref{eq6.8bbpe}) gives
\begin{eqnarray}
-\partial_3\bftau_3 &=& \i\omega\R{\bf v} + \partial_\alpha\bigl({\bfS}_\alpha\bftau_3 + {\bfT}_\alpha{\bf E}_0 - \frac{1}{\i\omega}{\bfU}_{\alpha \beta}' \partial_\beta {\bf v}  + \frac{1}{\i\omega}{\bfV}_\alpha\bfpartiala^t {\bf H}_0
+\frac{1}{\i\omega}{\bfU}_{\alpha \beta}'{\bf h}_\beta +\frac{1}{\i\omega}{\bfV}_\alpha J_3^e\bigr)+ {\bf f}\label{eqe48}\\
&=&\partial_\alpha{\bfS}_\alpha\bftau_3+\partial_\alpha{\bfT}_\alpha{\bf E}_0 +\bigl(\i\omega\R-\frac{1}{\i\omega}\partial_\alpha{\bfU}_{\alpha \beta}' \partial_\beta \bigr){\bf v} 
+\frac{1}{\i\omega}\partial_\alpha{\bfV}_\alpha\bfpartiala^t {\bf H}_0 +\frac{1}{\i\omega}\partial_\alpha {\bfU}_{\alpha \beta}'{\bf h}_\beta  +\frac{1}{\i\omega}\partial_\alpha {\bfV}_\alpha J_3^e+{\bf f}.\nonumber
\end{eqnarray}
Substituting equations  (\ref{eqe37}) and (\ref{eq94f}) into (\ref{Aeqhh00ape}) yields 
\begin{eqnarray}
\partial_3 {\bf E}_0 &=&\i\omega\bfmu_1 {\bf H}_0 +\i\omega \bfmu_3\mu_{33}^{-1}\bigl(- \bfmu_3^t {\bf H}_0  +\frac{1}{\i\omega}\bfpartialb^t {\bf E}_0+\frac{1}{\i\omega} J_3^m\bigr)-  {\bf J}_0^m\label{eqe50}\\
&&+\bfpartiala\Biggl((\eepp)^{-1}\biggl(- (\beepp)^t {\bf E}_0  +\frac{1}{\i\omega}\bfpartiala^t {\bf H}_0+\frac{1}{\i\omega}{\bf D}_{3\alpha}^t\Bigl({\bfU}_{\alpha \beta} \partial_\beta {\bf v} - {\bfU}_{\alpha \beta}{\bf h}_\beta\Bigr)
-{\bf D}_{3k}^t{\bfC}_{k 3}{\bfC}_{33}^{-1}\bftau_3+\frac{1}{\i\omega} J_3^e\biggr)\Biggr)\nonumber\\
&=&-\bfpartiala(\eepp)^{-1}{\bf D}_{3k}^t{\bfC}_{k 3}{\bfC}_{33}^{-1}\bftau_3+\bigl(\bfmu_3\mu_{33}^{-1}\bfpartialb^t - \bfpartiala(\eepp)^{-1}(\beepp)^t \bigr){\bf E}_0
+\frac{1}{\i\omega}\bfpartiala(\eepp)^{-1}{\bf D}_{3\alpha}^t{\bfU}_{\alpha \beta} \partial_\beta {\bf v}\nonumber\\
&&+\Bigl(\i\omega\bigl(\bfmu_1- \bfmu_3\mu_{33}^{-1}\bfmu_3^t\bigr)+\frac{1}{\i\omega}\bfpartiala(\eepp)^{-1}\bfpartiala^t\Bigr) {\bf H}_0
+ \bfmu_3\mu_{33}^{-1} J_3^m-  {\bf J}_0^m-\frac{1}{\i\omega}\bfpartiala(\eepp)^{-1}{\bf D}_{3\alpha}^t{\bfU}_{\alpha \beta} {\bf h}_\beta+\frac{1}{\i\omega}\bfpartiala(\eepp)^{-1} J_3^e.\nonumber
\end{eqnarray}
Substituting equations  (F.14), (\ref{eqe37}), (\ref{eq94f}) and (\ref{eqe32})   into equation (\ref{Aeqee00ape}) gives
\begin{eqnarray}
 \partial_3 {\bf H}_0&=&\i\omega \bfcalE_1  {\bf E}_0+
 \i\omega \bfcalE_3 E_3
 +\bfpartialb \Bigl(\mu_{33}^{-1}\bigl(- \bfmu_3^t {\bf H}_0  +\frac{1}{\i\omega}\bfpartialb^t {\bf E}_0+\frac{1}{\i\omega} J_3^m\bigr)\Bigr)
 \nonumber\\
&& 
 -{\bf D}_{1k}^t\Bigl({\bfC}_{k\beta} \partial_\beta {\bf v}+{\bfC}_{k3} \partial_3 {\bf v}+\i\omega{\bfC}_{kl}\bigl({\bf D}_{1l}{\bf E}_0+{\bf D}_{3l}E_3\bigr)- {\bfC}_{kl}{\bf h}_l\Bigr) 
  - {\bf J}_0^e\nonumber\\
 &=&\i\omega \bfcalE_1  {\bf E}_0 + \i\omega(\eepp)^{-1} \bigl(\bfcalE_3-{\bf D}_{1m}^t{\bfC}_{ml}{\bf D}_{3l}\bigr) \biggl(- (\beepp)^t {\bf E}_0  +\frac{1}{\i\omega}\bfpartiala^t {\bf H}_0+\frac{1}{\i\omega}{\bf D}_{3k}^t\Bigl({\bfU}_{k \beta} \partial_\beta {\bf v} - {\bfU}_{k l}{\bf h}_l\Bigr)
\nonumber\\&&-{\bf D}_{3k}^t{\bfC}_{k 3}{\bfC}_{33}^{-1}\bftau_3+\frac{1}{\i\omega} J_3^e\biggr)
 +\bfpartialb \Bigl(\mu_{33}^{-1}\bigl(- \bfmu_3^t {\bf H}_0  +\frac{1}{\i\omega}\bfpartialb^t {\bf E}_0+\frac{1}{\i\omega} J_3^m\bigr)\Bigr)\nonumber\\
&&  -{\bf D}_{1k}^t\biggl({\bfC}_{k\beta} \partial_\beta {\bf v}+{\bfC}_{k3}{\bfC}_{33}^{-1}\bigl(-\i\omega{\bf I}'\bftau_3
-\i\omega{\bfC}_{3l}{\bf D}_{1l}'{\bf E}_0 -{\bfC}_{3\beta}'\partial_\beta {\bf v} 
-(\eepp)^{-1}{\bfC}_{3l}{\bf D}_{3l}\bfpartiala^t {\bf H}_0\nonumber\\
&&-(\eepp)^{-1}{\bfC}_{3l}{\bf D}_{3l} J_3^e+{\bfC}_{3m}'{\bf h}_m  \bigr)+\i\omega{\bfC}_{kl}{\bf D}_{1l}{\bf E}_0- {\bfC}_{kl}{\bf h}_l\biggr) 
  - {\bf J}_0^e\nonumber\\
 &=& \i\omega\bigl({\bf D}_{1k}^t{\bfC}_{k3}{\bfC}_{33}^{-1}{\bf I}' -(\eepp)^{-1} \beep {\bf D}_{3k}^t{\bfC}_{k 3}{\bfC}_{33}^{-1}\bigr) \bftau_3  
 +\Bigl(\i\omega\bigl(\beepo - (\eepp)^{-1} \beep (\beepp)^t \bigr)+\frac{1}{\i\omega}\bfpartialb \mu_{33}^{-1}\bfpartialb^t\Bigr)  {\bf E}_0 \nonumber \\
 &&+\bigl((\eepp)^{-1} \beep{\bf D}_{3k}^t{\bfU}_{k \beta}  - {\bf D}_{1k}^t({\bfC}_{k \beta}-{\bfC}_{k3}{\bfC}_{33}^{-1}{\bfC}_{3 \beta}') \bigr)\partial_\beta{\bf v} 
 +\Bigl((\eepp)^{-1} \beepp\bfpartiala^t -\bfpartialb\mu_{33}^{-1}\bfmu_3^t \Bigr) {\bf H}_0   \nonumber\\
 &&-\bigl((\eepp)^{-1} \beep{\bf D}_{3k}^t{\bfU}_{km}  - {\bf D}_{1k}^t({\bfC}_{km}-{\bfC}_{k3}{\bfC}_{33}^{-1}{\bfC}_{3m}') \bigr){\bf h}_m 
 + (\eepp)^{-1} \beepp J_3^e+\frac{1}{\i\omega}\bfpartialb\mu_{33}^{-1}J_3^m- {\bf J}_0^e, \label{Aeqee00apea}
\end{eqnarray}
with
\begin{eqnarray}
\beepo&=&\bfcalE_1 -{\bf D}_{1k}^t({\bfC}_{k l} {\bf D}_{1l}-{\bfC}_{k3}{\bfC}_{33}^{-1}{\bfC}_{3 l} {\bf D}_{1l}')=\bfcalE_1 -{\bf D}_{1k}^t{\bfU}_{k l} {\bf D}_{1l}- (\eepp)^{-1}{\bf D}_{1k}^t{\bfC}_{k3}{\bfC}_{33}^{-1}{\bfC}_{3 l}{\bf D}_{3l} (\beepp)^t.
\end{eqnarray}
Using the following relations 
\begin{eqnarray}
{\bf D}_{1k}^t{\bfC}_{k3}{\bfC}_{33}^{-1}{\bf I}' -(\eepp)^{-1} \beep {\bf D}_{3k}^t{\bfC}_{k 3}{\bfC}_{33}^{-1}&=&
\bigl({\bf D}_{1k}^t-(\eepp)^{-1} \beepp {\bf D}_{3k}^t\bigr){\bfC}_{k 3}{\bfC}_{33}^{-1},\\
\beepo - (\eepp)^{-1} \beep (\beepp)^t&=&\bfcalE_1 -{\bf D}_{1k}^t{\bfU}_{k l} {\bf D}_{1l}-(\eepp)^{-1}\beepp(\beepp)^t,\\
(\eepp)^{-1} \beep{\bf D}_{3k}^t{\bfU}_{k m}  - {\bf D}_{1k}^t({\bfC}_{k m}-{\bfC}_{k3}{\bfC}_{33}^{-1}{\bfC}_{3 m}') &=&-\bigl({\bf D}_{1k}^t-(\eepp)^{-1} \beepp{\bf D}_{3k}^t\bigr){\bfU}_{k m},
\end{eqnarray}
equation (\ref{Aeqee00apea}) can be rewritten as
\begin{eqnarray}
 \partial_3 {\bf H}_0&=&\i\omega\bigl({\bf D}_{1k}^t-(\eepp)^{-1} \beepp {\bf D}_{3k}^t\bigr){\bfC}_{k 3}{\bfC}_{33}^{-1}\bftau_3
  + \Bigl(\i\omega\bigl(\bfcalE_1 -{\bf D}_{1\alpha}^t{\bfU}_{\alpha \beta} {\bf D}_{1\beta}-\beepp(\eepp)^{-1}(\beepp)^t\bigr)+\frac{1}{\i\omega}\bfpartialb \mu_{33}^{-1}\bfpartialb^t\Bigr) {\bf E}_0\nonumber\\
  &&-\bigl({\bf D}_{1\alpha}^t-(\eepp)^{-1} \beepp{\bf D}_{3\alpha}^t\bigr){\bfU}_{\alpha \beta}\partial_\beta{\bf v} 
  +\Bigl((\eepp)^{-1} \beepp\bfpartiala^t -\bfpartialb\mu_{33}^{-1}\bfmu_3^t \Bigr) {\bf H}_0\nonumber\\
&&+\bigl({\bf D}_{1\alpha}^t-(\eepp)^{-1} \beepp{\bf D}_{3\alpha}^t\bigr){\bfU}_{\alpha \beta}{\bf h}_\beta   
+ (\eepp)^{-1} \beepp J_3^e+\frac{1}{\i\omega}\bfpartialb\mu_{33}^{-1}J_3^m- {\bf J}_0^e.  \label{eqe55}
\end{eqnarray}
Equations  (\ref{eqe48}), (\ref{eqe50}), (\ref{eqe32}) and (\ref{eqe55}) can be cast in the form of matrix-vector wave equation  (1) in the main paper,
with the wave vector $\bq =\bq ({\bf x},\omega)$ and 
 the source vector $\bd =\bd ({\bf x},\omega)$ defined as
\begin{eqnarray}\label{eq6mvpe}
\bq  = \begin{pmatrix}- \bftau_3 \\ {\bf E}_0 \\ {\bf v}\\  {\bf H}_0 \end{pmatrix},\quad
  \bd  = \begin{pmatrix} 
{\bf f}+\frac{1}{\i\omega}\partial_\alpha( {\bfU}_{\alpha \beta}'{\bf h}_\beta)  - \frac{1}{\i\omega}\partial_\alpha \bigl((\eepp)^{-1}{\bfU}_{\alpha \beta}{\bf D}_{3\beta} J_3^e\bigr)\\
\frac{1}{\i\omega}\bfpartiala\bigl((\eepp)^{-1}(J_3^e-{\bf D}_{3\alpha}^t{\bfU}_{\alpha\beta} {\bf h}_\beta)\bigr)-  {\bf J}_0^m+\bfmu_3\mu_{33}^{-1} J_3^m\\
{\bfC}_{33}^{-1}\bigl({\bfC}_{3m}'{\bf h}_m-(\eepp)^{-1}{\bfC}_{3l}{\bf D}_{3l} J_3^e\bigr)\\
- {\bf J}_0^e   
+ (\eepp)^{-1} \beepp J_3^e+\frac{1}{\i\omega}\bfpartialb(\mu_{33}^{-1}J_3^m)+ ({\bf D}_{1\alpha}')^t{\bfU}_{\alpha\beta}{\bf h}_\beta
\end{pmatrix}
\end{eqnarray}
and the  operator matrix $\bfAhat=\bfAhat({\bf x},\omega)$ having the form defined in equation  (2) in the main paper, with
\begin{eqnarray}
\bfAhat_{11} &=& \begin{pmatrix}
\bfAhat_{11}^{11} & \bfahat_{11}^{12}\\
\bfahat_{11}^{21} & \bfahat_{11}^{22}\end{pmatrix},\quad
\bfAhat_{12} = \begin{pmatrix}
\bfAhat_{12}^{11} & \bfahat_{12}^{12}\\
\bfahat_{12}^{21} & \bfahat_{12}^{22}\end{pmatrix},\\
\bfAhat_{21} &=& \begin{pmatrix}
\bfAhat_{21}^{11} & \bfahat_{21}^{12}\\
\bfahat_{21}^{21} & \bfahat_{21}^{22}\end{pmatrix},\quad
\bfAhat_{22} = \begin{pmatrix}
\bfAhat_{22}^{11} & \bfahat_{22}^{12}\\
\bfahat_{22}^{21} & \bfahat_{22}^{22}\end{pmatrix},
\end{eqnarray}
where
\begin{eqnarray}
\bfAhat_{11}^{11}&=&-\partial_\alpha{\bfS}_\alpha=
-\partial_\alpha\bigl({\bfC}_{\alpha 3}+(\eepp)^{-1}{\bfU}_{\alpha \beta}{\bf D}_{3\beta}{\bf D}_{3k}^t{\bfC}_{k 3}\bigr){\bfC}_{33}^{-1},\\
\bfahat_{11}^{12}&=&\partial_\alpha{\bfT}_\alpha=-\partial_\alpha{\bfU}_{\alpha \beta}\bigl({\bf D}_{1\beta}-(\eepp)^{-1}{\bf D}_{3\beta} (\beepp)^t\bigr),\\
\bfahat_{11}^{21}&=&\bfpartiala(\eepp)^{-1}{\bf D}_{3k}^t{\bfC}_{k 3}{\bfC}_{33}^{-1},\\
\bfAhat_{11}^{22}&=&\bfmu_3\mu_{33}^{-1}\bfpartialb^t -\bfpartiala (\eepp)^{-1}(\beepp)^t ,\\
%
\bfAhat_{12}^{11}&=&\i\omega\R-\frac{1}{\i\omega}\partial_\alpha{\bfU}_{\alpha \beta}' \partial_\beta, \\
\bfahat_{12}^{12}&=&\frac{1}{\i\omega}\partial_\alpha{\bfV}_\alpha\bfpartiala^t=-\frac{1}{\i\omega}\partial_\alpha(\eepp)^{-1}{\bfU}_{\alpha \beta}{\bf D}_{3\beta}\bfpartiala^t,\\
\bfahat_{12}^{21}&=&\frac{1}{\i\omega}\bfpartiala(\eepp)^{-1}{\bf D}_{3\alpha}^t{\bfU}_{\alpha \beta} \partial_\beta,\\
\bfAhat_{12}^{22}&=&\i\omega\bigl(\bfmu_1- \bfmu_3\mu_{33}^{-1}\bfmu_3^t\bigr)+\frac{1}{\i\omega}\bfpartiala(\eepp)^{-1}\bfpartiala^t,\\
%
\bfAhat_{21}^{11}&=&\i\omega \bigl({\bfC}_{33}^{-1} - (\eepp)^{-1} {\bfC}_{33}^{-1}{\bfC}_{3l}{\bf D}_{3l}{\bf D}_{3k}^t{\bfC}_{k 3}{\bfC}_{33}^{-1} \bigr),\\
\bfahat_{21}^{12}&=&-\i\omega {\bfC}_{33}^{-1}{\bfC}_{3l}{\bf D}_{1l}' =-\i\omega {\bfC}_{33}^{-1}{\bfC}_{3l}\bigl({\bf D}_{1l}  - (\eepp)^{-1}{\bf D}_{3l} (\beepp)^t\bigr),\\
\bfahat_{21}^{21}&=&-\i\omega({\bf D}_{1k}')^t{\bfC}_{k 3}{\bfC}_{33}^{-1}=-\i\omega\bigl({\bf D}_{1k}^t-(\eepp)^{-1} \beepp {\bf D}_{3k}^t\bigr){\bfC}_{k 3}{\bfC}_{33}^{-1},\\
\bfAhat_{21}^{22}&=&\i\omega\bigl(\bfcalE_1 -\beepp(\eepp)^{-1}(\beepp)^t-{\bf D}_{1\alpha}^t{\bfU}_{\alpha \beta} {\bf D}_{1\beta}\bigr)+\frac{1}{\i\omega}\bfpartialb\mu_{33}^{-1}\bfpartialb^t,\\
%
\bfAhat_{22}^{11}&=& - {\bfC}_{33}^{-1}{\bfC}_{3\beta}'\partial_\beta=- {\bfC}_{33}^{-1}\bigl({\bfC}_{3\beta}+ (\eepp)^{-1}{\bfC}_{3l}{\bf D}_{3l}{\bf D}_{3\alpha}^t{\bfU}_{\alpha \beta}\bigr)\partial_\beta, \\
\bfahat_{22}^{12}&=& - (\eepp)^{-1}{\bfC}_{33}^{-1}{\bfC}_{3l}{\bf D}_{3l}\bfpartiala^t,\\
\bfahat_{22}^{21}&=&-\bigl({\bf D}_{1\alpha}^t-(\eepp)^{-1} \beepp{\bf D}_{3\alpha}^t\bigr){\bfU}_{\alpha \beta}\partial_\beta,\\
\bfAhat_{22}^{22}&=&(\eepp)^{-1} \beepp\bfpartiala^t -\bfpartialb\mu_{33}^{-1}\bfmu_3^t.\label{eq232}  
\end{eqnarray}
%


\section{Seismoelectric waves}

Equations (G.18) $-$ (G.25) in Appendix G in the main paper form the starting point for deriving a seismoelectric
matrix-vector wave equation 
for the quantities $\bftau_3^b$, $p^f$, ${\bf E}_0$,  ${\bf v}^s$, $w_3$ and ${\bf H}_0$ in vector $\bq$. 
We separate the derivatives in the $x_3$-direction from the lateral
derivatives, according to
\begin{eqnarray}
-\partial_3\bftau_3^b &=&
\i\omega \rho^b{\bf v}^s
+\i\omega \rho^f(\bfdelta_\alpha w_\alpha+\bfdelta_3 w_3)
+\partial_\alpha\bftau_\alpha^b+
{\bf f}^b,
\label{Eeq6.8bbv}\\
%
\partial_3  p^f &= &
\i\omega \rho^f\bfdelta_3^t{\bf v}^s-\frac{\eta}{k}\bigl(w_3-{L} E_3\bigr)+ f_3^f,\label{Eeq5.8mv}\\
%
\partial_3 {\bf E}_0 &=&\i\omega
 \mu {\bf H}_0 +
\bfpartiala E_3-
 {\bf J}_0^m,\label{eqhh00}\\
%
\partial_3 {\bf v}^s&=&-  {\bfC}_{33}^{-1}\Bigl(\i\omega\bftau_3^b
+\i\omega\frac{ C}{ M}\bfdelta_3 p^f + {\bfC}_{3\beta}\partial_\beta{\bf v}^s
- {\bfC}_{3l}\hh_l
\Bigr),\label{Eeq6.19bbv}
\end{eqnarray}
\begin{eqnarray}
\partial_3  w_3& =&  \frac{\i\omega}{ M}p - \frac{ C}{ M}\bigl( \bfdelta_\beta^t \partial_\beta{\bf v}^s+ \bfdelta_3^t \partial_3{\bf v}^s\bigr)
-\partial_\beta  w_\beta+\frac{C}{M}\bfdelta_l^t \hh_l+\qf,\label{Eeq5.14mv}\\
%
%
 \partial_3 {\bf H}_0&=&
\i\omega {\cal E}  {\bf E}_0 -\frac{\eta}{k}{L}\bfgamma_\alpha w_\alpha
+\bfpartialb  H_3-
{\bf J}_0^e,\label{eqee00}
\end{eqnarray}
with $2\times 1$ unit vector $(\bfgamma_\alpha)_\beta=\delta_{\alpha\beta}$. The field components  $\bftau_\alpha^b$,  $w_\alpha$, $E_3$ and $H_3$ need to be eliminated.
Using equations (G.19) and (G.20),
we can eliminate the terms $\bfdelta_\alpha w_\alpha$ and
$\partial_\alpha\bftau_\alpha^b$ from
equation (\ref{Eeq6.8bbv}), yielding
\begin{eqnarray}\label{Eeq6.8bcv}
-\partial_3\bftau_3^b &=&
\i\omega \rho^b{\bf v}^s
-\i\omega \rho^f\bfdelta_\alpha\Bigl( -\i\omega \rho^f\frac{k}{\eta}\bfdelta_\alpha^t{\bf v}^s +\frac{k}{\eta}\partial_\alpha{p^f} -{L}\bfgamma_\alpha^t{\bf E}_0\Bigr)
+\i\omega \rho^f\bfdelta_3 w_3
\nonumber\\
&&-\frac{1}{\i\omega}\partial_\alpha \Bigl(
 {\bfC}_{\alpha\beta}\partial_\beta {\bf v}^s + {\bfC}_{\alpha 3}\partial_3 {\bf v}^s
+\i\omega\frac{ C}{ M}\bfdelta_\alpha p^f\Bigr) + {\bf f}^b+\i\omega \rho^f\frac{ k}{\eta}\bfdelta_\alpha 
f_\alpha^f+\frac{1}{\i\omega}\partial_\alpha\bigl({\bfC}_{\alpha l}\hh_l\bigr),
\end{eqnarray}
or, upon substitution of equation (\ref{Eeq6.19bbv}),
\begin{eqnarray}\label{Eeq6.8bdv}
-\partial_3\bftau_3^b &=&
\partial_\alpha\bigl( {\bfC}_{\alpha 3}
 {\bfC}_{33}^{-1}\bftau_3^b\bigr)
-\i\omega \rho^f\frac{ k}{\eta}
\bfdelta_\alpha\partial_\alpha{p^f}
-\frac{1}{\i\omega}\partial_\alpha\Bigl(
\frac{\i\omega}{ M} {\bfu}_\alpha p^f + {\bfU}_{\alpha\beta}\partial_\beta {\bf v}^s
\Bigr)\nonumber\\
&&+\i\omega\Bigl( \rho^b{\bf I}_3
+\i\omega( \rho^f)^2\frac{k}{\eta}\bfdelta_\alpha\bfdelta_\alpha^t\Bigr) {\bf v}^s
+\i\omega \rho^f\bfdelta_3 w_3 +\i\omega \rho^f{L}\bfdelta_\alpha\bfgamma_\alpha^t{\bf E}_0 
+ {\bf f}^b+\i\omega \rho^f\frac{k}{\eta}\bfdelta_\alpha f_\alpha^f+\frac{1}{\i\omega}\partial_\alpha\bigl({\bfU}_{\alpha l}\hh_l\bigr),
\end{eqnarray}
with  ${\bf I}_3$ being a $3\times 3$ identity matrix and
\begin{eqnarray}
 {\bfU}_{\alpha l}&=& {\bfC}_{\alpha l}- {\bfC}_{\alpha 3} {\bfC}_{33}^{-1} {\bfC}_{3l},\\
 {\bfu}_{l}&=&C(\bfdelta_l-  {\bfC}_{l3} {\bfC}_{33}^{-1}\bfdelta_3),
\end{eqnarray}
where ${\bfU}_{\alpha 3}={\bf O}$,  ${\bfu}_{3}={\bf 0}$ and
\begin{eqnarray}\label{Esymstif3v}
 {\bfU}_{\alpha\beta}^t= {\bfU}_{\beta\alpha}
\end{eqnarray}
on account of $ {\bfC}_{jl}^t= {\bfC}_{lj}$.
Using equation (G.23), we eliminate $ E_3$ from equation
(\ref{Eeq5.8mv}), yielding
\begin{eqnarray}\label{Eeq5.8mvb}
\partial_3  p^f &=&
\i\omega \rho^f\bfdelta_3^t{\bf v}^s-\frac{\eta}{ k}
\Bigl(1-\frac{1}{\i\omega{\cal E}}\frac{\eta}{ k}{L}^2\Bigr)
 w_3
+\frac{1}{\i\omega{\cal E}}\frac{\eta}{ k}{L}
\bfpartiala^t
 {\bf H}_0
+\frac{1}{\i\omega{\cal E}}\frac{\eta}{ k}{L}
J_3^e+ f_3^f.
\end{eqnarray}
Using equation (G.23), we eliminate $ E_3$ from
equation (\ref{eqhh00}), yielding
\begin{eqnarray}\label{eqhh001}
\partial_3 {\bf E}_0 &=&
\bfpartiala\Bigl(\frac{1}{\i\omega{\cal E}} \frac{\eta}{ k}{L} w_3\Bigr)
+\i\omega \mu {\bf H}_0 +
\bfpartiala\Bigl(\frac{1}{\i\omega{\cal E}}
\bfpartiala^t {\bf H}_0\Bigr)
- {\bf J}_0^m
+\bfpartiala\Bigl(\frac{1}{\i\omega{\cal E}} J_3^e\Bigr).
\end{eqnarray}
Using equations (\ref{Eeq6.19bbv}) and (G.19), we
eliminate  the terms $\partial_3{\bf v}^s$ and $\partial_\beta{w}_\beta$ from
equation (\ref{Eeq5.14mv}), according to
\begin{eqnarray}\label{Eeq5.14mx}
\partial_3  w_3 &=&
\frac{ C}{ M}\bfdelta_3^t {\bfC}_{33}^{-1}\Bigl(\i\omega\bftau_3^b
+\i\omega\frac{ C}{ M}\bfdelta_3 p^f \Bigr)+ \frac{\i\omega}{ M} p^f+\partial_\beta\Bigl( \frac{k}{\eta}\partial_\beta{p^f} -{L}\bfgamma_\beta^t{\bf E}_0
-\i\omega \rho^f\frac{k}{\eta}\bfdelta_\beta^t{\bf v}^s \Bigr)\nonumber\\
&&  - \frac{ 1}{ M} {\bfu}_\beta^t \partial_\beta{\bf v}^s
+\qf+\frac{1}{M}{\bfu}_\alpha^t\hh_\alpha
-\partial_\beta\Bigl( \frac{ k}{\eta} f_\beta^f\Bigr).
\end{eqnarray}
Using equations (G.19) and (G.25), we eliminate
 $ w_\alpha$ and $ H_3$ from equation (\ref{eqee00}),
yielding
\begin{eqnarray}\label{eqee001}
 \partial_3 {\bf H}_0&=&
 {L}\bfgamma_\alpha 
\Bigl(  \partial_\alpha{p^f}
-\frac{\eta}{ k}{L}\bfgamma_\alpha^t{\bf E}_0 -\i\omega \rho^f\bfdelta_\alpha^t{\bf v}^s \Bigr)
+\i\omega {\cal E}  {\bf E}_0
+\bfpartialb \Bigl(\frac{1}{\i\omega \mu}\bfpartialb^t
 {\bf E}_0\Bigr) - {\bf J}_0^e +\bfpartialb
\Bigl(\frac{1}{\i\omega \mu} J_3^m\Bigr)-{L}\bfgamma_\alpha f_\alpha^f.
\end{eqnarray}
Equations (\ref{Eeq6.8bdv}), (\ref{Eeq5.8mvb}), (\ref{eqhh001}),  (\ref{Eeq6.19bbv}),
 (\ref{Eeq5.14mx}) and (\ref{eqee001}) can be cast in the form of matrix-vector wave equation  (1) in the main paper, 
 with the wave vector $\bq=\bq(\bx,\omega)$ 
 and the source vector $\bd =\bd ({\bf x},\omega)$  defined as
\begin{eqnarray}\label{Eeq6mv}
\bq  = \begin{pmatrix}- \bftau_3^b \\ p^f \\ {\bf E}_0\\
{\bf v}^s\\ w_3 \\ {\bf H}_0
\end{pmatrix},\quad
  \bd  = \begin{pmatrix}
 {\bf f}^b+\i\omega \rho^f\frac{k}{\eta}\bfdelta_\alpha f_\alpha^f+
\frac{1}{\i\omega}\partial_\alpha\bigl({\bfU}_{\alpha \beta}\hh_\beta\bigr)\\
\frac{1}{\i\omega{\cal E}}\frac{\eta}{ k}{L}
J_3^e
+ f_3^f\\
- {\bf J}_0^m+
\bfpartiala\bigl(\frac{1}{\i\omega{\cal E}}
 J_3^e\bigr) \\
 {\bfC}_{33}^{-1}{\bfC}_{3l}\hh_l\\
\qf+\frac{1}{M}{\bfu}_\alpha^t\hh_\alpha-\partial_\beta\bigl( \frac{ k}{\eta} f_\beta^f\bigr)\\
- {\bf J}_0^e +\bfpartialb
\bigl(\frac{1}{\i\omega \mu} J_3^m\bigr)-{L}\bfgamma_\alpha
f_\alpha^f
\end{pmatrix}
\end{eqnarray}
and the operator matrix $\bfAhat=\bfAhat({\bf x},\omega)$ having the form of equation  (2) in the main paper, with
\begin{eqnarray}
\bfAhat_{11} &=& \begin{pmatrix}
\bfAhat_{11}^{11} & \bfahat_{11}^{12} & \bfAhat_{11}^{13}\\
\bfahat_{11}^{21} & \bfAhat_{11}^{22} & \bfahat_{11}^{23}\\
\bfAhat_{11}^{31} & \bfahat_{11}^{32} & \bfAhat_{11}^{33}
\end{pmatrix},\quad
\bfAhat_{12} = \begin{pmatrix}
\bfAhat_{12}^{11} & \bfahat_{12}^{12} & \bfAhat_{12}^{13}\\
\bfahat_{12}^{21} & \bfAhat_{12}^{22} & \bfahat_{12}^{23}\\
\bfAhat_{12}^{31} & \bfahat_{12}^{32} & \bfAhat_{12}^{33}
\end{pmatrix},\label{eqS172}\\
\bfAhat_{21} &=& \begin{pmatrix}
\bfAhat_{21}^{11} & \bfahat_{21}^{12} & \bfAhat_{21}^{13}\\
\bfahat_{21}^{21} & \bfAhat_{21}^{22} & \bfahat_{21}^{23}\\
\bfAhat_{21}^{31} & \bfahat_{21}^{32} & \bfAhat_{21}^{33}
\end{pmatrix},\quad
\bfAhat_{22} = \begin{pmatrix}
\bfAhat_{22}^{11} & \bfahat_{22}^{12} & \bfAhat_{22}^{13}\\
\bfahat_{22}^{21} & \bfAhat_{22}^{22} & \bfahat_{22}^{23}\\
\bfAhat_{22}^{31} & \bfahat_{22}^{32} & \bfAhat_{22}^{33}
\end{pmatrix},\label{eqS173}
\end{eqnarray}
where
\begin{eqnarray}
\bfAhat_{11}^{11}&=& -\partial_\alpha {\bfC}_{\alpha 3}
 {\bfC}_{33}^{-1},\\
\bfahat_{11}^{12}&=&-\i\omega \rho^f\frac{k}{\eta}\bfdelta_\alpha \partial_\alpha-
\partial_\alpha\frac{1}{ M}{\bfu}_\alpha,\\
\bfAhat_{11}^{13}&=&\i\omega \rho^f
{L}\bfdelta_\alpha\bfgamma_\alpha^t,\\
%
\bfAhat_{12}^{11}&=&-\frac{1}{\i\omega}\partial_\alpha   {\bfU}_{\alpha\beta}\partial_\beta+\i\omega\Bigl( \rho^b{\bf I}_3
+\i\omega( \rho^f)^2\frac{k}{\eta}\bfdelta_\alpha\bfdelta_\alpha^t\Bigr),\\
%
\bfahat_{12}^{12}&=&\i\omega \rho^f\bfdelta_3,\\
\bfahat_{12}^{21}&=&\i\omega \rho^f\bfdelta_3^t,\\
\bfAhat_{12}^{22}&=&-\frac{\eta}{ k}
\Bigl(1-\frac{1}{\i\omega{\cal E}}\frac{\eta}{ k}{L}^2\Bigr),\\
\bfahat_{12}^{23}&=&\frac{1}{\i\omega{\cal E}}\frac{\eta}{k}{L}
\bfpartiala^t,\\
\bfahat_{12}^{32}&=& \bfpartiala\frac{1}{\i\omega{\cal E}}
\frac{\eta}{ k}{L},\\
\bfAhat_{12}^{33}&=&\i\omega  \mu{\bf I}_2 +
\bfpartiala\frac{1}{\i\omega{\cal E}}
\bfpartiala^t,\\
%
\bfAhat_{21}^{11}&=&\i\omega {\bfC}_{33}^{-1},\\
\bfahat_{21}^{12}&=&-\i\omega\frac{ C}{ M} {\bfC}_{33}^{-1}\bfdelta_3,\\
\bfahat_{21}^{21}&=&-\i\omega\frac{ C}{ M}\bfdelta_3^t {\bfC}_{33}^{-1},\\
\bfAhat_{21}^{22}&=& \i\omega\frac{ C^2}{ M^2}\bfdelta_3^t {\bfC}_{33}^{-1}\bfdelta_3 +\frac{\i\omega}{ M}+\partial_\beta \frac{ k}{\eta}\partial_\beta,\\
%
\bfahat_{21}^{23}&=&-\partial_\beta{L}
\bfgamma_\beta^t,\\
\bfahat_{21}^{32}&=&{L}\bfgamma_\alpha
\partial_\alpha,\\
\bfAhat_{21}^{33}&=&\i\omega {\cal E}{\bf I}_2+
\bfpartialb\frac{1}{\i\omega \mu}\bfpartialb^t
-\frac{\eta}{ k}{L}^2\bfgamma_\alpha\bfgamma_\alpha^t
=\Bigl(\i\omega {\cal E}-\frac{\eta}{ k}{L}^2\Bigr){\bf I}_2+
\bfpartialb\frac{1}{\i\omega \mu}\bfpartialb^t,\\
%
\bfAhat_{22}^{11}&=&- {\bfC}_{33}^{-1}
 {\bfC}_{3\beta}\partial_\beta,\\
\bfahat_{22}^{21}&=& -\i\omega\partial_\beta \rho^f \frac{k}{\eta}\bfdelta_\beta^t-\frac{1}{ M} {\bfu}_\beta^t
\partial_\beta,\\
\bfAhat_{22}^{31}&=&-\i\omega \rho^f{L}\bfgamma_\alpha \bfdelta_\alpha^t,
\end{eqnarray}
where ${\bf I}_2$ is a $2\times 2$ identity matrix. 
The submatrices that have not been listed here are zero.
Using this in equations (\ref{eqS172}) and (\ref{eqS173}), we obtain
%
\begin{eqnarray}
\bfAhat_{11}&=&\begin{pmatrix} 0 & 0 & -\partial_1
\frac{ S}{ K_c} &\frac{ \rho^f}{\rho^E}
\partial_1-\partial_1
\frac{2 C N}{ M K_c} & \i\omega \rho^f{L} & 0\\
0 & 0 & -\partial_2
\frac{ S}{ K_c}& \frac{ \rho^f}{\rho^E}
\partial_2-\partial_2
\frac{2 C N}{ M K_c} & 0 &\i\omega \rho^f{L}\\
-\partial_1 & -\partial_2&0&0&0&0\\
0 & 0& 0 &0&0&0\\
0 & 0& 0 &0&0&0\\
0 & 0& 0 &0&0&0
\end{pmatrix},
\label{eqa11pr}
\end{eqnarray}

\begin{eqnarray}
\bfAhat_{12}&=&\left(\begin{matrix}
\i\omega\bigl( \rho^b-\frac{( \rho^f)^2}{\rho^E}\bigr)
-\frac{1}{\i\omega}\bigl( \partial_1 \nu_1\partial_1+
\partial_2 N\partial_2\bigr)
& -\frac{1}{\i\omega}\bigl(\partial_2  N\partial_1+ \partial_1  \nu_2\partial_2 \bigr) 
\\
-\frac{1}{\i\omega}\bigl(\partial_2  \nu_2\partial_1+ \partial_1  N\partial_2 \bigr)
& \i\omega\bigl( \rho^b-\frac{( \rho^f)^2}{\rho^E}\bigr)-\frac{1}{\i\omega}\bigl(
\partial_1 N\partial_1+ \partial_2 \nu_1\partial_2\bigr) \\
 0 &0 \\
 0 & 0 \\
 0 & 0  \\
 0 & 0 
\end{matrix}\right.\nonumber\\
&&\nonumber\\
%
&&\left.\begin{matrix}
0 & 0&0&0\\
0 & 0&0&0\\
 \i\omega \rho^b & \i\omega \rho^f&0&0 \\
  \i\omega \rho^f & \i\omega\rho^E\bigl(1+
 \frac{\rho^E}{{\cal E}}{L}^2\bigr)&
\frac{\rho^E}{{\cal E}}{L}
\partial_1
&\frac{\rho^E}{{\cal E}}{L}
\partial_2\\
0 &-\partial_1\frac{\rho^E}{{\cal E}}{L}
&\i\omega \mu-\frac{1}{\i\omega}\partial_1
\frac{1}{{\cal E}}\partial_1& -\frac{1}{\i\omega}\partial_1
\frac{1}{{\cal E}}\partial_2\\
0 & -\partial_2\frac{\rho^E}{{\cal E}}{L} &-\frac{1}{\i\omega}\partial_2
\frac{1}{{\cal E}}\partial_1
&\i\omega \mu-\frac{1}{\i\omega}\partial_2
\frac{1}{{\cal E}}\partial_2
\end{matrix}\right),
\label{eqa12pr}\end{eqnarray}

\begin{eqnarray}
\bfAhat_{21}&=&\begin{pmatrix}
\frac{\i\omega}{ N} & 0 & 0 & 0 & 0&0\\
0 & \frac{\i\omega}{ N} & 0 & 0 & 0&0\\
0 & 0 & \frac{\i\omega}{ K_c} & -\frac{\i\omega  C}{ M K_c} & 0&0 \\
0 & 0 & -\frac{\i\omega  C}{ M K_c} &\i\omega\bigl(\frac{  C^2}{ M^2 K_c}
+\frac{1}{ M}\bigr)-\partial_\beta
\frac{1}{\i\omega\rho^E}\partial_\beta & -\partial_1{L}&
-\partial_2{L}\\
0&0&0&{L}\partial_1 & \i\omega  \varepsilon-\sigma
-\frac{1}{\i\omega}\partial_2
\frac{1}{ \mu}\partial_2&
\frac{1}{\i\omega}\partial_2
\frac{1}{ \mu}\partial_1\\
0&0&0&{L}\partial_2 & \frac{1}{\i\omega}\partial_1
\frac{1}{ \mu}\partial_2
&\i\omega  \varepsilon-\sigma
-\frac{1}{\i\omega}\partial_1\frac{1}{ \mu}\partial_1
\end{pmatrix},
\nonumber\\&&
\label{eqa21pr}
\end{eqnarray}

\begin{eqnarray}
\bfAhat_{22}&=&\begin{pmatrix} 0 & 0 &
-\partial_1 & 0 & 0 & 0
\\
0 & 0 & -\partial_2 & 0 & 0 & 0
\\
-\frac{ S}{ K_c}\partial_1 &
-\frac{ S}{ K_c}\partial_2&0 & 0 & 0 & 0\\
\partial_1\frac{ \rho^f}{\rho^E}
-\frac{2 C N}{ M K_c}\partial_1 &
\partial_2\frac{ \rho^f}{\rho^E}
-\frac{2 C N}{ M K_c}\partial_2  & 0 & 0 & 0 & 0\\
-\i\omega \rho^f{L} & 0 & 0 & 0 & 0 & 0 \\
0 & -\i\omega \rho^f{L} & 0 & 0 & 0 & 0
\end{pmatrix},
\label{eqa22pr}
\end{eqnarray}
with $\rho^E$, $\nu_1$, $\nu_2$, $S$ and $K_c$ defined in section \ref{sec3}.

\end{footnotesize}

\end{document}